\documentclass[a4paper,fleqn,usenatbib]{mnras}
\usepackage{mathptmx}
\usepackage{pdflscape}
\usepackage{mathtools,cuted}
\usepackage{ulem}

\usepackage[T1]{fontenc}
\usepackage{ae,aecompl}

\usepackage{graphicx}	
\usepackage{amsmath}	
\usepackage{amssymb}	

\title[Beyond-Newtonian dynamics with Kerr-like primaries]{Beyond-Newtonian dynamics of a planar circular restricted three-body problem with Kerr-like primaries}

\author[De et al.]{
Shounak De,$^{1}$\thanks{E-mail: sd868@cam.ac.uk}
Suparna Roychowdhury,$^{2}$\thanks{E-mail: suparna@sxccal.edu}
and Roopkatha Banerjee$^{2}$
\\
$^{1}$Department of Applied Mathematics and Theoretical Physics, University of Cambridge, Wilberforce Road, Cambridge CB3 0WA, UK\\
$^{2}$Department of Physics, St. Xavier's College, 30 Park Street, Kolkata 700016, India\\
}

\date{Accepted XXX. Received YYY; in original form ZZZ}

\pubyear{2019}

\begin{document}
\label{firstpage}
\pagerange{\pageref{firstpage}--\pageref{lastpage}}
\maketitle

\begin{abstract}
The dynamics of the planar circular restricted three-body problem with Kerr-like primaries in the context of a beyond-Newtonian approximation is studied. The beyond-Newtonian potential is developed by using the Fodor-Hoenselaers-Perj\'es procedure. An expansion in the Kerr potential is performed and terms up-to the first non-Newtonian contribution of both the mass and spin effects are included. With this potential, a model for a test particle of infinitesimal mass orbiting in the equatorial plane of the two primaries is examined. The introduction of a parameter, $\epsilon$, allows examination of the system as it transitions from the Newtonian to the beyond-Newtonian regime. The evolution and stability of the fixed points of the system as a function of the parameter $\epsilon$ is also studied. The dynamics of the particle is studied using the Poincar\'e map of section and the Maximal Lyapunov Exponent as indicators of chaos. Intermediate values of $\epsilon$ seem to be the most chaotic for the two cases of primary mass-ratios ($=0.001,0.5$) examined. The amount of chaos in the system remains higher than the Newtonian system as well as for the planar circular restricted three-body problem with Schwarzschild-like primaries for all non-zero values of $\epsilon$.
\end{abstract}

\begin{keywords}
methods-numerical -- chaos -- gravitation
\end{keywords}



\section{Introduction}

	In the field of modern celestial mechanics and dynamical astronomy, one of the most intriguing and important problems is the dynamics of few bodies, in particular being the case of a circularly restricted three body problem \citep{Szebehely67}. This problem has been applied in various fields in astronomy, like planetary dynamics, galactic and stellar cluster dynamics and even molecular dynamics. Currently, with the advent of LIGO and the detection of gravitational waves from binary black hole mergers \citep{Abbott2016a, Abbott2016b}, the investigation of such systems in strong gravitational fields have become a field of intense research once again. The black holes involved in these discoveries span a mass range of 10 M$_\odot$ to 100 M$_\odot$, and are all consistent to have initially formed from the death of massive stars.

	There is also strong observational evidence that a different class of super massive black holes (SMBHs), with masses ranging from 10$^5$ to 10$^{10}$ M$_\odot$, are residing in almost all centres of galaxies \citep{Beckmann12}. It is expected that some of these SMBHs will pair up as binaries as their host galaxies merge \citep*{Begelman80}. In fact, there is ample evidence of several active galaxies with double nucleus \citep{Komossa03, Muller15}. It is also speculated that the eventual inspiral and merger of some of these SMBH binaries constitutes a prime gravitational wave source for the planned LISA observatory \citep{Amaro-Seoane12}. In addition, there is also increasing evidence that there are Kerr black hole binaries which are merging \citep{Gwak19, Ruffini16, Beckmann12} and are sources of gravitational radiation.
	
	In such binary black hole mergers which also accrete, the investigation of the chaotic dynamics of test particles within accretion discs or inside the halo surrounding these compact objects has become a subject of prime importance \citep*{Levin00, Schnittman01, Cornish02, Cornish03, Hartl05, Gopakumar05, Wu07, Wu08, Wu10, Wu11, Wu15a, Wu15b, Zhong10a, Wang11, Li19, Mei13, Luo17, Huang14a, Huang14c, Huang16}. Some authors have also studied the numerical schemes and techniques which can be used for such non-linear, chaotic problems along with the dynamics of these systems \citep{Zhong10b, Wu15b, Luo17}. Investigations of such dynamics of charged particles moving under the influence of magnetic and strong gravitational fields of a single compact object have already been studied in some detail within the general relativistic framework \citep*{Kopacek10, Kopacek14, Kopacek15, Takahashi09, Kovar08, Kovar10}. Such studies have been extended to the motion of test particles under the influence of the relativistic gravitational field of accreting black holes \citep*{Semerak10, Semerak12, Semerak13, Witzany15, Vogt03} and also for motion under the influence of gravity produced by an extended body \citep*{Letelier97, Vieira99, Wu06a, deCastro11}.
	
	On the other hand, escaping particles from dynamical systems has also been a subject under focus for decades. Especially the issue of escape in Hamiltonian systems is directly related to the problem of chaotic scattering which has been an active field of research over the last decade and it still remains an open area \citep*{Benet98, Benet96, Bleher90, Bleher89, Bleher98, Churchill75, Contoupoulos90, Contoupoulos92, Eckhardt88, Motter02, Ott93, Seoane06}. It is well known that some types of Hamiltonian systems have a finite energy of escape. For lower values of the particle energy, the equipotential surfaces of these systems are closed and therefore escape is impossible. For energies above the escape energy, these surfaces open and exit channels emerge through which  particles can escape to infinity. There is a comprehensive body of work on such ``open'' or ``leaking'' Hamiltonian systems \citep*[e.g.][]{Barrio09, Contopoulos12, Ernst14, Kandrup99, Lai11, Navarro01, Siopis95a, Siopis95b, Siopis96, Zotos14a, Zotos14b, Zotos15a, Zotos15b, Zotos16a, Zotos17a}. However, it is needless to say that this list of citations is neither complete nor exhaustive. It is just indicative of the body of work that has happened in these fields and is still continuing.

The restricted three-body problem (RTBP) is an excellent example of such open Hamiltonian systems with escape \citep*[e.g.][]{Winter94a, Winter94b}. Over the last few decades, a large number of studies have been devoted to the classification of orbits in the RTBP. It all started with the pioneering works of \citet*{Nagler04, Nagler05} where initial conditions of orbits were classified as bounded, escaping or collisional. Moreover, bounded orbits were further classified into orbital families by taking into account the type of motion of the test particle around the primary sources. Such classifications have also been done in the context of planetary systems, Earth-Moon system and Saturn-Titan system \citep*{deAssis14, Zotos16a}. In this context, it is important to mention that a simplified modification of the RTBP is the Hill approximation which focuses on the vicinity of the secondary source \citep*[e.g.][]{Hill86, Petit86, Petit87, Steklain06, Steklain09}. This facilitates  for the study of the motion of test particles in the neighborhood of the Lagrange (equilibrium) points $L_1$ and $L_2$. At this point it should be mentioned that the Hill approximation is valid only when the mass of the secondary is much smaller than the mass of the primary body. One can directly obtain the Hill model from the classical RTBP by translating the origin to the center of the secondary body and also by re-scaling the coordinates suitably. The Hill problem was proved to be non-integrable by \citet*{Meletlidou01}, and is chaotic, as shown by \citet*{Simo00}. Subsequently, thorough numerical investigations of this problem were performed by carrying out a systematic classification of the initial conditions of the orbits \citep*{Zotos17a}. More precisely, the initial conditions of the orbits were classified into four categories: (i) non escaping regular orbits; (ii) trapped chaotic orbits; (iii) escaping orbits; and (iv) collisional orbits. In addition, the issue of equilibrium points in circular restricted three body problem (CRTBP) has also been studied widely and in great detail (see \citet*{Henon97} and references there in). The discovery of the Trojan asteroids around the Lagrangian points L$_4$ and L$_5$ in the Sun-Jupiter system \citep*{Murray99}, and the recent observations of asteroids around L$_4$ for the Sun-Earth system \citep*{Connors11}, has added a great impetus to theoretical studies on the subject. Moreover, the dynamics of non-conservative RTBP have also been investigated extensively, like the case of CRTBP with gravitational radiation \citep*{Schnittman10}, an elliptic restricted three-body problem \citep*{Wang16} and that of a dissipative CRTBP with drag forces \citep*{Wang18}.

	One of the first attempts at studying the relativistic CRTBP under the assumptions of low velocities and weak gravity was made by \citet*{Krefetz67} in the year 1967. He looked at the post-Newtonian equations for the first time using the Einstein-Infeld-Hoffmann (EIH) formalism \citep*{Einstein38}. Since then this problem has been studied by several authors where they have attempted to present the Lagrangian explicitly \citep*{Contopoulos76}. Some authors have also tried to explore the deviations to the Lagrangian points due to the post-Newtonian corrections \citep*{Maindl96}. In addition, analytical solutions were also attempted in the GR regime using the EIH approximation up-to the first order \citep*{Yamada10}. Recently, as one of the first studies of chaotic orbits in the post-Newtonian CRTBP, \citet*{Huang14a} explored the influence of the distance of separation between the two primaries. They observed that if the primary bodies are close enough, the post-Newtonian dynamics is qualitatively quite different, particularly where some Newtonian bounded orbits become unstable.

In more recent studies, several authors have formulated this problem using pseudo-Newtonian potentials developed for non-rotating Schwarzschild-like (Paczy\'{n}sky-Witta potential) \citep*{Paczynski80} and rotating Kerr-like primaries \citep*{Artemova96, Semerak99, Mukhopadhyay02} to avoid the complications of a post-Newtonian formulation. Subsequently, detailed studies of orbits and the dynamics of test particles around a single Schwarzschild primary and a binary system, as well as Kerr like primaries have been made in recent years with the idea of investigating the chaotic and unstable nature of orbits in the relativistic regime. In a very recent study, \citet*{Dubeibe16} used the Fodor-Hoenselaers-Perj\'es (FHP) procedure \citep*{Fodor89} (taking into account the corrections made by \citet*{Sotiriou04}) to derive an approximate potential for the gravitational field of two uncharged spin-less particles modeled as sources with multi-pole moment, $m$. In this work, they have explored the dynamics of a massless test particle using the Poincar\'{e} section and the Lyapunov exponent as indicators of chaos. As they have mentioned, this potential is not ad-hoc as other pseudo-Newtonian potentials but rather it is exactly derived from the multipolar structure of the sources. In our current study, we also follow a similar route and use the FHP procedure to derive the multipolar structure of a spinning binary system. Subsequently, we construct a beyond-Newtonian potential to imitate the gravitational effects of this system on a test particle in the CRTBP scheme.

The paper is organized as follows. In the next section, we present the formulation of the gravitational beyond-Newtonian potential of each Kerr-like source using the FHP procedure. Next, we present the Lagrangian and the equations of motion of a test particle in context to CRTBP. In the subsequent section, we present a detailed analysis of the Hill curves or the zero velocity surfaces as the system makes a gradual transition from the FHP beyond-Newtonian approximation to the classical regime through a parameter $\epsilon$ in the beyond-Newtonian potential. Here we also present a detailed analysis of the orbits and a discussion on the fixed points of this system along with their stability as a function of the parameter $\epsilon$. The classification of the nature of orbits is made using Poincar\'e surfaces of section and the variational method for the calculation of the largest Lyapunov exponent, as done by several previous authors. In the next section, we present a comparison between the dynamics of a test particle around a binary system of Schwarzchild and spinning primaries. Finally, in the last section we conclude with a summary of our main results and present certain new directions that we intend to investigate in the near future.

\section{Formulation of Beyond-Newtonian Potential for Kerr Binary}

\begin{figure}
\includegraphics[scale=0.7]{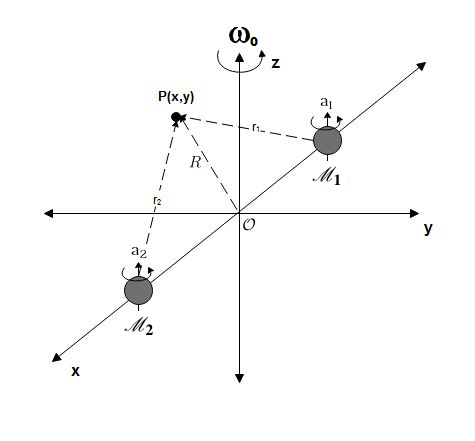}
\caption{The configuration of the two primaries, $\mathbf{\mathcal{M}_1}$ and $\mathbf{\mathcal{M}_2}$, in the centre-of-mass frame which is rotating about the z-axis with angular frequency $\mathbf{\omega_0}$ ($=1$). A test particle with infinitesimal mass $\mathbf{P}$ is placed at an arbitrary position in the equatorial plane.}
\label{diagram}
\end{figure}

The version of CRTBP we consider consists of two massive, spinning primaries with masses $\mathcal{M}_1$ and $\mathcal{M}_2$ and intrinsic angular momenta $a_1$ and $a_2$, at positions $X_1$ and $X_2$, respectively, describing a circular orbit in the $x-y$ plane about their common centre of mass (taken to be the origin $\mathcal{O}$). The centre-to-centre distance remains fixed and remains sufficiently far apart, while the orbital angular velocity is $\omega_0$. The aim is to set up the beyond-Newtonian potential (up to the first non-Newtonian term) for this CRTBP system and consequently write down the Euler-Lagrange equations of motion of a test particle under the influence of this potential. The schematic of the configuration is illustrated in figure (\ref{diagram}). 

To simulate the dynamics of the CRTBP at hand, we employ the Fodor-Hoenselaers-Perj\'es (FHP) procedure to perform an expansion in the mass and rotation potential of each primary up to the first non-Newtonian term. This essentially generates first-order general relativistic effects, the dynamics of which is analyzed at length in the following section. The beyond-Newtonian potential for the system is then constructed by virtue of a superposition of the potentials corresponding to the two primaries, modelled to describe circular orbits around their common centre of mass. We then write down the Lagrangian and consequently the equations of motion for a test particle under the influence of such a potential. 

\subsection{Beyond-Newtonian potential}
\label{sec:PNpot} 

We shall now briefly outline the steps involved in the FHP procedure leading to the construction of the beyond-Newtonian potential for the problem at hand. The FHP algorithm involves the decomposition of the Einstein field equation in the so-called Ernst formalism. In this formalism, the field equations of GR are reduced to a pair of complex equations by virtue of introducing the complex potentials $\zeta$ and $\Psi$. These complex potentials are further defined in terms of two new potentials $\xi$ and $\varsigma$ through the relations
\begin{align}
\zeta = \frac{1 - \xi}{1 + \xi} \,, \quad \Psi = \frac{\varsigma}{1 + \xi} \,.
\label{1.01}
\end{align}
The field potentials satisfy \citep*{Ernst68I, Ernst68II}
\begin{align}
(\xi \xi^* - \varsigma \varsigma^* - 1) \, \nabla^2 \xi = 2 (\xi^* \nabla \xi - \varsigma^* \nabla \varsigma ) \cdot \nabla \xi \,, \label{1.02} \\
(\xi \xi^* - \varsigma \varsigma^* - 1) \, \nabla^2 \varsigma = 2 (\xi^* \nabla \xi - \varsigma^* \nabla \varsigma ) \cdot \nabla \varsigma \,.
\label{1.03}
\end{align} 
The above set of equations are an alternative representation of the Einstein-Maxwell field equations. As a matter of fact, they could be interpreted as the generalization of Laplace's equation for the Papapetrou's metric describing the space-time around a stationary and axisymmetric source
\begin{align}
ds^2 = - F (dt - \omega d\phi)^2 + F^{-1} [e^{2 \gamma}(d \rho^2 + dz^2) + \rho^2 d \phi^2] \,,
\label{1.04}
\end{align}
where the metric coefficients $F$, $\omega$, and $\gamma$ depend only on the Weyl-Papapetrou co-ordinates $\rho$ and $z$. These metric functions can be reformulated in terms of the Ernst complex potentials \citep*{Sotiriou04} $\zeta$ and $\Psi$ and described by the associated Einstein-Maxwell field equations (\ref{1.02}) and (\ref{1.03}). \\
The new set of field potentials $\xi$ and $\varsigma$ are related to the classical gravitational and electromagnetic potentials in the following way
\begin{align}
\xi = \Phi_M + i \, \Phi_J \,, \quad \varsigma = \Phi_E + i \, \Phi_H \,,
\label{1.04}
\end{align}
where $\Phi_M, \Phi_J, \Phi_E,$ and $\Phi_H$ represent the mass, angular momentum, electrostatic and magnetic potentials, respectively. As our massive, spinning primaries do not possess electromagnetic fields, we set $\Phi_E = \Phi_H =0$, which from (\ref{1.01}) implies $\varsigma = \Psi = 0$. The seminal work of \citet*{Geroch70} and \citet*{Hansen74} allows us to determine the multipolar moments of asymptotically flat spacetimes. In this prescription, the induced 3-metric $h_{i j}$ is mapped by virtue of a conformal transformation $h_{i j} \rightarrow \tilde{h}_{i j} = \Omega^2(x) h_{i j}$ onto a conformal metric $\tilde{h}_{i j}$. This conformal factor $\Omega$ satisfies the conditions
\begin{align}
\Omega\big|_{\Lambda} = \tilde{D}_i  \Omega\big|_{\Lambda} = 0,\quad \tilde{D}_i \tilde{D}_j \Omega\big|_{\Lambda} = 2 h_{i j}\big|_{\Lambda}, 
\end{align}
where $\tilde{D}$ denotes the covariant derivative on the induced surface and $\Lambda$ denotes the point added due to conformal compactification. Essentially, $\Omega$ transforms the potential $\xi$ into $\tilde{\xi} = \Omega^{-1/2} \xi$ with the explicit transformation being $\Omega = r^{\prime 2} = \rho^{\prime 2} + z^{\prime 2}$. The relation between the primed and unprimed Weyl-Papapetrou coordinates are 
\begin{align}
\rho^{\prime} = \frac{\rho}{\rho^2 + z^2},\quad z^{\prime} = \frac{z}{\rho^2 + z^2} \,,
\end{align}
with $\phi$ remaining unchanged. This helps in mapping the infinity to the origin of the primed coordinates $(\rho^{\prime}, z^{\prime}) = (0,0)$. Besides, the potential $\tilde{\xi}$ can be expressed as a power series expansion in $\rho^{\prime}$ and $z^{\prime}$ as
\begin{align}
\tilde{\xi} = \sum_{i, j=0}^{\infty} a_{i j} \rho^{\prime\, i} z^{\prime\, j}
\end{align}
with the coefficients $a_{i j}$ determined by recursive relations presented explicitly in \cite{Sotiriou04}. Following this procedure, one can deduce approximate relations for the gravitational potential $\xi$, in terms of the parameters of the primary once its gravitational multiple moments $P_i$ are known. Thus, we apply this outlined prescription to a massive, spinning primary whose multipolar structure we take to be: 
\begin{align}
P_0 = m\,, \quad P_1 = i m a\,, \quad P_i = 0 \,\quad \textrm{for} \,\, i \geq 2 \,,
\end{align}
such that $m$ and $a$ denote the mass and angular momentum of the source, respectively. 

We now aim to set up the beyond-Newtonian potential (up-to the first non-Newtonian term) for the CRTBP system at hand and consequently write down the Euler-Lagrange equations of motion of a test particle under the influence of this potential. For clarity, we restate the conditions and assumptions of the CRTBP model we are trying to construct: 
\begin{itemize}
\item The two primaries, with masses $\mathcal{M}_1$ and $\mathcal{M}_2$ and intrinsic angular momenta $a_1$ and $a_2$, at positions $X_1$ and $X_2$, respectively, describe a circular orbit about their common centre of mass (taken to be the origin $\mathcal{O}$). The centre-to-centre distance $r$ remains fixed and remains sufficiently far apart, while the orbital angular velocity is $\omega_0$.
\item A beyond-Newtonian potential describing the primaries is constructed assuming that the principle of superposition holds: that the total gravitational potential of the system is a linear sum of the mass and rotation potentials (up to first order effects) of the individual sources. 
\item A test particle of mass $\mathcal{M}$, that is very small compared to the primaries, now moves under the effect of this beyond-Newtonian potential in the $z = 0$ orbital plane of the primaries. The motion of this test particle has no effect on the primaries whatsoever. 
\item The convention $G = \mathcal{M} = \omega_0 = r = 1$ is used throughout the analysis hereon (further details on this choice of units has been discussed extensively in section (\ref{sec:dyn})).
\end{itemize}

In accordance with the above conditions and following the preceding discussion on the FHP formalism, we now construct the beyond-Newtonian potential $\Omega$ describing the primaries of our CRTBP model from the reconstructed potential $\xi$ describing a single source. Keeping explicitly the factors of $c$ to show the corresponding order-wise contributions, we have the beyond-Newtonian potential for our system:
\begin{align}
\Omega (x,y) &= -\sum_{i=1}^{2}\frac{\mathcal{M}_i}{r_i} + \frac{1}{2c^4} \sum_{i=1}^2 \frac{\mathcal{M}_i^3}{r_i^3} \nonumber \\
&+ \frac{1}{c^2}\sum_{i=1}^{2} \frac{\mathcal{M}_i a_i}{r_i^2} \cos \theta_i + \frac{1}{2c^4}\sum_{i=1}^2 \frac{\mathcal{M}_i a_i^2}{r_i^3}(3\cos^2 \theta_i - 1)
\label{pnp}
\end{align}
where the primaries are stationed at positions $X_1 = (x_1,0)$ and $X_2 = (x_2,0)$ respectively, and $r_{1,2} = \sqrt{(x - x_{1,2})^2 + y^2}$. We note that the first two terms of equation (\ref{pnp}) describe the mass potential and the next two terms represent the rotation potential of the binary system upto first order corrections respectively. Also, following the FHP procedure, we see the potential that is constructed is written in terms of powers of $1/c^2$. The 1st order corrections to the Newtonian potentials, both in mass as well as for rotation, are retained and the higher order terms are dropped since their contribution is smaller compared to the leading order (by appropriate factors of $1/c^2$). The term `beyond-Newtonian' is designated to these 1st order corrections to the Newtonian potentials that arise in our final form of the potential, as seen in equation~(\ref{pnp}).

Moreover, in order to observe the transition of the system from the Newtonian regime to a beyond-Newtonian one, we introduce a parameter $\epsilon$, such that,\[\frac{1}{c^2}\rightarrow\frac{1}{c^2}\epsilon\]
with $\epsilon\in[0,1]$ using the fact that $\frac{1}{c^2}\rightarrow0$ reduces equation~(\ref{pnp}) to the Newtonian case. That is, the $\epsilon = 0$ classical limit is essentially the Newtonian problem that models non-spinning binaries composed of weak gravitational sources as found in say, planetary systems and binary stars which are not in close contact with each other. On the other hand, the $\epsilon = 1$ beyond-Newtonian case models departures from Newtonian behaviour that can be found in compact spinning binaries constituted of strong gravitational sources, for example black-hole and compact binaries. The parameter $\epsilon$ can thereby be thought of as a knob that slowly ``turns on'' corrections (both in the mass and rotation potentials as seen from equation~(\ref{pnp})) to the Newtonian potential as we gradually go from the classical limit $\epsilon = 0$ to the beyond-Newtonian regime $\epsilon = 1$.

\section{Dynamics of a test particle}\label{sec:dyn}
In order to simplify the numerical simulation of the three dimensional system described in the section above, we confine ourselves to the plane of the two primaries. We adopt a modified version of the Szebehely convention to de-dimensionalize the problem. Numerous types of scaling transformations have had applications in literature \citep*{Huang14a, Huang14b, Su16}. For example, studies of chaotic dynamics of asteroids in planetary systems scale primaries to the solar mass. However, for our problem, the absolute masses of the two primaries are irrelevant and do not reveal any new physical information about the system. Therefore, with $\mathcal{M}_1+\mathcal{M}_2=\mathcal{M}$ and $a_1+a_2=a$, we define a dimensionless mass $\mu_1=\mathcal{M}_2/\mathcal{M}$ and dimensionless spin $\mu_2=a_2/a$. Applying the scaling relations described above, we enforce the sum of the masses of the two primaries and the distance between the two to be unity. This has been enforced by adopting geometrized units, $G=1$ and $c=1$, with distance and time now having the dimension of mass (this choice of units has been discussed in detail in the next paragraph). Additionally, this scaling also ensures that the sum of the spins of the primaries be unity. Thus, applying the above discussed scaling relations we obtain:
\begin{equation}
\begin{array}{lcl}
\mathcal{M}_1=1-\mu_1&;&\mathcal{M}_2=\mu_1 \\
a_1=1-\mu_2&;&a_{2}=\mu_2 \\
\end{array}
\end{equation}

At this point, it is worthwhile to note that different system of units have been used in literature for simplifying the respective problem, both analytically and numerically. The choice of units always mostly depend on the length scales, masses and the time-scales involved. As a result of this, the speed of light $c$ can assume different values. For example, in planetary systems, setting $G=1$, the unit of mass to be the sum of masses, the unit of distance to be the semi-major axis of the secondary body (which is set to unity) and using Kepler's second law, the speed of light assumes different values like $c$ = 22946.5 for the case of Sun-Jupiter, and $c$ = 10065.3 in the case of Sun-Earth \citep*{Lhotka15}. However, while studying the dynamics of test particles around compact objects under the circular restricted three body scheme (CRTBP) in post-Newtonian (PN) treatments \citep{Einstein38}, the speed of light $c$ surfaces as a parameter which measures the order of the PN contributions. For ease in numerical simulations, $c$=1 is later enforced and $a$, which is the separation between the parent bodies, becomes an important parameter for the first post-Newtonian (1-PN) order effect. Thus, this choice of unit and relevant scaling transformations facilitates the study of how the separation between the primaries affect the dynamics of this system \citep*{Huang14a}. Another variation to this post-Newtonian three body scheme was recently studied by \citet*{Dubeibe17} who used $c=10000$ in his calculations. To show this, one can use the Sun-Earth system as an example (for details refer to \citet*{Klavcka08}). It was shown here that this value of $c$, the choice of units and relevant scaling transformations, as opposed to $c=1$ in an earlier work by \citet*{Huang14a}, facilitates a better conservation of the Jacobi integral of motion numerically. This is due to the fact that the contributions of the higher order PN terms vary depending on the formulation and thus a truncation brings about a non-conservation of the Jacobi integral (discussed in detail later). Recently, \citet*{Deng20} used different values of $c$ to indicate perturbations from the PN contribution, which were used to find an optimal method for the calculation of eccentric anomaly.

However, in our study, the Jacobi integral of motion is a constant. Thus, our choice of the value of $c$ is to just facilitate the simplification of the system, both algebraically and numerically. As we had noted earlier, the beyond-Newtonian effects are scaled by a factor of $1/c^2$ which is taken care of by the introduction of the parameter $\epsilon$ in our system of units. Hence, $c=10000$ will scale down the beyond-Newtonian terms by a factor of $10^{-8}$, which can be compensated by suitably adjusting the range  of $\epsilon$, since it is a free parameter in our system. Thus it can be concluded that the nature of the dynamics of the system will not be affected by the choice of the value of $c$, as has been verified by our simulations too.

The separation between the two primaries is then scaled as, 
\begin{equation}
\begin{array}{lcl}
x_{1}=-\mu_1&;&x_{2}=1-\mu_1
\end{array}
\end{equation}

Thus, $\mu_{1},\mu_{2}\in[0,\frac{1}{2}]$ are the only two control parameters for the system. Applying the earlier described scaling and putting $c=1$, the potential becomes:
\begin{align}
\Omega(x,y)=&-\bigg(\frac{1-\mu_1}{r_1}+\frac{\mu_1}{r_2} \bigg)+\frac{1}{2}\epsilon^2\bigg(\frac{(1-\mu_1)^3}{r_1^3}+\frac{\mu_1^3}{r_2^3} \bigg) \nonumber \\
	&+\epsilon\bigg(\frac{(1-\mu_1)(1-\mu_2)}{r_1^2}\cos\theta_1+\frac{\mu_1\mu_2}{r_2^2}\cos\theta_2\bigg) \nonumber \\
	& +\frac{1}{2}\epsilon^2\bigg\{\frac{(1-\mu_1)(1-\mu_2)^2}{r_1^3}\bigg(3\cos^2\theta_1-1\bigg) \nonumber \\
	& +\frac{\mu_1\mu_2^2}{r_2^3}\bigg(3\cos^2\theta_2-1\bigg)\bigg\}
\end{align}
The Lagrangian for the system may be constructed as follows:
\begin{equation}
\mathcal{L}=\frac{V^2+2A+R^2}{2}-\Omega(x,y)
\label{lag}
\end{equation}
where $V=\sqrt{\dot{x}^2+\dot{y}^2}$ represents the magnitude of the velocity of the test particle, $R=\sqrt{x^2+y^2}$ the position of the test particle with respect to the centre of mass in the non-inertial rotating frame and $A = \dot{y} x - \dot{x} y$. Thus, the Euler-Lagrange equations of motion are:
\begin{align}
\ddot{x} &=  2 \dot{y}+x-\bigg[\frac{(1-\mu_{1})}{r_{1}^3}\bigg( x+\mu_{1} \bigg)+\frac{\mu_{1}}{r_{2}^3}\bigg(x+\mu_{1}-1\bigg)\bigg] \nonumber \\
		  & -\epsilon\bigg\{ \frac{(1-\mu_{1})(1-\mu_{2})}{r_{1}^4}\bigg[y\sin\theta_{1}-2\cos\theta_{1}\bigg(x+\mu_{1}\bigg)\bigg] \nonumber \\		  
		  &  +\frac{\mu_{1}\mu_{2}}{r_{2}^4}\bigg[y\sin\theta_{2}-2\cos\theta_{2}\bigg(x+\mu_{1}-1\bigg) \bigg]  \bigg\} \nonumber \\
		  & -\frac{3}{2}\epsilon^2\bigg\{\frac{(1-\mu_{1})(1-\mu_{2})^2}{r_{1}^5}\bigg[y\sin2\theta_{1}-\bigg(3\cos^2\theta_{1}-1\bigg) \nonumber \\
		  & \bigg(x+\mu_{1}\bigg) \bigg] -\frac{(1-\mu_{1})^3}{r_{1}^5}\bigg(x+\mu_{1}\bigg)+\frac{\mu_{1}\mu_{2}^2}{r_{2}^5}\bigg[y\sin2\theta_{2} \nonumber \\
		  & -\bigg(3\cos^2\theta_{2}-1\bigg)\bigg(x+\mu_{1}-1\bigg) \bigg]-\frac{\mu_{1}^3}{r_2^5}\bigg(x+\mu_{1}-1\bigg) \bigg\}  
\label{eqn:eqnOfMotionfor_x}
\end{align}
\begin{align} 
\ddot{y} =& -2\dot{x}+y\bigg[1-\bigg( \frac{1-\mu_{1}}{r_{1}^3}+\frac{\mu_{1}}{r_{2}^3} \bigg)\bigg]+\epsilon\bigg\{\frac{(1-\mu_{1})(1-\mu_{2})}{r_{1}^4} \nonumber \\
		  & \bigg[\bigg(x+\mu_{1}\bigg)\sin\theta_{1}+2y\cos\theta_{1} \bigg]+\frac{\mu_{1}\mu_{2}}{r_{2}^4 } \bigg[\bigg(x+\mu_{1}-1\bigg) \nonumber \\
		  & \sin\theta_{2}+2y\cos\theta_{2} \bigg]\bigg\} +\frac{3}{2}\epsilon^2\bigg\{ \frac{(1-\mu_{1})(1-\mu_{2})^2}{r_{1}^5}\bigg[\bigg(x+\mu_{1}\bigg) \nonumber \\
		  &  \sin2\theta_{1}+\bigg(3\cos^2\theta_{1}-1\bigg)y\bigg] +\frac{(1-\mu_{1})^3}{r_{1}^5}y+\frac{\mu_{1}\mu_{2}^2}{r_{2}^5} \nonumber\\
		  & \bigg[\bigg(x+\mu_{1}-1\bigg)\sin2\theta_{2}+\bigg(3\cos^2\theta_{2}-1\bigg)y \bigg] +\frac{\mu_{1}^3}{r_{2}^5}y\bigg\} 
\label{eqn:eqnOfMotionfor_y}
\end{align}
where,
\begin{align*}
r_{1}&=\sqrt{(x+\mu_{1})^2+y^2} \\
r_{2}&=\sqrt{(x+\mu_{1}-1)^2+y^2}\\
\theta_{1}&=\tan^{-1}[y/(x+\mu_{1})] \\
\theta_{2}&=\tan^{-1}[y/(x+\mu_{1}-1)]
\end{align*}
The Jacobi integral for the above system is given by,
\begin{equation}
J(x,y,\dot{x},\dot{y})=(x^2+y^2)-2\Omega(x,y)-(\dot{x}^2+\dot{y}^2)=C_{j}
\label{jacobian}
\end{equation} 
where $C_{j}$ is a constant of motion for the given system and is called the Jacobian constant.

Here we note that the Lagrangian for our system, as stated in equation~(\ref{lag}), has terms only up to the quadratic order in velocity $V$ of the test particle as a result of which the Jacobian constant (equation (\ref{jacobian})) is exactly derived. This is in contrast to the post-Newtonian (PN) framework where the Jacobian does not remain conserved and consequently limits the extent of dynamical studies. The reasoning behind this has to do with the relations between the PN Lagrangian and Hamiltonian approaches at the same PN order. Additionally, it also depends on the relations between the approximately truncated as well as the exactly non-truncated Euler-Lagrange equations of motion for this PN Lagrangian approach. The equivalence between the Lagrangian and Hamiltonian approaches at the same PN order was established in \citet*{Damour01, Damour02, deAndrade01, Levi14}. However, recent contradictions of the same have been discussed in \citet{Wu15a, Wu15b}; \citet*{Wang15, Chen16}; \citet{Huang16}. It has been shown by \citet{Li19, Li20} that the approximately truncated Euler-Lagrange equations of motion for this PN Lagrangian approach have different dynamical behaviours of order and chaos than its exactly non-truncated counterpart. As a result, the reasons why the Jacobian constant cannot be conserved in the PN approach is because (a) some higher-order PN terms are truncated when the Euler-Lagrange equations of motion are derived from this PN Lagrangian approach, and (b) some higher-order PN terms are still truncated when the Hamiltonian (corresponding to the Jacobian constant) is derived from this PN Lagrangian approach. If the truncated higher-order PN terms are large, as in the case of strong gravitational fields of compact objects, the Jacobian constant shows a poor accuracy; while it shows a better accuracy if the same truncated terms are comparatively smaller, as in the case of weak gravitational fields found in our Solar system. It should be expected that for our potential (equation (\ref{pnp})), the Lagrangian and Hamiltonian approaches at the same beyond-Newtonian order are not equivalent in general. This inequivalence should also be true for the approximately truncated as well as the exactly non-truncated Euler-Lagrangian equations of motion for our beyond-Newtonian Lagrangian approach. However, the equations of motion (\ref{eqn:eqnOfMotionfor_x}) and (\ref{eqn:eqnOfMotionfor_y}), the corresponding Hamiltonian and the Jacobian constant (\ref{jacobian}) are exactly derived and have no terms truncated from the beyond-Newtonian Lagrangian (equation (\ref{lag})) because it has no higher-order terms with respect to the test particle velocity $V$. As a result, the Jacobian constant, given by equation (\ref{jacobian}), is said to be exactly derived. 

\begin{figure*}
\centering
\includegraphics[scale=0.5]{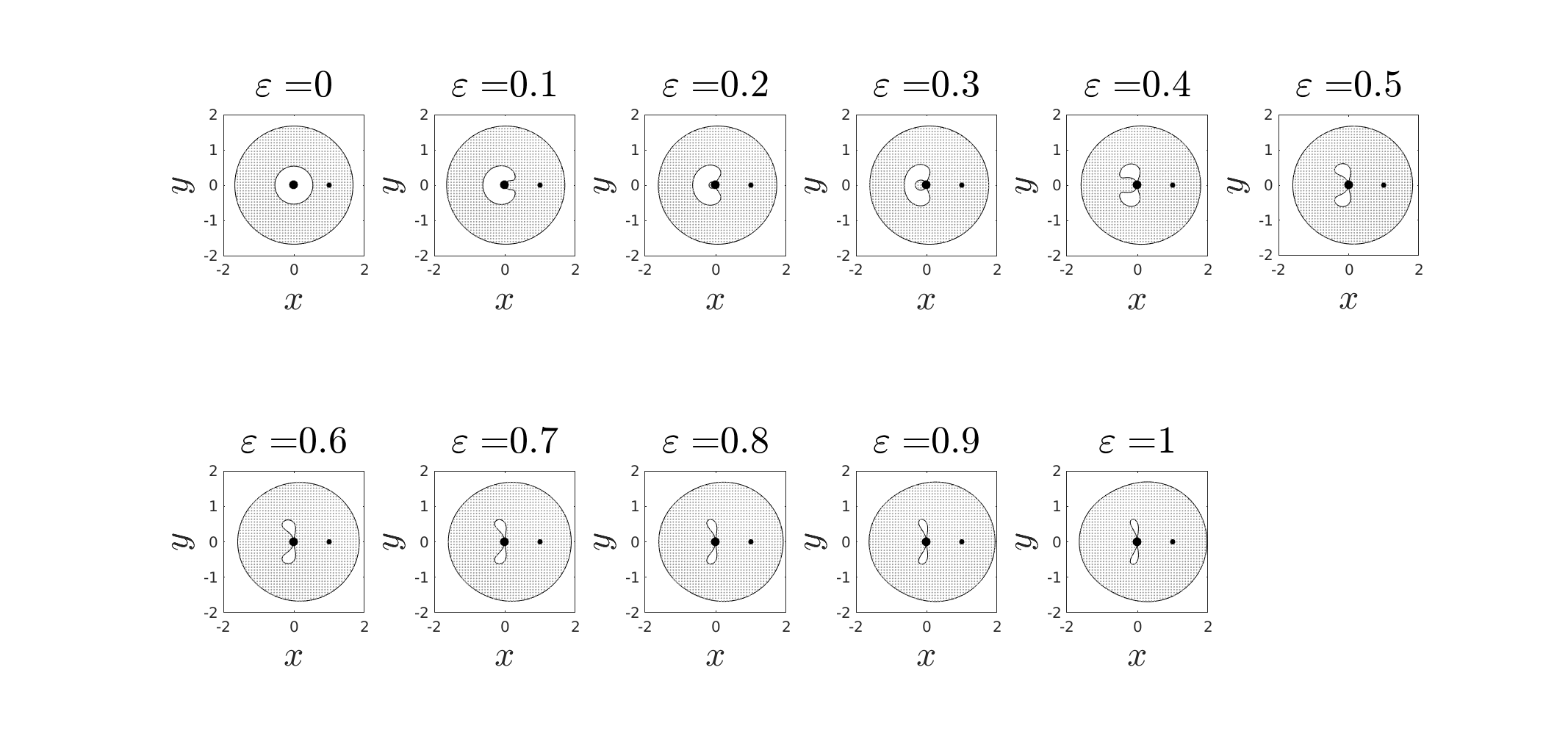}
\caption{Evolution of the Hill curves for $\mu_{1}=\mu_{2}=0.001$ and $C_{j}=4.0$ with the parameter $\epsilon$. The white regions of the plot represents the points in the X-Y plane are energetically allowed, while the dotted regions are energetically disallowed, for the test particle whose Jacobian $C_{j}=4.0$. The larger black dot on the left represents the position of the mass $\mathcal{M}_1$ and the smaller black dot on the right represents the position of mass $\mathcal{M}_2$ in each of the plots.}
\label{fig:HillCurve0001WithEpsilon}

\includegraphics[scale=0.5]{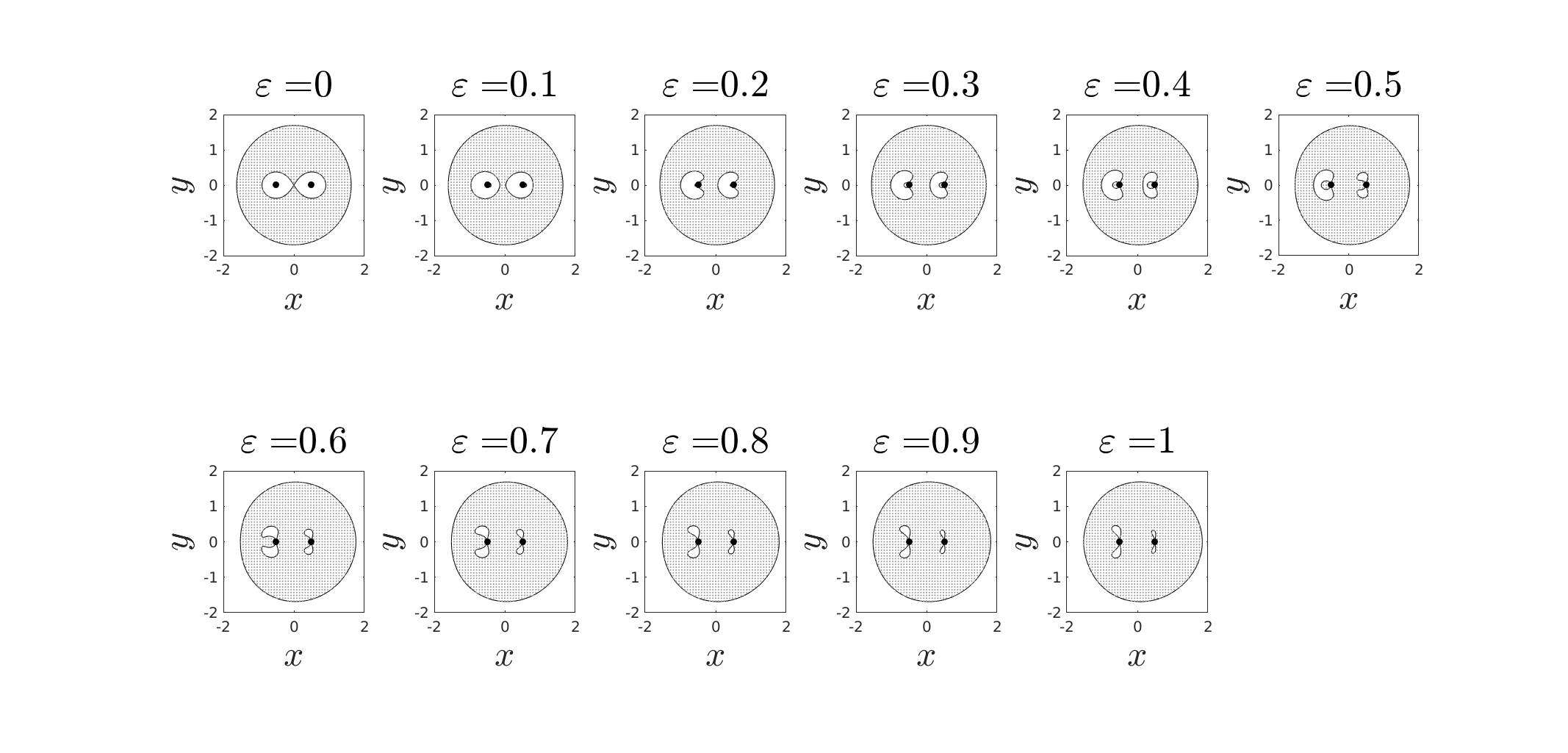}
\caption{Evolution of the Hill curves for $\mu_{1}=\mu_{2}=0.5$ and $C_{j}=4.0$ with the parameter $\epsilon$. The white regions of the plot represents the points in the X-Y plane are energetically allowed, while the dotted regions are energetically disallowed, for the test particle whose Jacobian $C_{j}=4.0$.  The black dot on the left represents the position of the mass $\mathcal{M}_1$ and the black dot on the right represents the position of mass $\mathcal{M}_2$ in each of the plots.}
\label{fig:HillCurve05WithEpsilon}
\end{figure*}
\subsection{Hill Curves}
The Hill curves or the zero-velocity curves for the system, for a set of chosen values of $C_{j}$, $\mu_{1}$, $\mu_{2}$ and $\epsilon$, divide the equatorial plane into regions where the motion of the body is energetically allowed and regions where the motion is energetically disallowed (for a discussion on zero-velocity curves refer to \citet*{Szebehely63} for a Newtonian CRTBP system and \citet*{Zotos18a} for a pseudo-Newtonian CRTBP with Schwarzschild like primaries). All points, where $(x^2+y^2)-2\Omega(x,y)-(\dot{x}+\dot{y}) > C_{j}$, are energetically allowed for the test particle, while all points, where $(x^2+y^2)-2\Omega(x,y)-(\dot{x}+\dot{y}) < C_{j}$ are energetically disallowed. The velocity of the test particle (as we shall calculate from equations (\ref{eqn:initial_conditions})) in the disallowed region is imaginary (will be calculated explicitly in the next subsection). The Hill curves of the system have an equation,
\begin{equation}
(x^2+y^2)-2\Omega(x,y)=C_{j} \,.
\end{equation}
Figures (\ref{fig:HillCurve0001WithEpsilon}) and (\ref{fig:HillCurve05WithEpsilon}) show the evolution of the Hill curves with the introduction of beyond-Newtonian effects for $\mu_{1}=\mu_{2}=0.001$ (or the biased-mass system) and $\mu_{1}=\mu_{2}=0.5$ (or the Copenhagen system) respectively. The beyond-Newtonian effects are introduced by increasing $\epsilon$ from $0.0$ to $1.0$ in steps of $0.1$. The equatorial plane is divided into three regions by the Hill curves -- a central region where the particle is energetically allowed but is bounded by the Hill curves, an unbounded energetically allowed region, and a disallowed region in-between them. Test particles with initial conditions in the unbound region may execute stable orbits around both the primaries or may escape to infinity, while test particles with initial positions in the enclosed and energetically allowed regions are `trapped' and cannot escape to infinity since they cannot cross the Hill curves. The energetically allowed regions are represented by white in figure (\ref{fig:HillCurve0001WithEpsilon}) and figure (\ref{fig:HillCurve05WithEpsilon}), while the dotted regions are energetically disallowed for the test particle. The two black dots represent the positions of the primaries $\mathcal{M}_1$ and $\mathcal{M}_2$ respectively. 

For the biased-mass system, the potential due to mass $\mathcal{M}_1$ dominates the Hill curves. The introduction of beyond-Newtonian effects distorts the curves of the Newtonian system, such that for all values of $\epsilon\gtrsim0.0865$, no trapped circular orbits exist. For the Copenhagen system, the chosen value of $C_j$ corresponds to the energy at the first Lagrange point $L_1$. As $\epsilon$ increases, the contribution of the spin becomes apparent and the enclosed allowed region becomes smaller. Circular trapped orbits around both the primaries exist for small values of $\epsilon$. For $\epsilon>0.1248$, circular orbits no longer exist around the primary $\mathcal{M}_2$ while for $\epsilon>0.134$, circular orbits no longer exist around the primary $\mathcal{M}_1$. Thus, for both systems, we choose our initial conditions in the unbounded energetically allowed region for the sake of consistency of initial conditions for all values of $\epsilon$, $\mu_{1}$ and $\mu_{2}$.

\subsection{Orbits}\label{sec:orbits}
Using the six stepped, fifth-order Runge-Kutta method implemented with the Dortmund-Prince algorithm, the equations of motion equations (\ref{eqn:eqnOfMotionfor_x},\ref{eqn:eqnOfMotionfor_y}) are integrated using time step $\tau=10$ for $n=3000$ iterations. For a preliminary investigation of the system, the following initial conditions are considered (similar to \citet{Dubeibe16} which investigates orbits for a system with Schwarzschild like primaries): $x_{0}=[3.0,3.5,3.75,4.0,4.25,4.5,4.75,5.0,5.25,5.5,6.0]$, $y_{0}=0.0$ and $\dot{x}_{0}=0.0$, with $C_{j}=4$. The value of $\dot{y}_{0}(x_{0},y_{0},\dot{x}_{0})$ is  calculated from the following equation:
\begin{align}
\dot{x}_{0}&=\frac{y_{0}}{r_{0}}\sqrt{(x^2+y^2)-2\Omega(x,y)-C_{j}} \nonumber \\
\dot{y}_{0}&=-\frac{x_{0}}{r_{0}}\sqrt{(x^2+y^2)-2\Omega(x,y)-C_{j}}
\label{eqn:initial_conditions}
\end{align}
where $r_{0}=\sqrt{x_{0}^2+y_{0}^2}$.
The orbits for a test particle for the biased mass and Copenhagen systems are investigated for $\epsilon\in[0,1]$ and the set of initial conditions mentioned in the paragraph above. Since the system is conservative, the Jacobi constant $C_{j}$ has to remain constant as the equations of motion are integrated. 

\begin{figure*}
	\begin{tabular}{ccc}
		\includegraphics[scale=0.22]{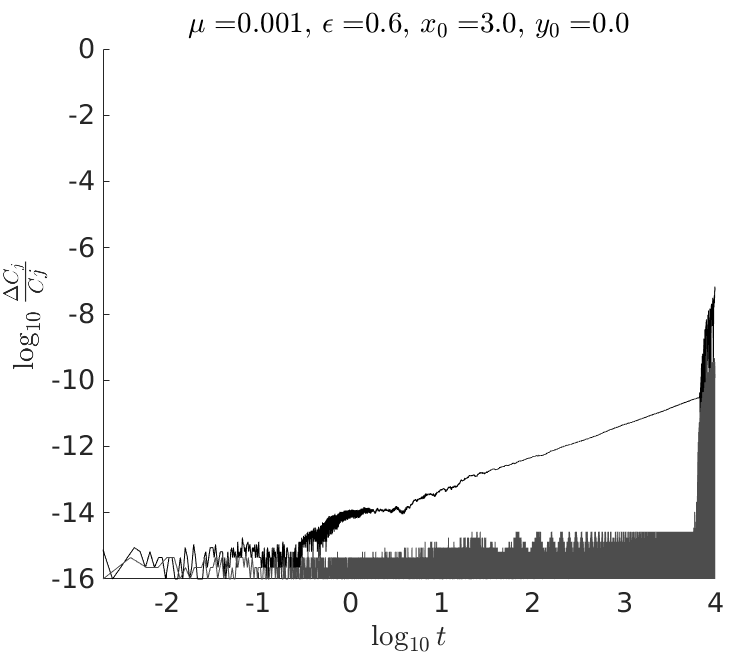}  &\includegraphics[scale=0.22]{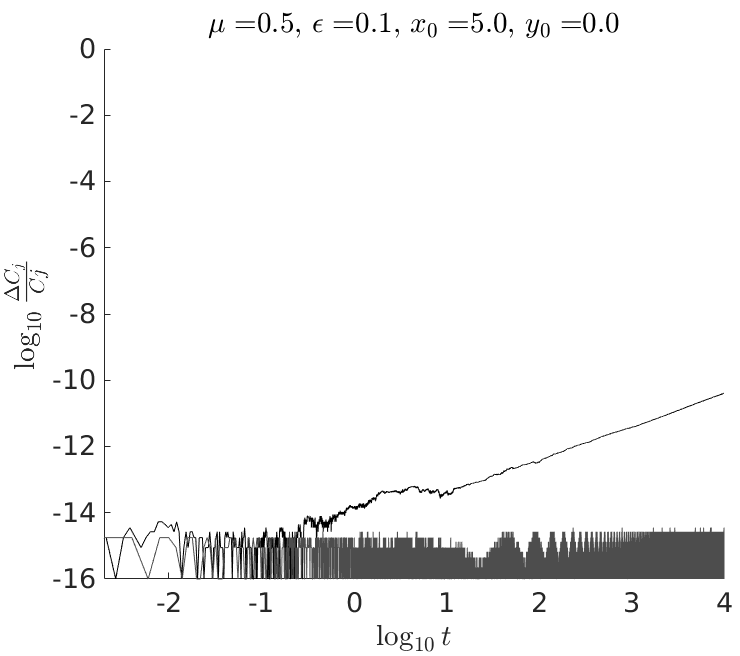}&\includegraphics[scale=0.22]{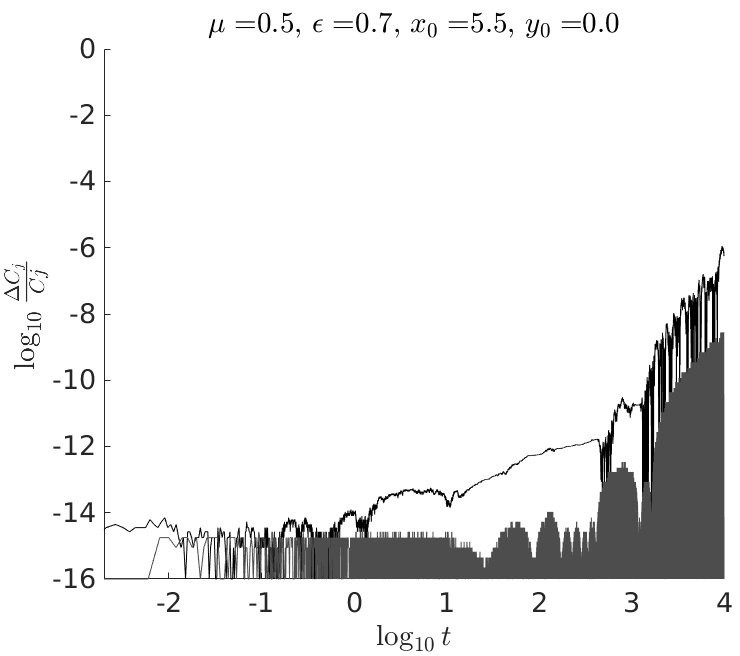} 
	\end{tabular}
	\caption{Plot of the log of the relative error in the Jacobi constant $C_j$ with log of time for the non-corrected Runge-Kutta (4,5) integrator using the Dormand-Prince algorithm (black) and velocity-corrected 4th order Runge-Kutta integrator (grey). The first plot from the left ($\mu=0.001$, $\epsilon=0.6$, $x_0=3.0$, $y_0=0.0$) shows the evolution of $C_j$ with time for a sticky initial condition, the centre plot ($\mu=0.5$, $\epsilon=0.1$, $x_0=5.0$, $y_0=0.0$) shows the evolution of $C_j$ with time for a stable initial condition, and the plot on the right ($\mu=0.5$, $\epsilon=0.7$, $x_0=5.5$, $y_0=0.0$) shows the evolution of $C_j$ with time for a chaotic initial condition.}
	\label{velCorr0001}
\end{figure*}

The integrator used, being non-symplectic in nature, usually does not conserve the Jacobian. The use of such integrators for conservative systems have been well studied and multiple corrective methods, such as the velocity correction method (\citet*{Ma08}; \citet{Wang16, Wang18, Deng20}), have been developed for better accuracy. In Figure (\ref{velCorr0001}), we have shown a comparison of the relative error in the Jacobi constant $C_j$ with time for both the non-corrected and velocity corrected integrators. It is observed that the accuracy in the conservation of $C_j$ for the velocity corrected method ranges from $10^{-16}$ -- $10^{-14}$ for stable orbits and goes up to $10^{-8}$ for chaotic and sticky orbits at large times ($> 5 \times 10^3$ years), as has been pointed out in \citet{Wang16, Wang18}. We also observe that our non-corrected integrator has a fairly similar accuracy at the start. However, the growth in error is faster at late times and reaches values of $10^{-10}$ for stable orbits and goes up to $10^{-8}$ for chaotic and sticky orbits. Hence, we conclude that the non-corrected fifth-order Runge-Kutta method is also of reasonable accuracy for the relevant time-periods of our investigation.

%

By observing their evolution, the orbits may be categorized as regular, sticky or escaping. Orbits are said to be sticky if they show regular behavior for a long period of time before their chaotic nature manifest \citep*{Dvorak99} and escaping if the particle directly escapes from the system without executing any regular orbits \citep{Contoupoulos90, Contoupoulos92}. We classify the stability of the initial conditions based on the number of iterations for which the orbit of the particle is stable. If the test particle executes stable orbits for 3000 iterations, it is classified as regular. If the orbits are stable for at-least 100 iterations before they escape from the system, they are classified as sticky. If the test particle reaches a distance of 50 times the separation between the two primaries within 1000 iterations, they are said to be escaping. The third column of the table in Appendix (\ref{tab:Details0001}) and Appendix (\ref{tab:Details05}) records the type of orbit for the test particle given some initial conditions for the biased-mass and Copenhagen system respectively.

For the biased mass system, among the initial conditions considered, orbits for $x_{0}=[3.5,3.75,4.0,4.25,4.5]$ are stable for all values of $\epsilon$. Most initial conditions are either sticky or escaping for non-zero values of $\epsilon$. But the interesting initial conditions are the ones where the intermediate values of $\epsilon$ are the most chaotic. The initial conditions $x_0=[5.0,5.25,5.5,6.0]$ show such behavior. For the Copenhagen system, orbits for $x_0=[3.5,3.75,4.0,4.5,4.75]$ are stable for all values of $\epsilon$. The initial condition $x_0=4.25$ destabilizes for $\epsilon>0.4$, implying a region of chaotic initial conditions interjects stable initial conditions in the phase space. Orbits for $x_0=[5.0,5.25]$ are either sticky or escaping for all values of $\epsilon$ except $\epsilon=[0.0,0.1]$. A stable orbit for $x_0=5.5$ exists only for $\epsilon=0.0$, while no stable orbits exist for $x_0=6.0.$ for any value of $\epsilon$. This implies that regions of initial conditions allowing stable orbits shrink as $\epsilon$ increases for the Copenhagen system. 

Contrary to expectation, $\epsilon=1.0$ does not result in the maximum number of sticky and escaping initial conditions in either of the systems. Instead, the intermediate values of $\epsilon$ have the most number of unstable initial conditions. For the biased mass system, $\epsilon=[0.2,0.4,0.6,0.7,0.8]$ have the least number of stable initial conditions, namely 6 out of the 11 investigated. $\epsilon=0.9$ has the least number of stable initial conditions for the Copenhagen system, namely 5 out of the 11 investigated. In contrast, $\epsilon=1.0$ has 9 and 6 initial conditions out of 11 for the biased mass and Copenhagen systems, respectively.  

\begin{figure*}
	\begin{tabular}{ccc}
		\includegraphics[scale=0.26]{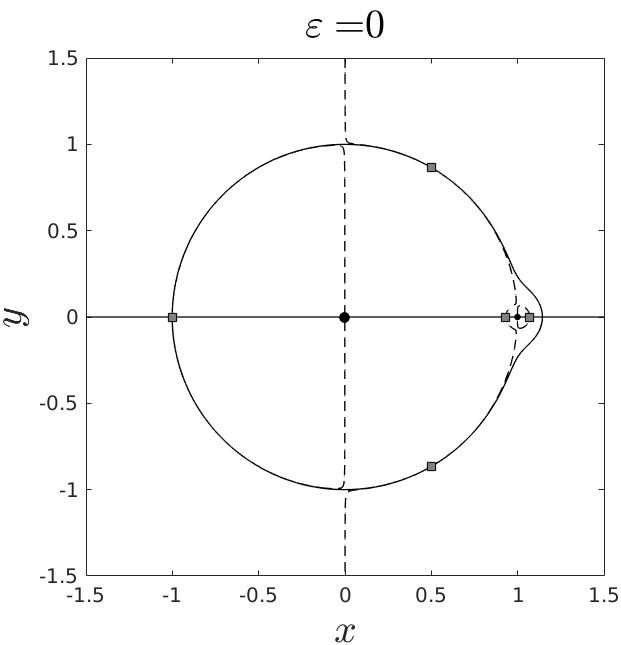}&\includegraphics[scale=0.26]{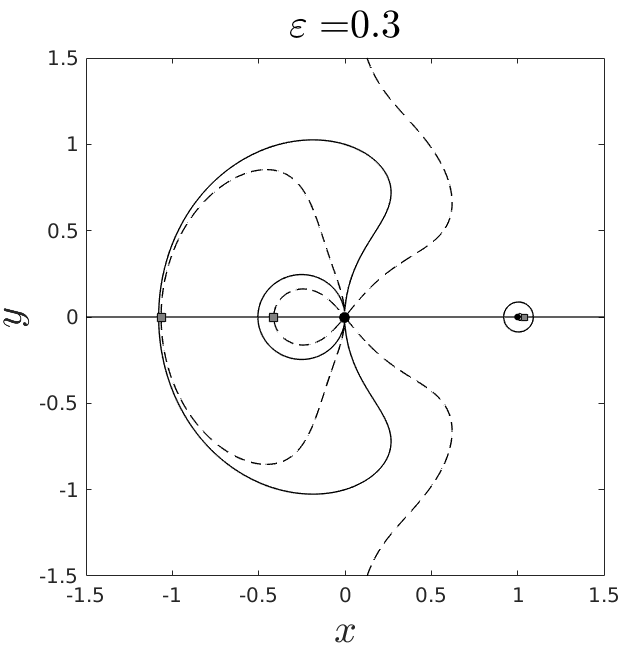}&\includegraphics[scale=0.26]{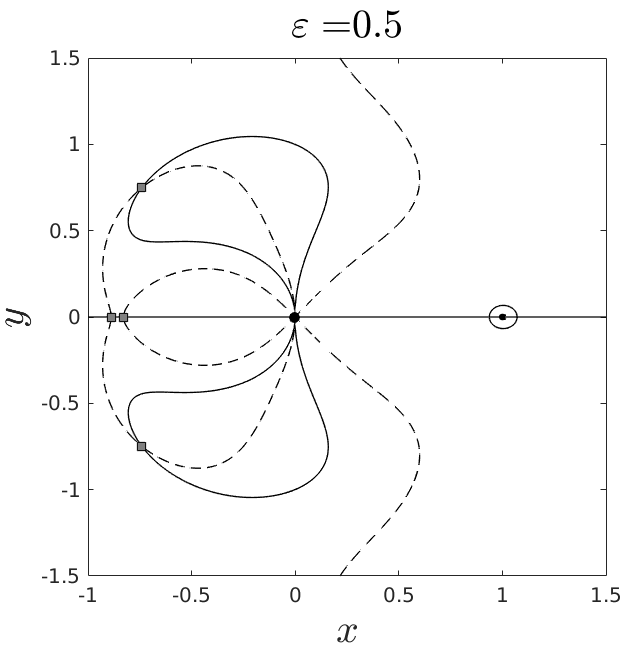} \\
		\includegraphics[scale=0.26]{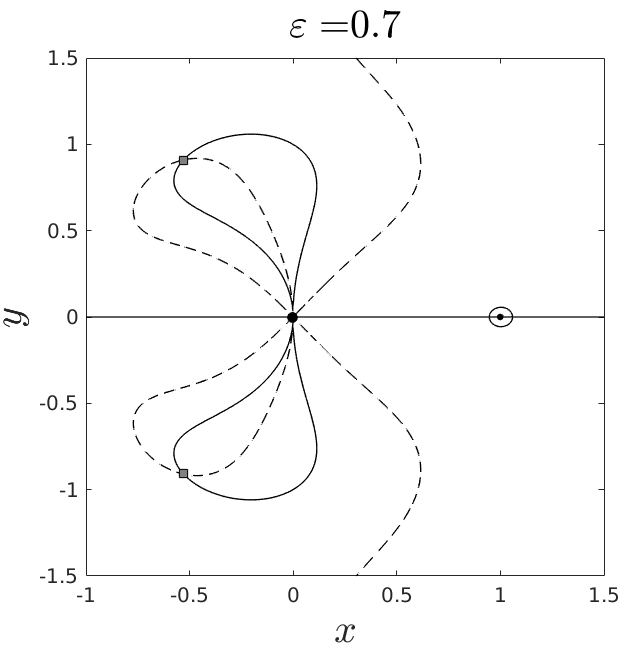}&\includegraphics[scale=0.26]{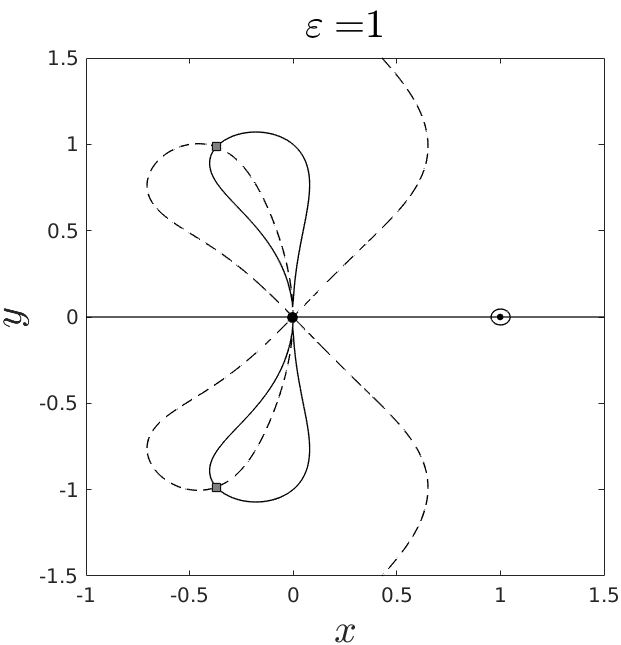}&
	\end{tabular}
	\caption{Locations of some of the equilibrium points of the biased-mass system ($\mu_1=\mu_2=0.001$), marked by grey squares, on the intersection of $\partial\Omega/\partial x=0$, marked by the dashed line, and $\partial\Omega/\partial y=0$, marked by the solid line, for $\epsilon=[0.0,0.3,0.5,0.7,1.0]$. For $\epsilon\neq0$, there are two non-collinear equilibrium points on either sides of the more massive primary which could not be shown on the plots due to their close proximity to it. The smaller primary has three more collinear equilibrium points, one of which lies between the two primaries. These too could not be marked on the plots due to their proximity to the primary.}
	\label{fixedPoint0001}
\end{figure*}
\begin{figure*}
	\begin{tabular}{ccc}
		\includegraphics[scale=0.26]{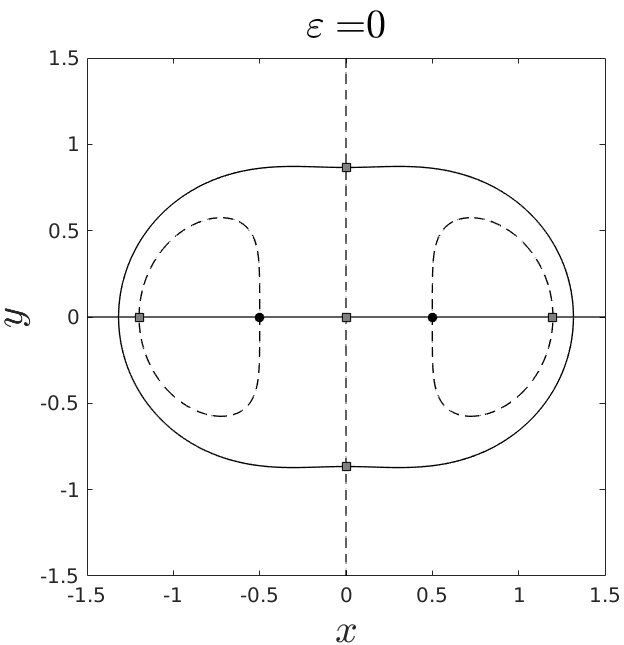}&\includegraphics[scale=0.26]{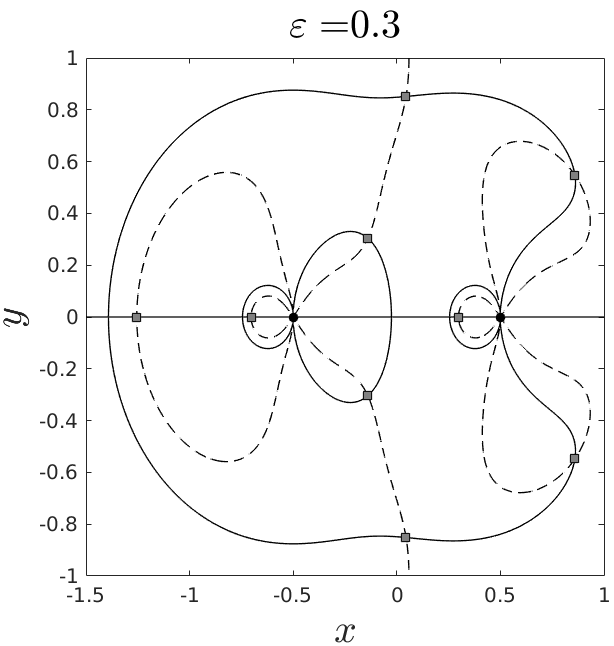}&\includegraphics[scale=0.26]{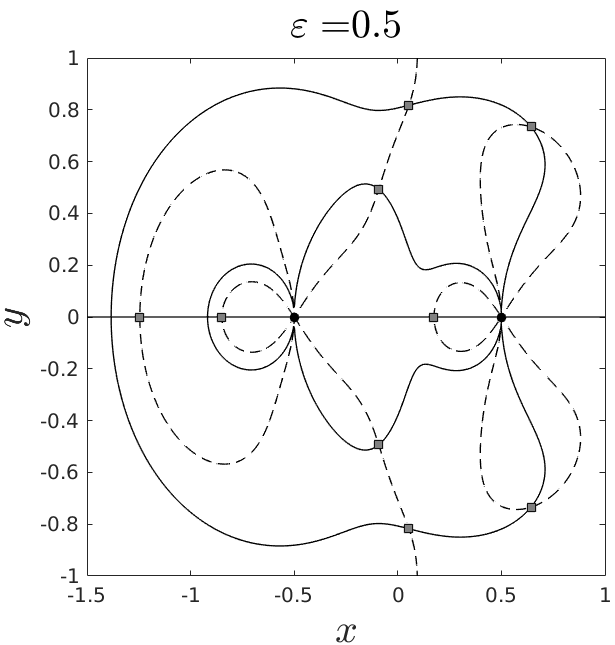} \\
		\includegraphics[scale=0.26]{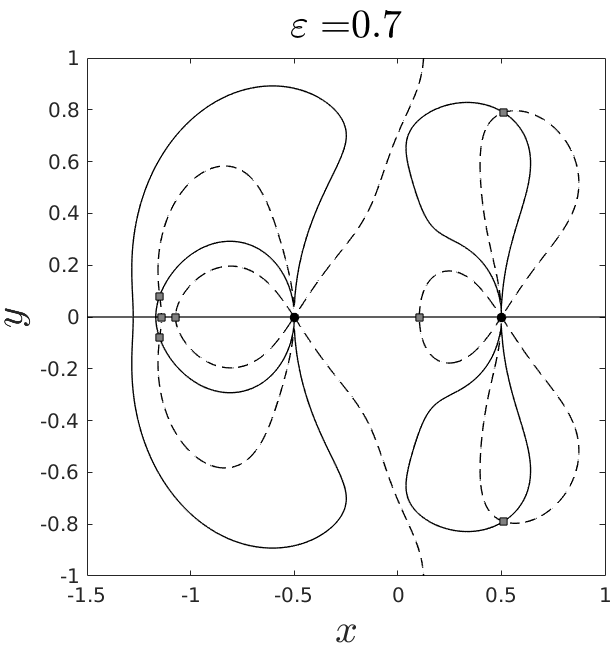}&\includegraphics[scale=0.26]{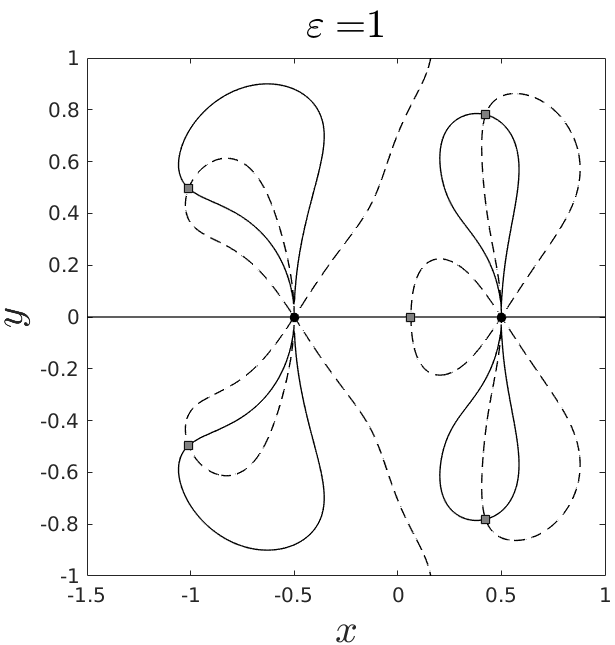}&
	\end{tabular}
	\caption{Locations of some of the equilibrium points of the Copenhagen system ($\mu_1=\mu_2=0.05$), marked by gray squares, on the intersection of $\partial\Omega/\partial x=0$, marked by the dashed line, and $\partial\Omega/\partial y=0$, marked by the solid line, for $\epsilon=[0.0,0.3,0.5,0.7,1.0]$. For $\epsilon\neq0$, there are two non-collinear equilibrium points on either sided of both primaries which could not be shown on the plots due to their close proximity to them.}
	\label{fixedPoint05}
\end{figure*}
\subsection{Fixed Points and their stability}

For a system having equilibrium points (fixed points), the necessary and sufficient conditions are:
\begin{equation}
\dot{x}=\dot{y}=\ddot{x}=\ddot{y}=0 
\end{equation}
Thus, the co-ordinates of the co-planar fixed points are determined by solving the following pair of partial differential equations (refer to equation~(\ref{pnp}) for the detailed expression of $\Omega(x,y)$) \citep*{Strogatz94}:
\begin{equation}
\frac{\partial\Omega(x,y)}{\partial x}=\frac{\partial\Omega(x,y)}{\partial y}=0
\label{omegaPartial}
\end{equation}    
The intersection of the curves for equations~(\ref{omegaPartial}) for a set of values of $\mu_1$, $\mu_2$ and $\epsilon$ gives us a set of fixed points for the system. Figures~(\ref{fixedPoint0001}) and~(\ref{fixedPoint05}) show the positions of the fixed points for $\epsilon=[0.0,0.3,0.5,0.7,1.0]$ for the biased-mass and Copenhagen systems respectively. It is evident that for both systems, the number of fixed points is highly dependent on the value of $\epsilon$, the details of which are enlisted in separate tables in Appendices~(\ref{tab:FixedPnts0001}) and (\ref{tab:FixedPnts05}). A summary of the salient features of the fixed points with respect to $\epsilon$ is presented below:
\begin{itemize}
	\item For $\epsilon=0$, both the biased-mass and Copenhagen systems reduce to their Newtonian counterparts. These systems have five fixed-points each, as expected.
	\item For $\epsilon=0.3$, the biased-mass system has five fixed points while the Copenhagen system has nine. 
	\item For $\epsilon=0.5$, the biased mass system has nine fixed points while the Copenhagen system has thirteen. The less massive primary in the biased system has three collinear fixed points.
	\item For both $\epsilon=0.7$ and $\epsilon=1.0$, the biased mass system has five fixed points. Only the collinear fixed points in either of the systems is beyond the less massive primary. However, the Copenhagen system has nine equilibrium points for both $\epsilon=0.7$ and $\epsilon=1.0$.
	\item Finally, the more massive primary in the biased mass system as well as both the primaries in the Copenhagen system have two non-collinear equilibrium points very near to it for values of $\epsilon\geq 0.3$ (not shown in Figures (\ref{fixedPoint0001}) and (\ref{fixedPoint05}) since they fall very close to the primaries).
\end{itemize} 
It is thus evident that the number of equilibrium points for both the biased-mass and the Copenhagen systems become maximum at intermediate values of $\epsilon$.

Now, moving on to the issue of stability of these fixed points, their linear stability may be determined 
by Taylor expanding the system's equations of motion around the fixed point ($x_0\,\text{,} y_0$) upto first order. In the perturbation equations, the time-independent coefficient matrix of variations is identified as 
\begin{equation}
A=\begin{bmatrix}
0&  0&  1& 0\\ 
0&  0&  0& 1\\ 
\frac{\partial^2\Omega_0}{\partial x^2}&\frac{\partial^2\Omega_0}{\partial x \partial y}  &0  &2 \\ 
\frac{\partial^2\Omega_0}{\partial y \partial x}& \frac{\partial^2\Omega_0}{\partial y^2} &  -2& 0
\end{bmatrix}
\label{charMatrix}
\end{equation}
where the subscript $0$, attached to the partial derivatives of second order of $\Omega$, denotes evaluation at the position of the equilibrium point ($x_0\,\text{,} y_0$). The necessary and sufficient condition that a fixed point is stable is that all the eigenvalues of matrix $A$ be purely imaginary. Applying this method to the fixed points, obtained by numerically solving equations~(\ref{omegaPartial}), we can conclude the following:
\begin{itemize}
	\item For $\epsilon=0$, the collinear fixed points for both the biased-mass and the Copenhagen systems are unstable while the triangular fixed points are stable.
	\item For $\epsilon=0.3$, none of the fixed points are stable for the biased mass system while two fixed points are stable for the Copenhagen system.
	\item For $\epsilon=0.5$, one fixed point is stable for the biased mass system while two are stable for the Copenhagen system.
	\item For $\epsilon=0.7$, no fixed point is stable for the biased mass system while one is stable for the Copenhagen system.
	\item For $\epsilon=1.0$, no fixed point is stable for either of the systems. 
\end{itemize} 
Thus, the evolution and stability of the fixed points of the system under consideration show non-trivial evolution with the parameter $\epsilon$. However, the knowledge about the basins of convergence along with the libration points is of prime importance since the attracting domains reflect some of the most intrinsic properties of the dynamical system. This has been a topic of intense research in recent years for many different dynamical systems such as the Hill problem \citep*{Douskos10}, the four-body problem \citep*[e.g.][]{Baltagiannis11, Kumari14, Zotos17c} and the pseudo-Newtonian planar circular restricted three body problem \citep*{Zotos17b, Zotos17d, Zotos18b}. We plan to investigate these aspects for our beyond-Newtonian potential in detail as part of our future work.

\section{Poincar\'e Map of Section}
\label{sec:PS}
\begin{figure*}
	\begin{tabular}{ccc}
		\includegraphics[scale=0.27]{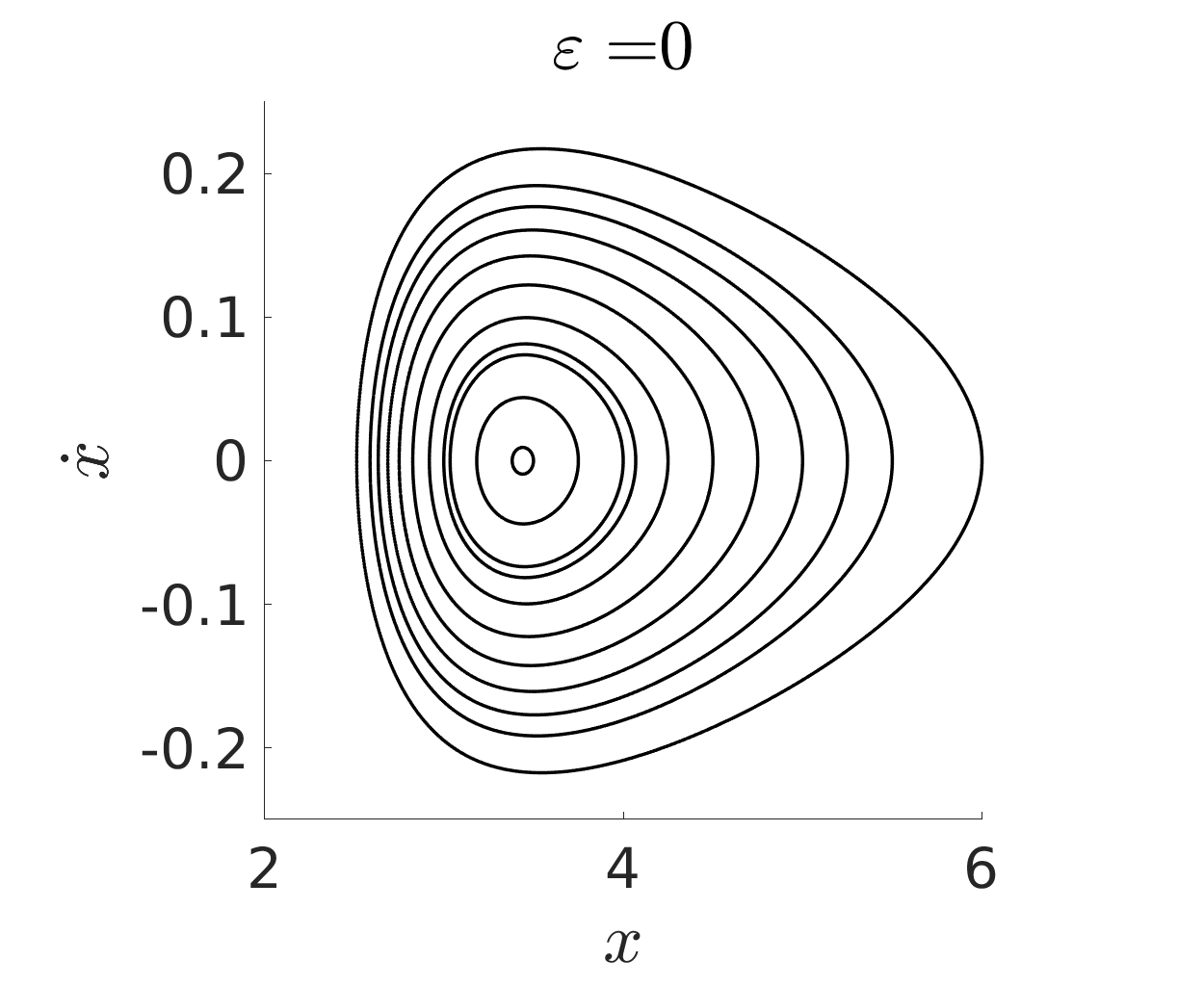}&\includegraphics[scale=0.27]{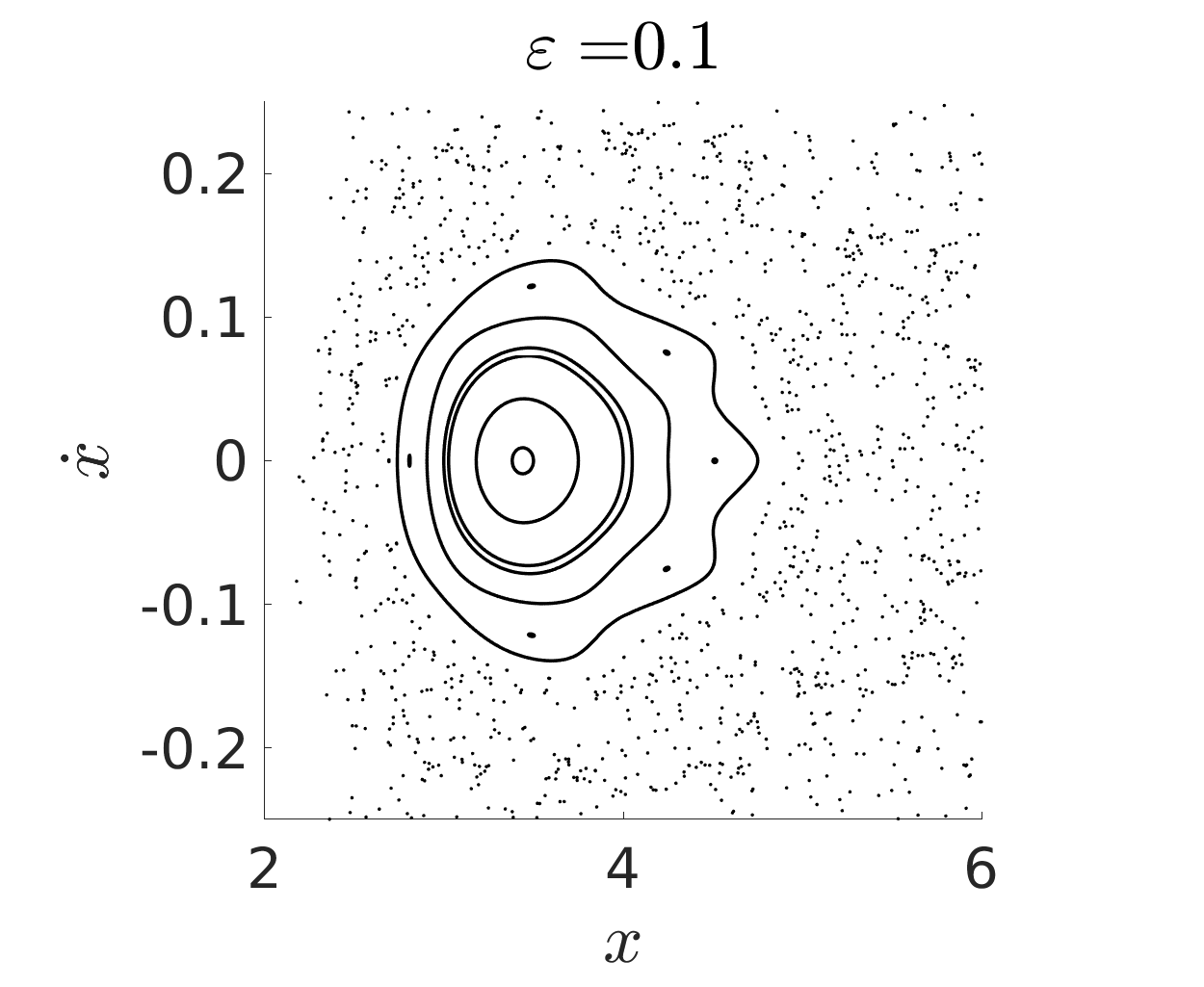}&\includegraphics[scale=0.27]{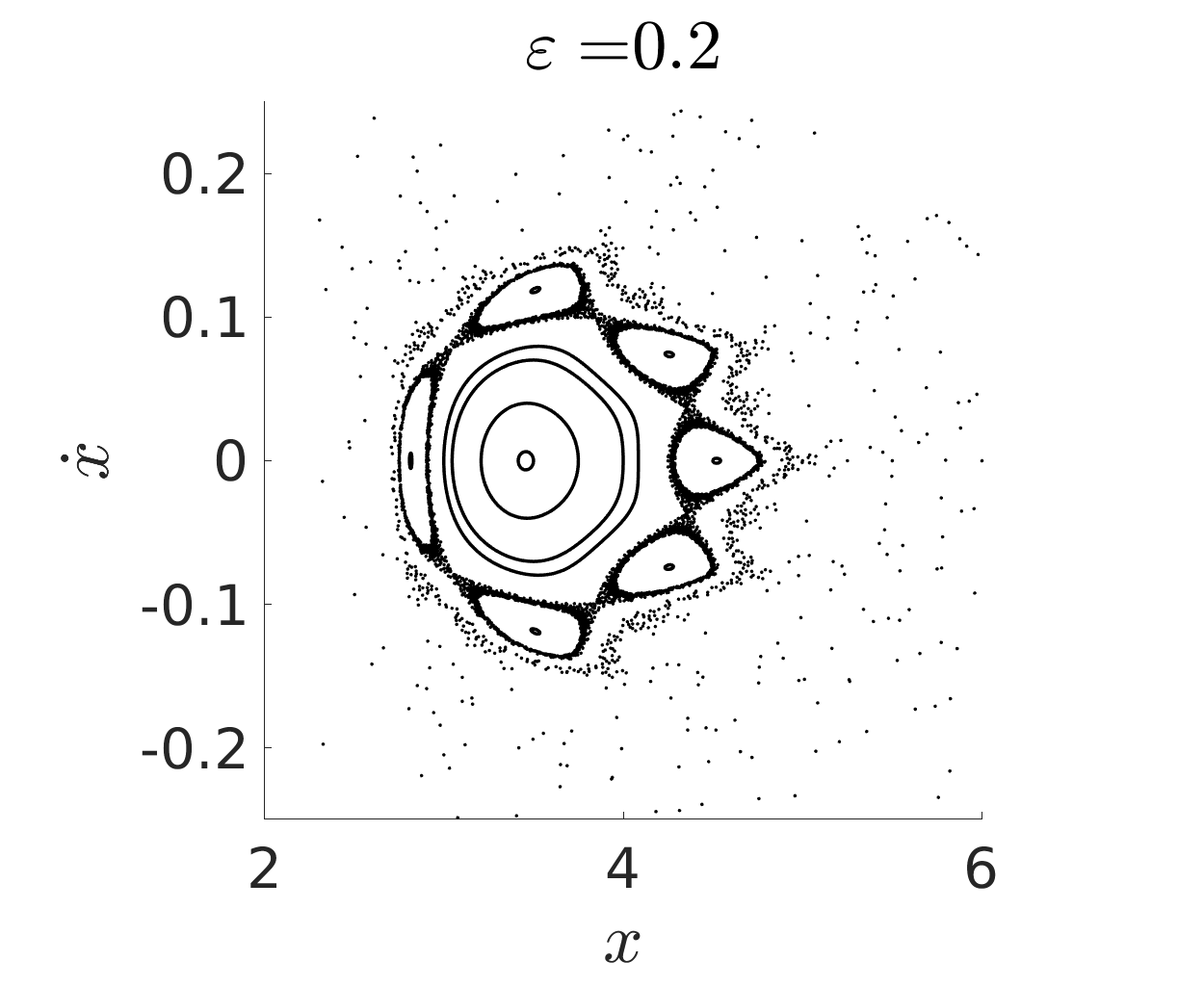} \\
		\includegraphics[scale=0.27]{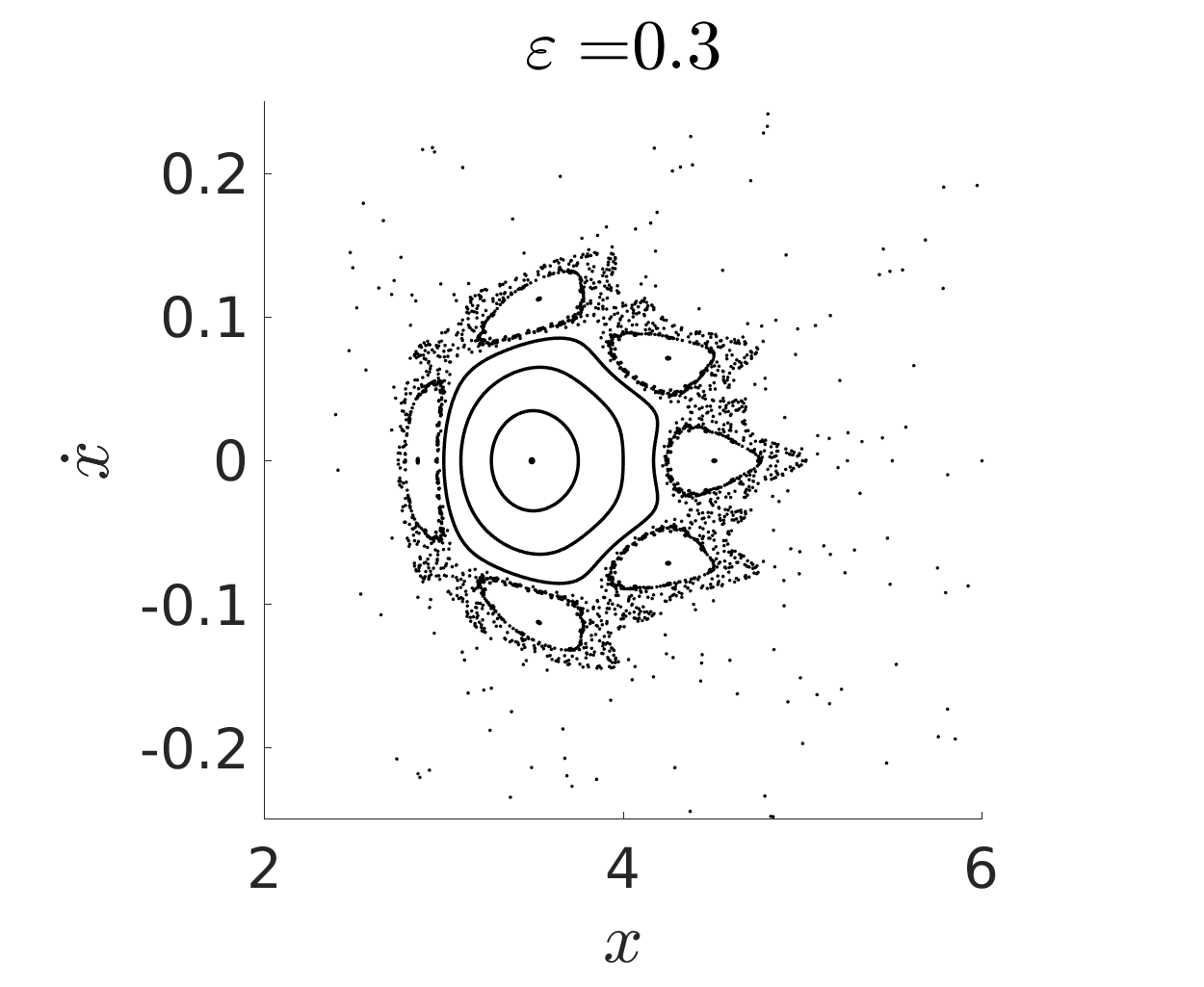}&\includegraphics[scale=0.27]{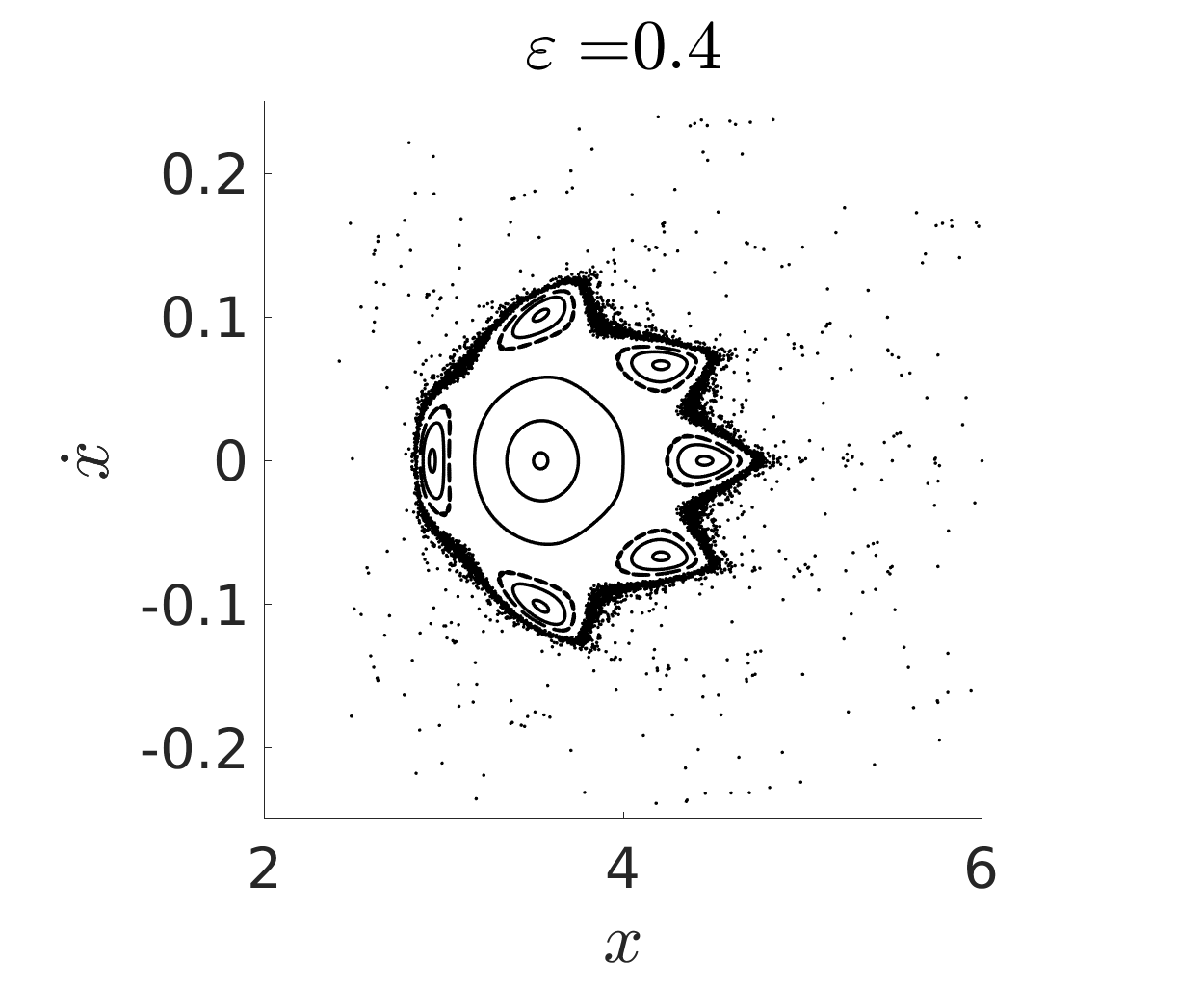}&\includegraphics[scale=0.27]{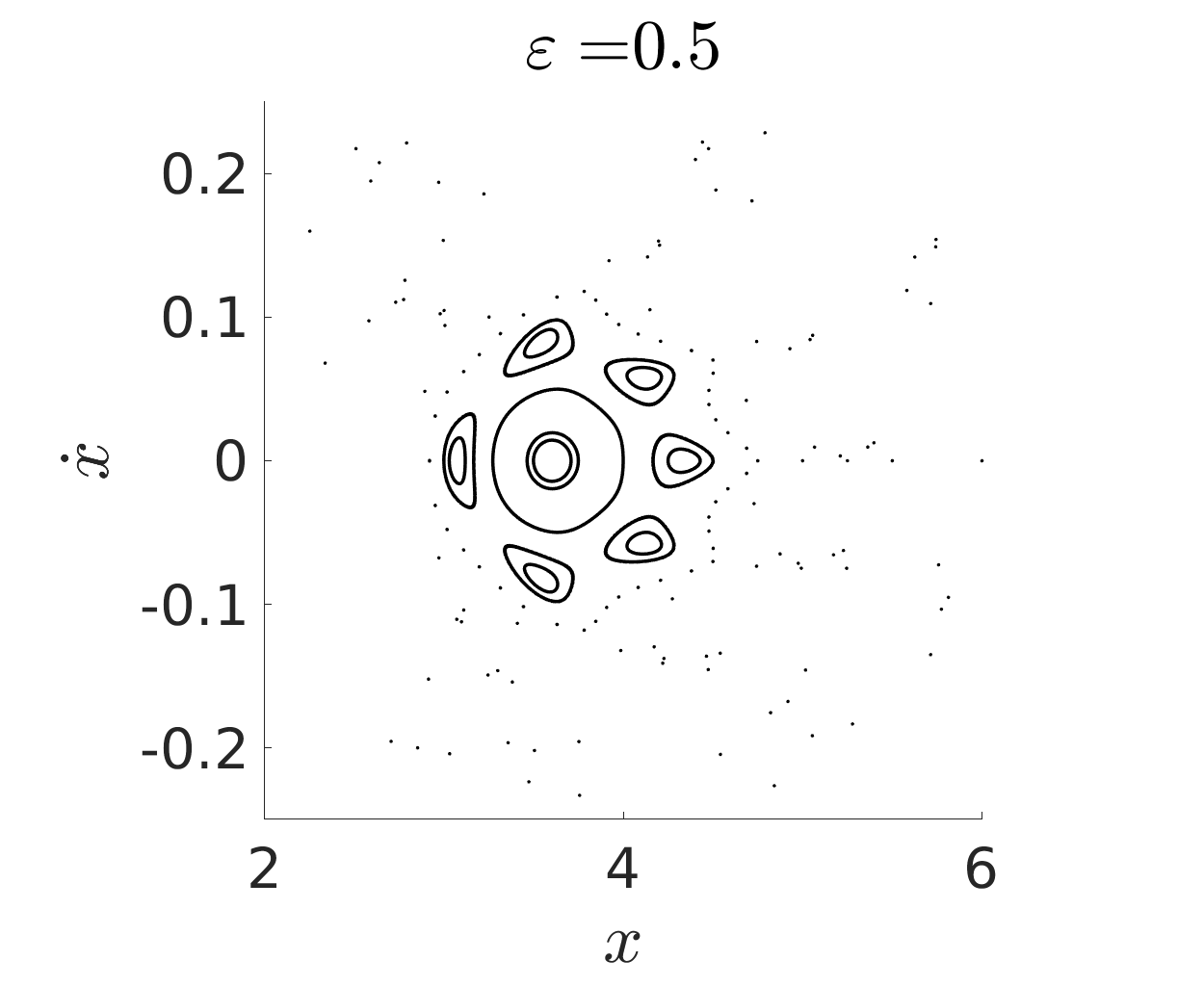} \\\includegraphics[scale=0.27]{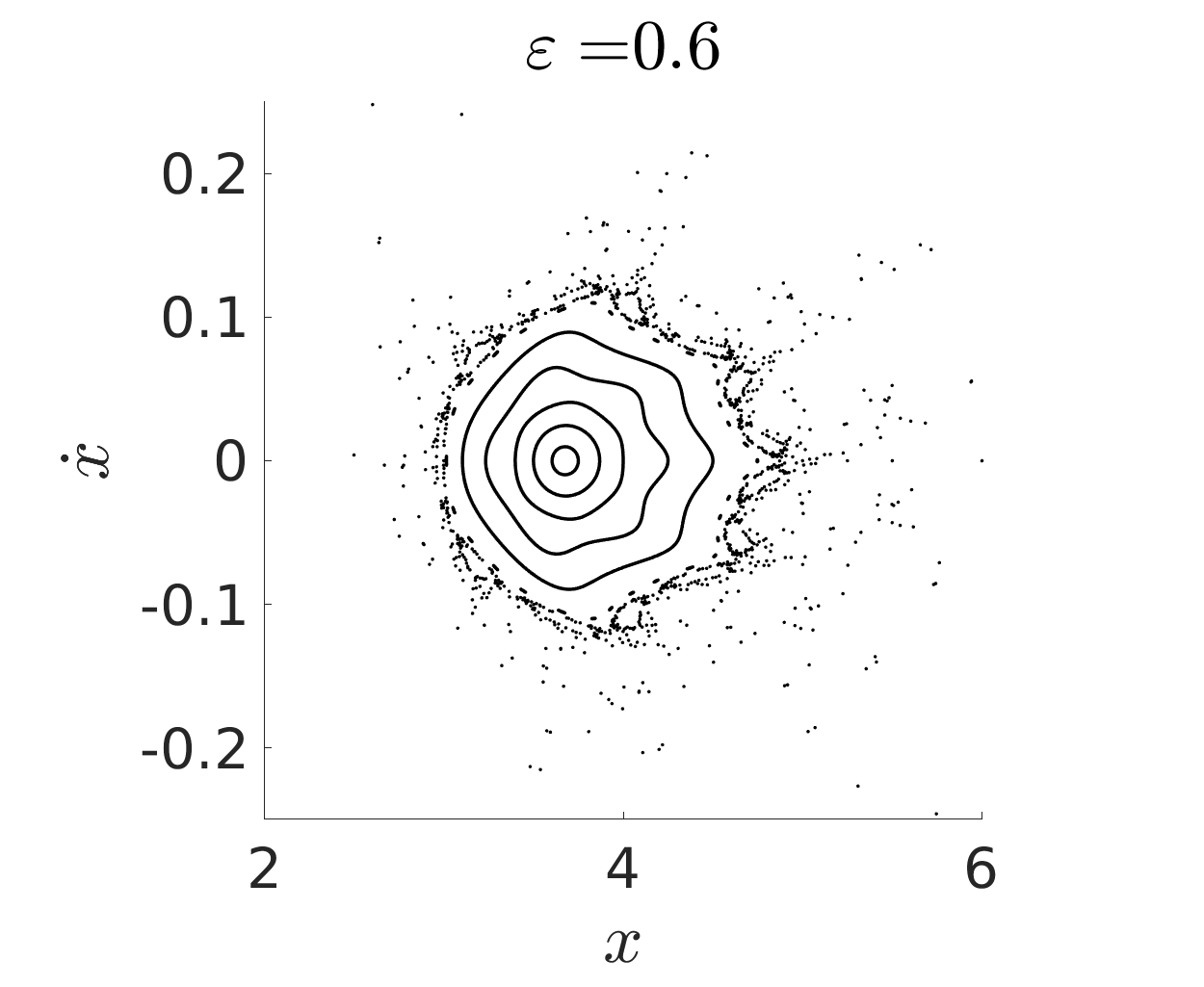}&\includegraphics[scale=0.27]{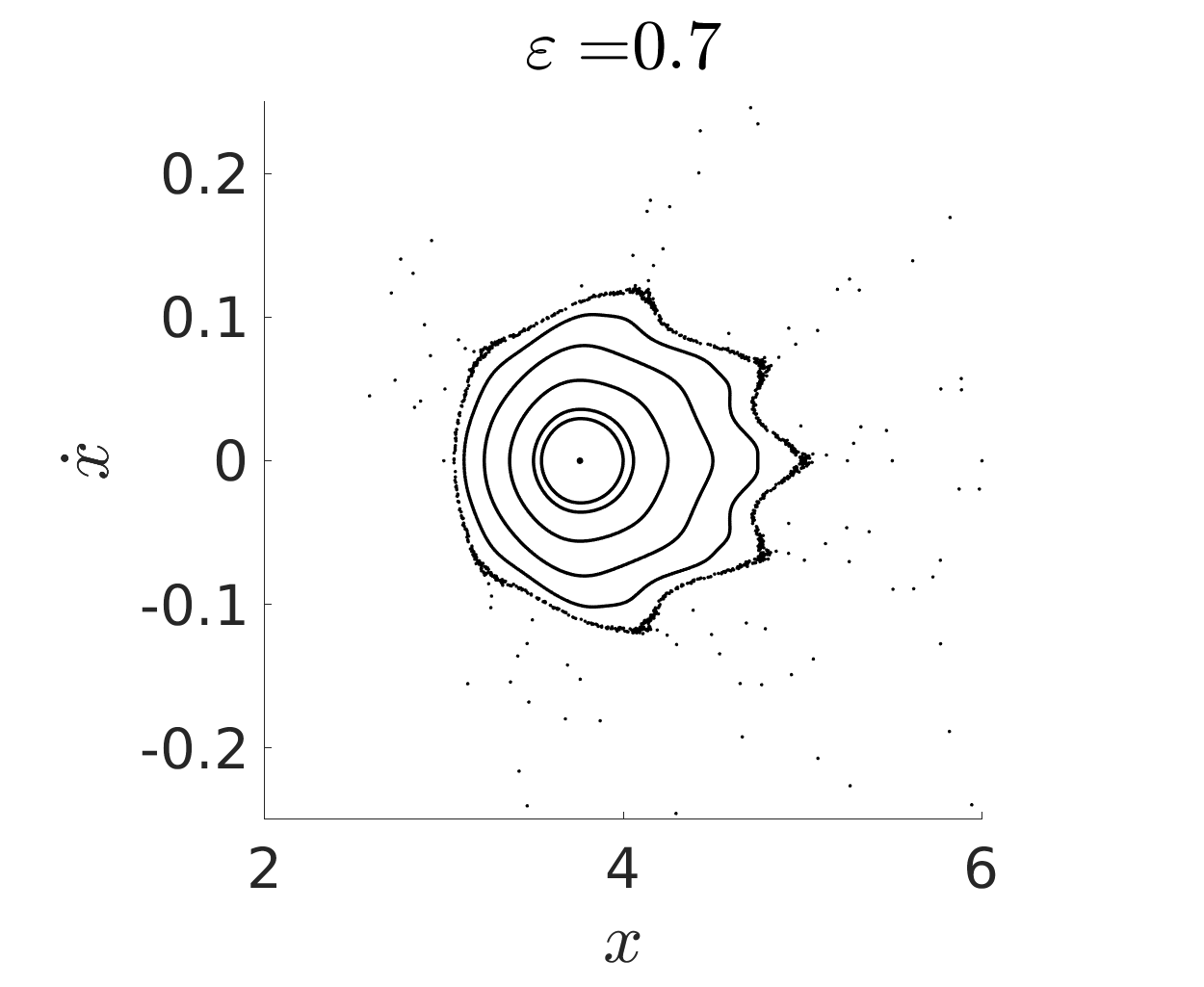}&\includegraphics[scale=0.27]{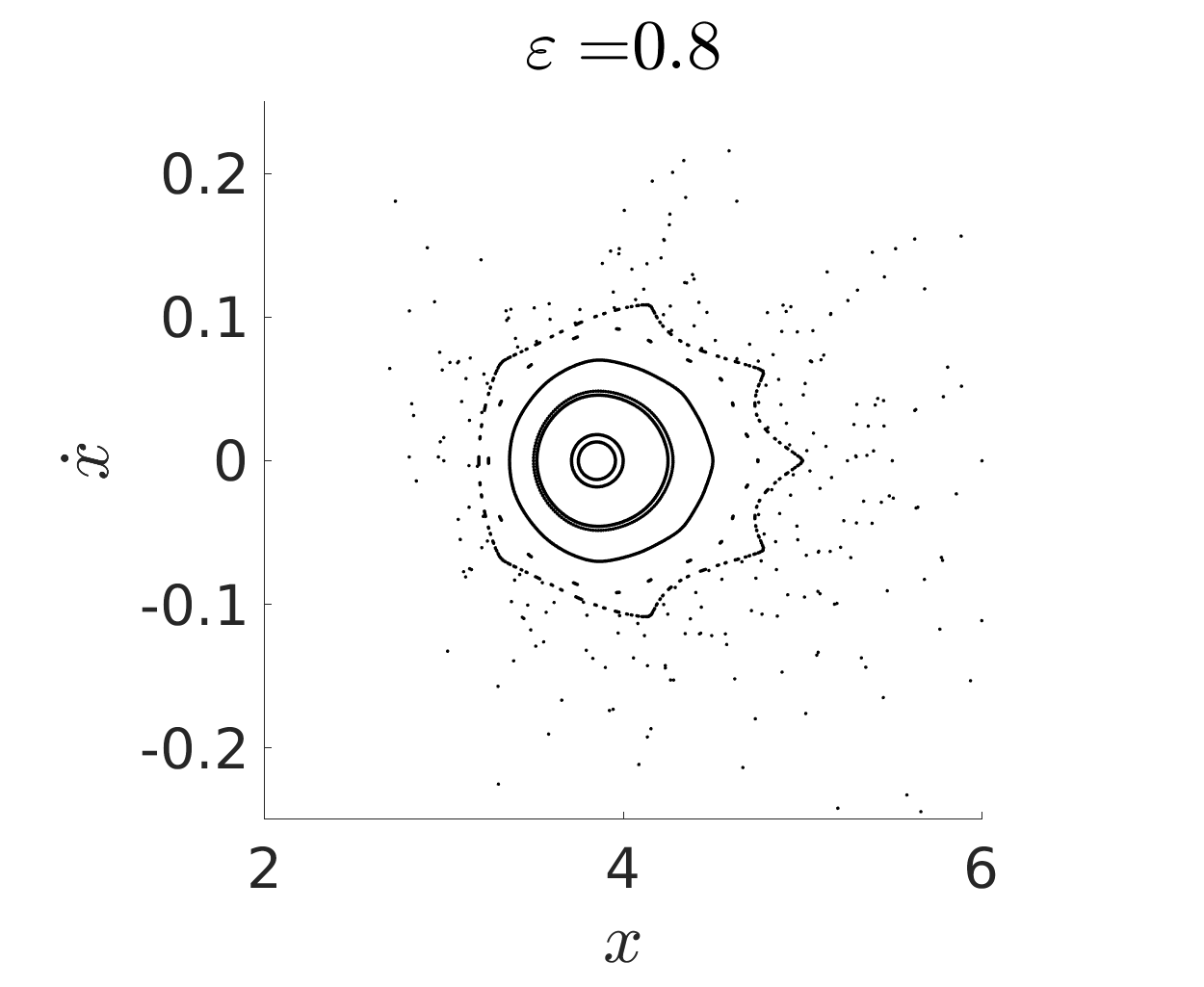} \\
		\includegraphics[scale=0.27]{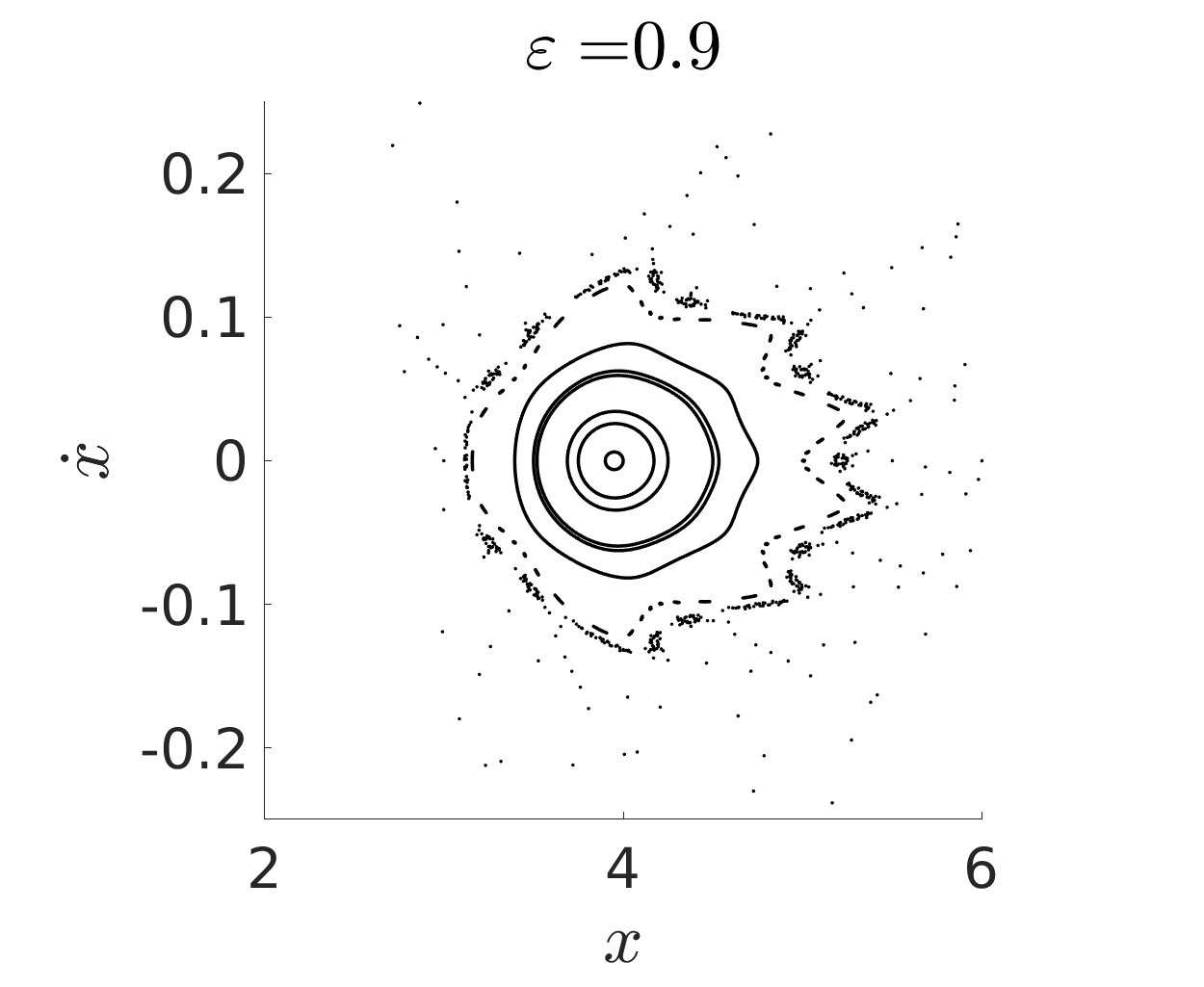}&\includegraphics[scale=0.27]{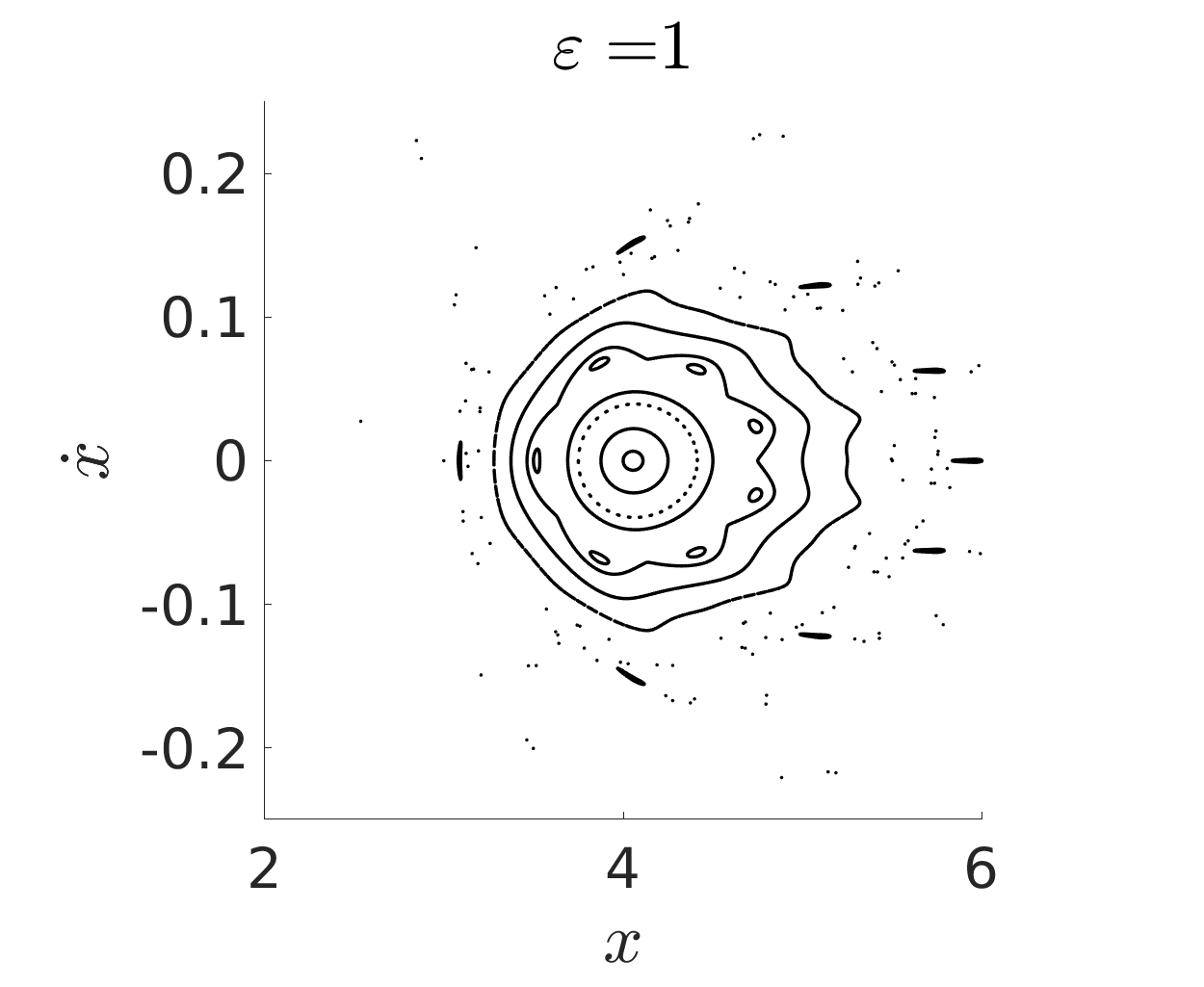}&\\
	\end{tabular}
	\caption{The Poincar\'e map for orbits with $x_0=[3.0,3.5,3.75,4.0,4.25,4.5,4.75,5.0,5.25,5.5,6.0]$ and $y_0=0.0$ as $\epsilon$ increases from 0 to 1 in steps of 0.1. The system is evolved for 3000 iterations in time-steps of $\tau=10$ and the Poincar\'e map of section for $y=0.0, \dot{y}>0.0$ is plotted for all 11 initial conditions for the biased-mass system ($\mu_{1}$=$\mu_{2}=0.001$).}
	\label{fig:PS_0.001}
\end{figure*}

\begin{figure*}
	\begin{tabular}{ccc}
		\includegraphics[scale=0.27]{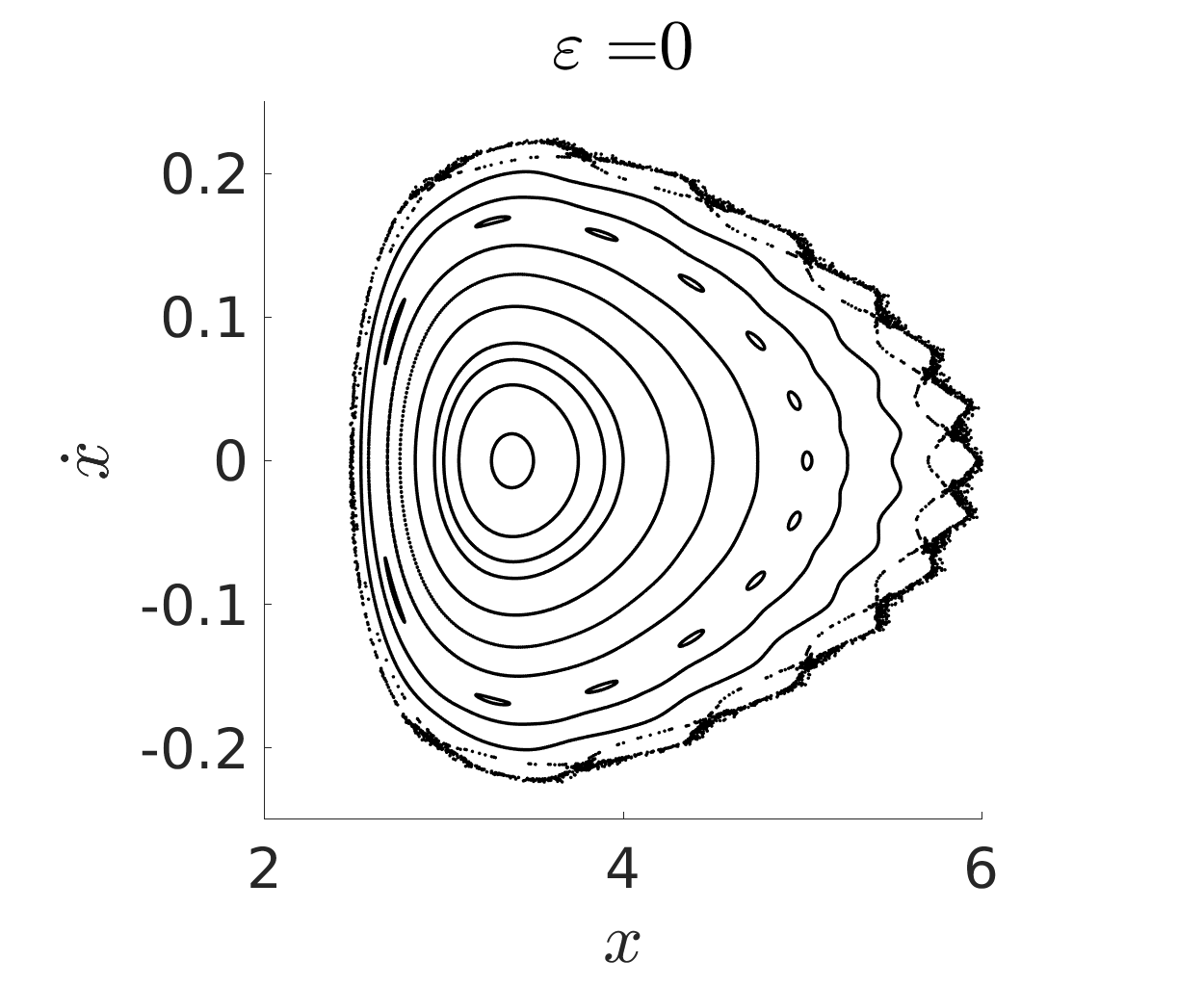}&\includegraphics[scale=0.27]{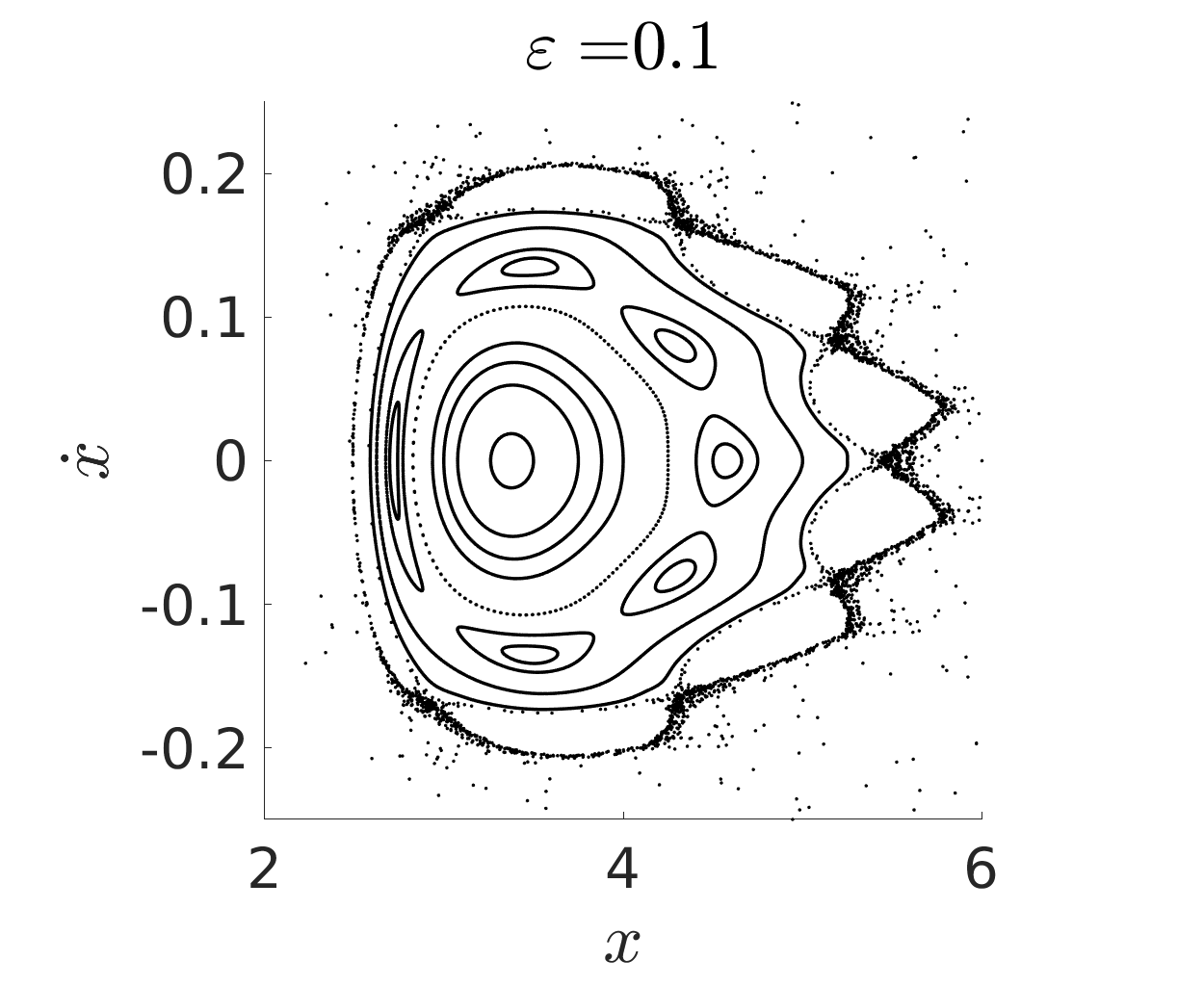}&\includegraphics[scale=0.27]{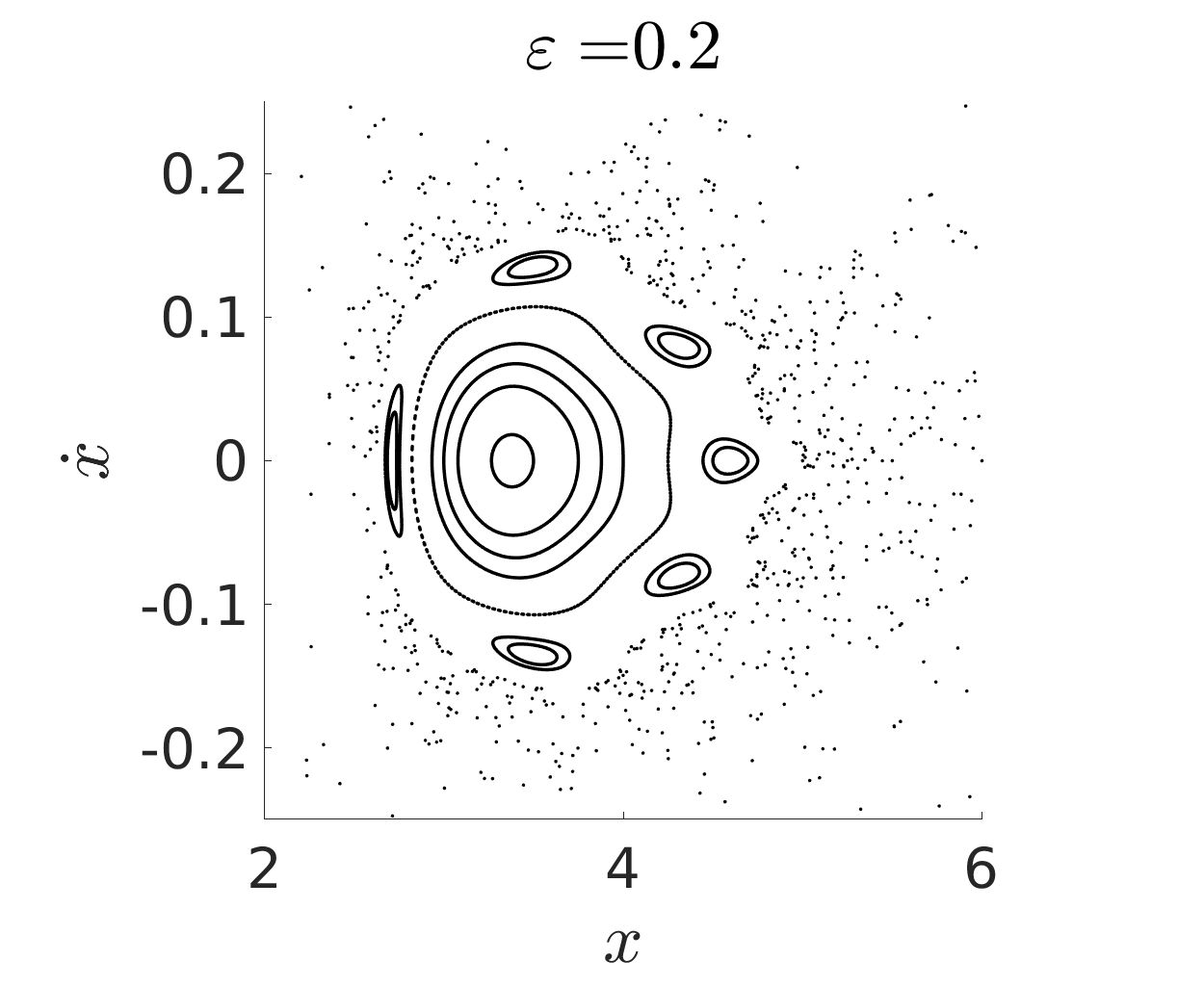} \\
		\includegraphics[scale=0.27]{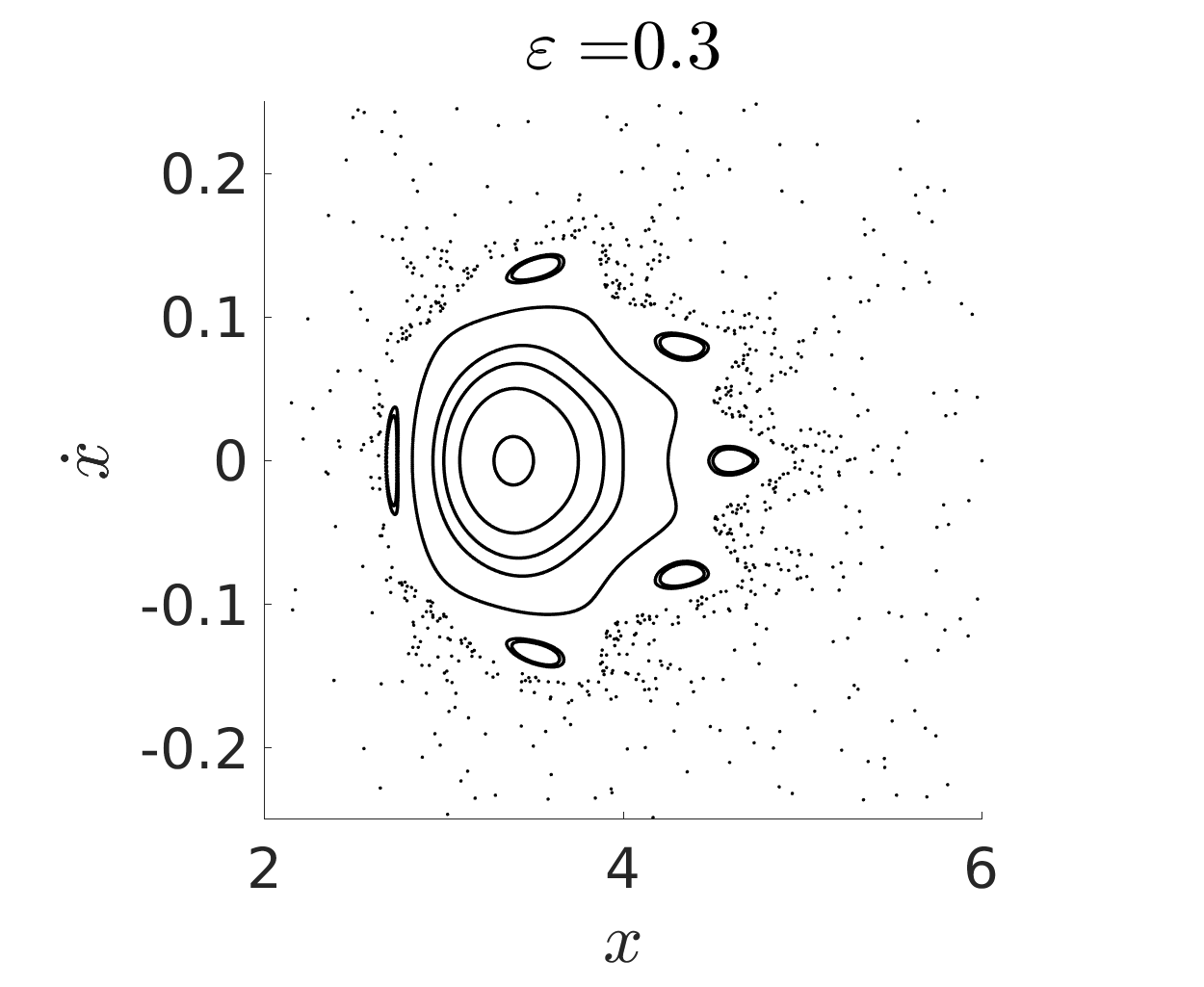}&\includegraphics[scale=0.27]{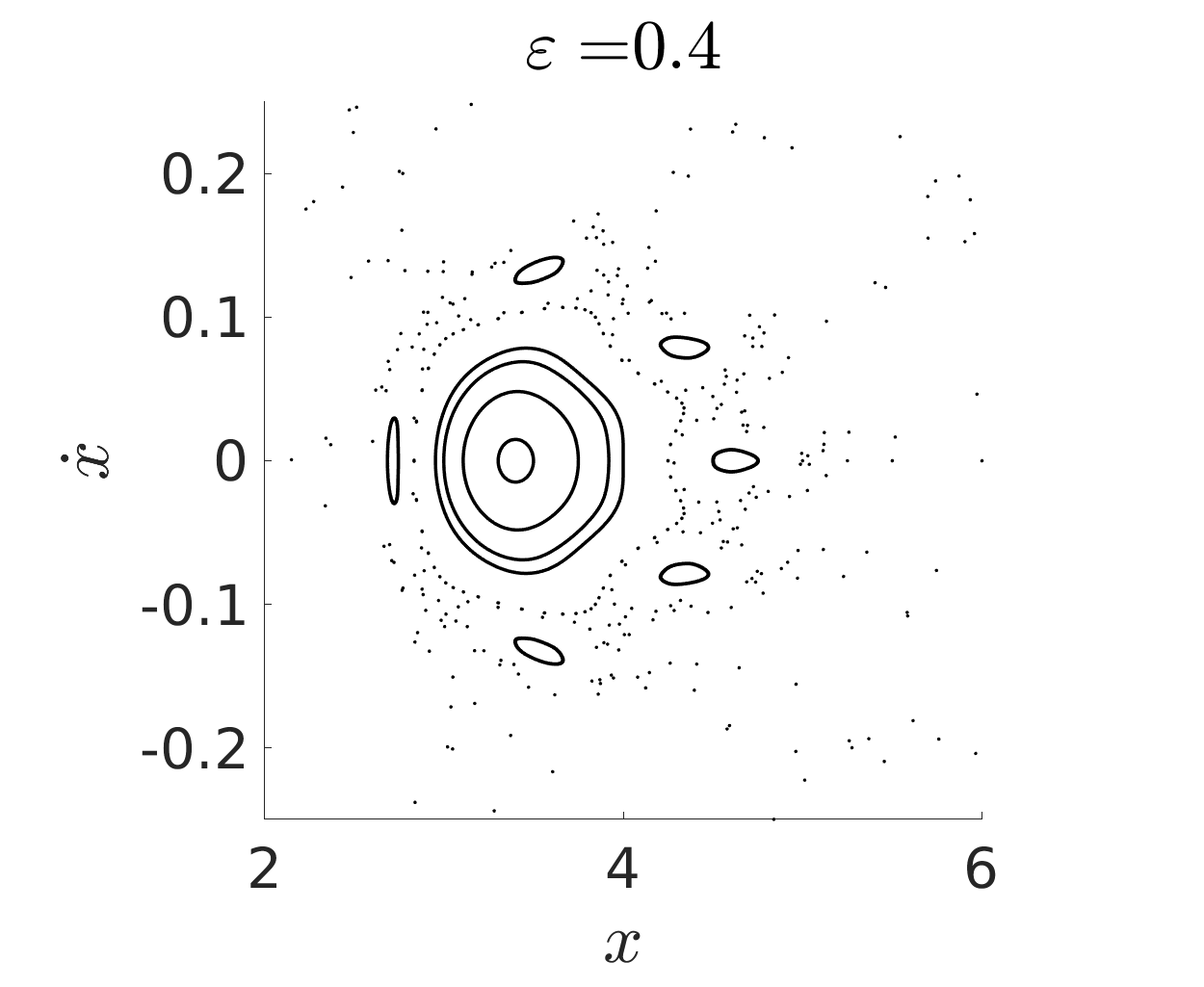}&\includegraphics[scale=0.27]{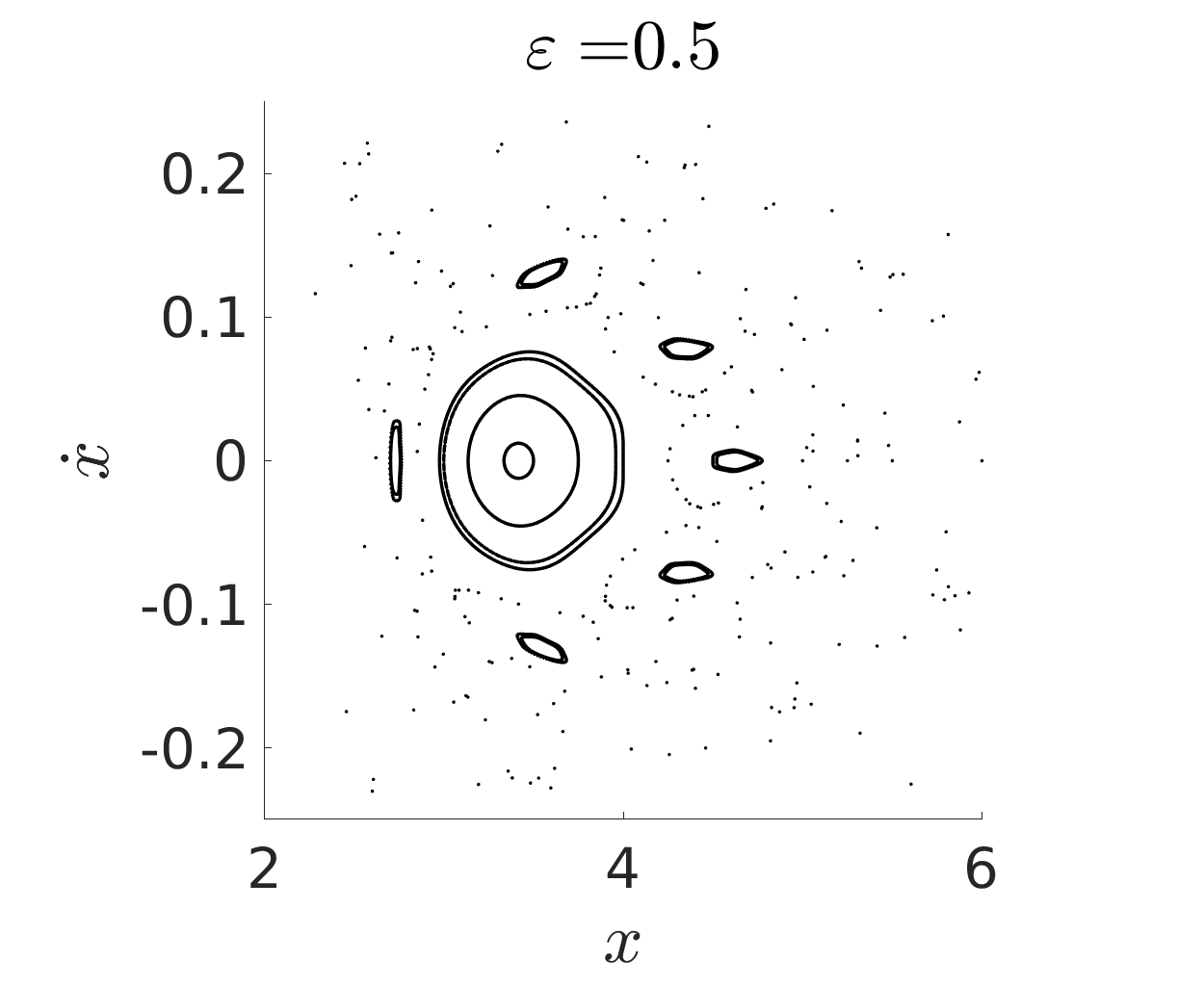} \\\includegraphics[scale=0.27]{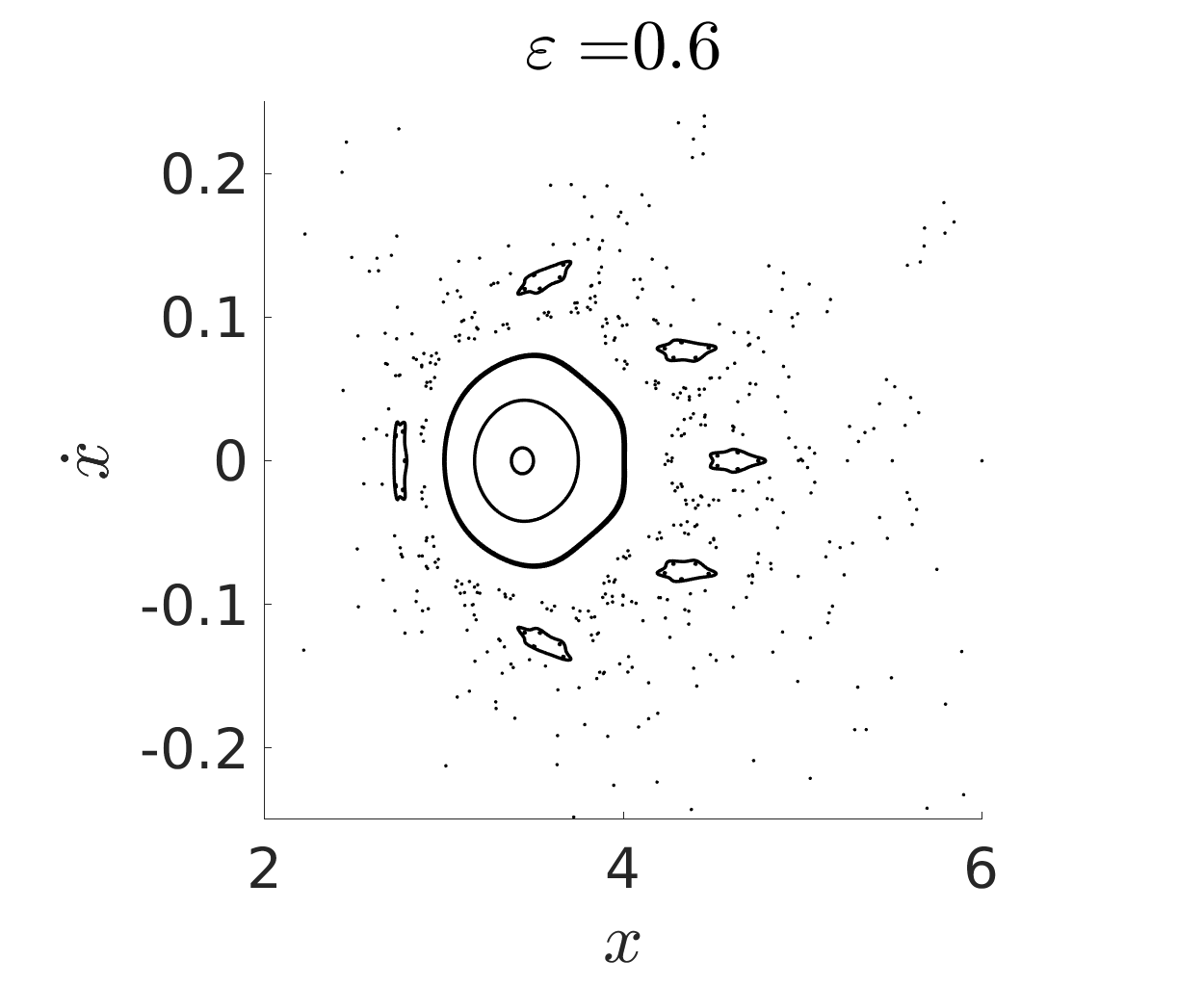}&\includegraphics[scale=0.27]{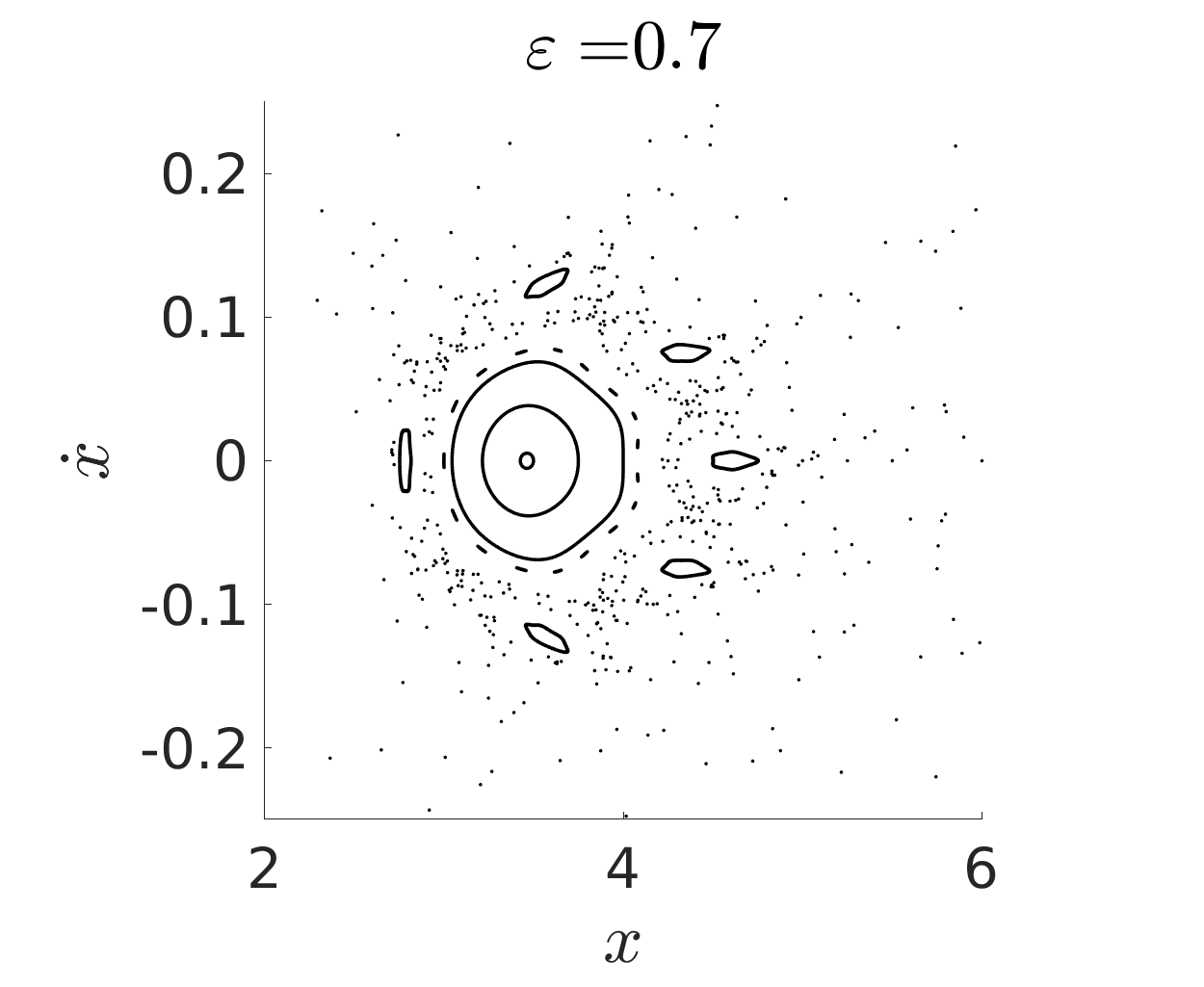}&\includegraphics[scale=0.27]{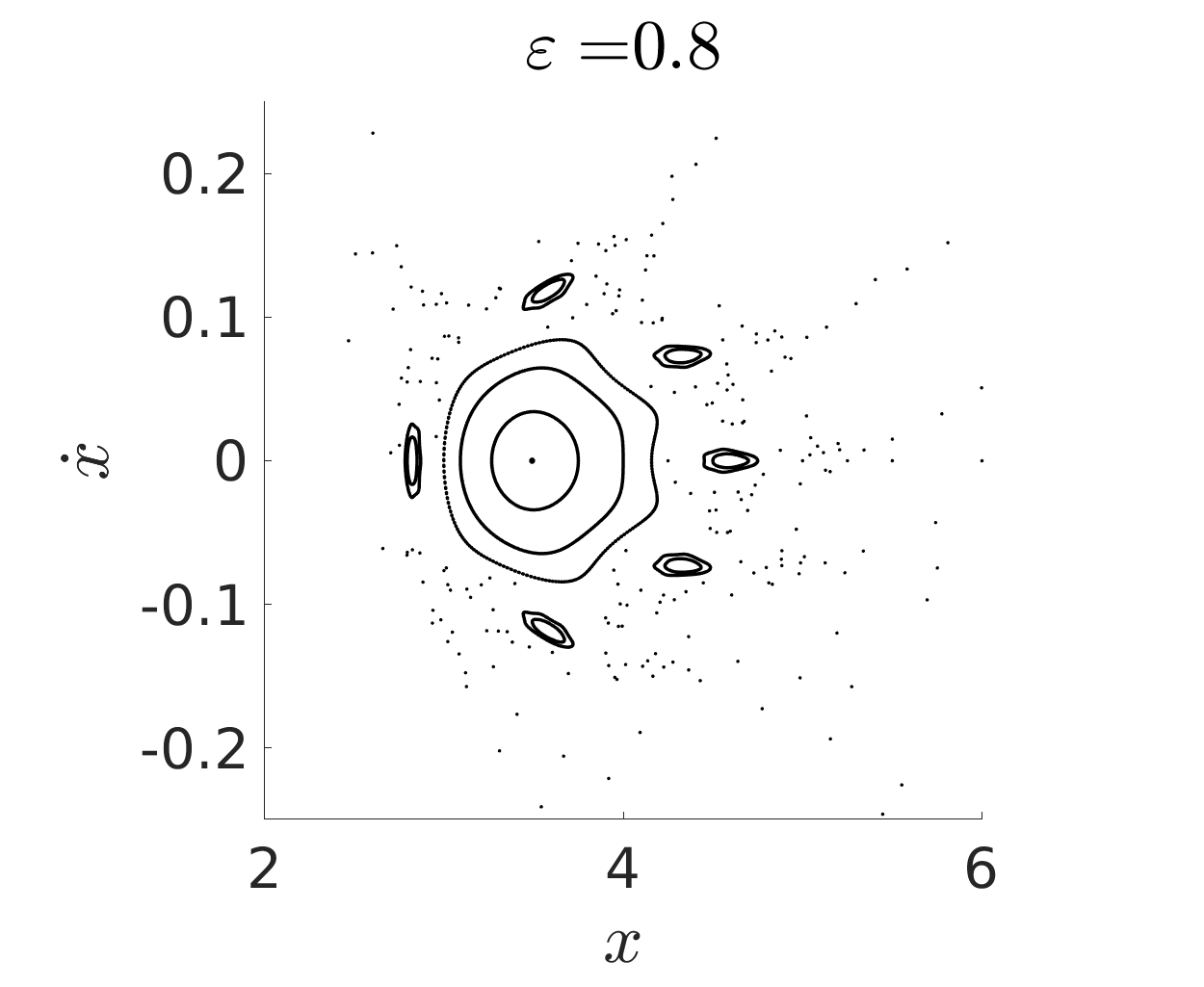} \\
		\includegraphics[scale=0.27]{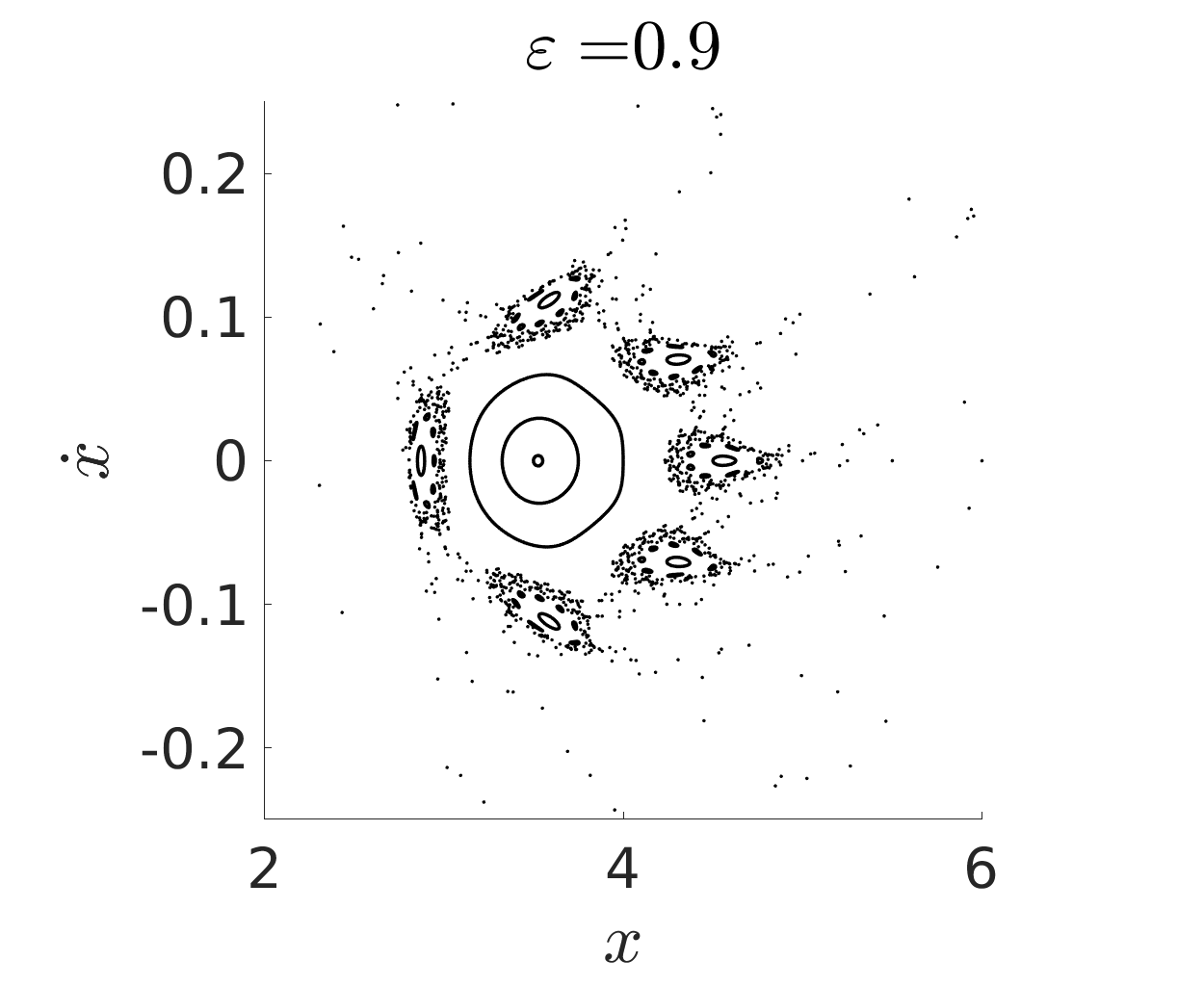}&\includegraphics[scale=0.27]{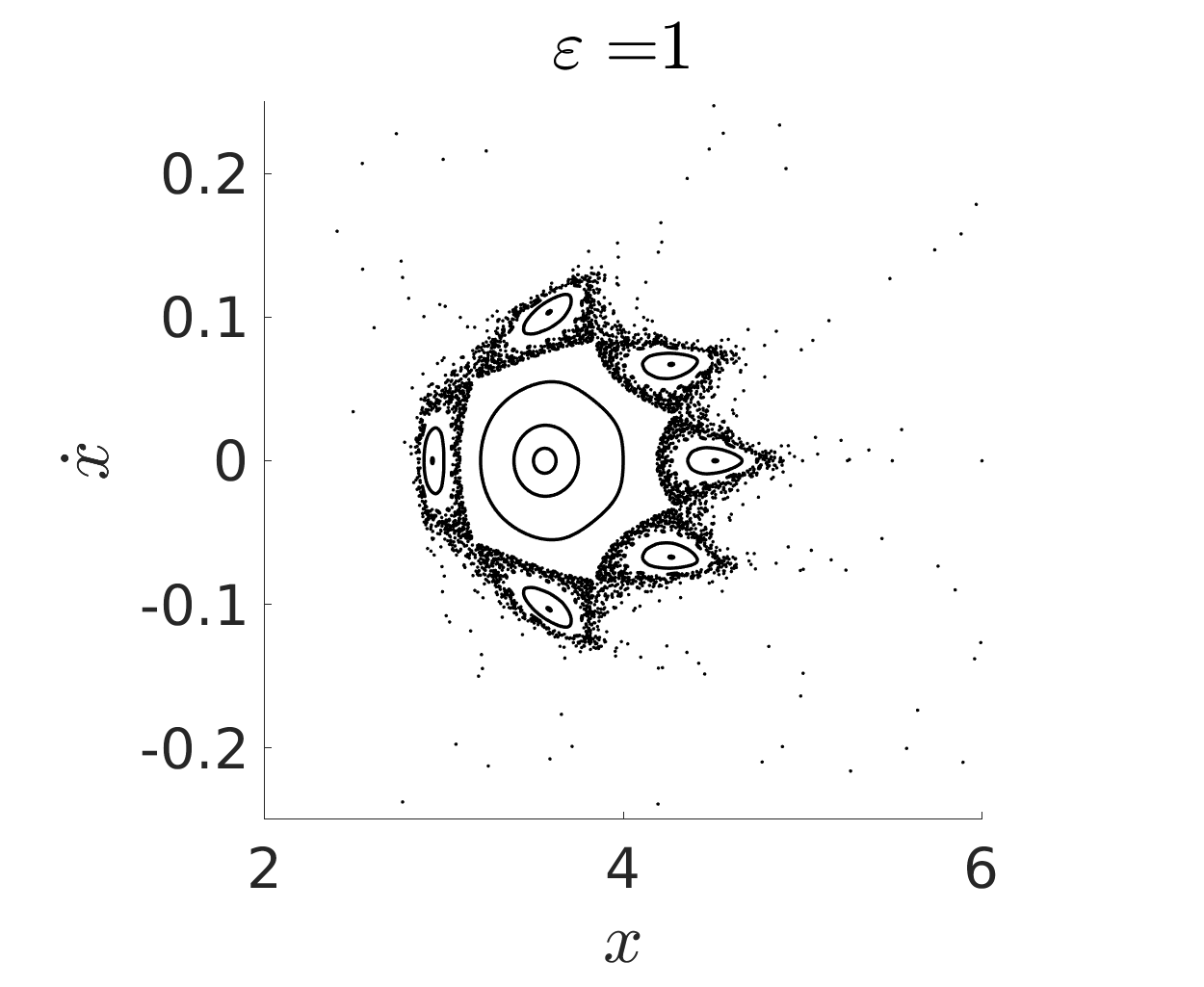}&\\
	\end{tabular}
	\caption{The Poincar\'e map for orbits with $x_0=[3.0,3.5,3.75,4.0,4.25,4.5,4.75,5.0,5.25,5.5,6.0]$ and $y_0=[0.0]$ as $\epsilon$ increases from 0 to 1 in steps of 0.1. The system is evolved for 3000 iterations in time-steps of $\tau=10$ and the Poincar\'e map of section for $y=0, \dot{y}<0$ is plotted for all 11 initial conditions for the Copenhagen system ($\mu_{1}$=$\mu_{2}=0.5$) }
	\label{fig:PS_0.5}
\end{figure*}

The Poincar\'e map, or the first return map, is a powerful and conventional tool for examining the motion of dynamical systems \citep{Tabor89, Parker89, Dubeibe16}. In order to construct the map, we evolve the system for 3000 iterations in time-steps of $\tau=10$ and plot the section of the orbit for $y=0.0, \dot{y}<0$. This is done for 11 initial conditions 
$x_0=[3.0,3.5,3.75,4.0,4.25,4.5,4.75,5.0,5.25,5.5,6.0]$ and $y_0=0.0$ while increasing $\epsilon$ from 0.0 to 1.0 in steps of 0.1. Figure (\ref{fig:PS_0.001}) and Figure (\ref{fig:PS_0.5}) show the evolution of the Poincar\'e map for the biased-mass and and Copenhagen systems respectively. The Poincar\'e map of a system primarily shows two types of structures: concentric Kolmogorov-Arnold-Moser (KAM) tori which represent bounded, quasi-periodic motions and a sea of chaotic points surrounding such tori. At the centre of each island of concentric tori is a point which corresponds to a stable, periodic and resonant orbit \citep*{Gidea07, Broer10, Huang14a}. The extent of the sea of scattered points in comparison to islands of the tori provides a visual representation of the extent of chaos in the system.

For the biased-mass system, the initial conditions $x_{0}=[3.5,3.75,4.0,4.25,4.5]$ show KAM tori on their Poincar\'e maps for all values of $\epsilon$, implying quasi-periodic orbits. For $x_{0}=[4.75,5.0,5.25,5.5,6.0]$, the destruction of their KAM tori implies chaotic or sticky orbits, as was observed in subsection (\ref{sec:orbits}). For the Copenhagen system, the Poincar\'e maps $x_0=[3.5,3.75,4.0,4.5,4.75]$ show KAM tori for all values of $\epsilon$. The torus for $x_0=3.0$ breaks up only for $\epsilon=0.9$, while the tori for $x_0=5.5$ and $x_0=[5.0,5.25]$ break up for $\epsilon>0.0$ and $\epsilon>0.1$ respectively. No KAM tori appear for $x_0=6.0$ for any value of $\epsilon$, implying that the initial condition is chaotic for all values of $\epsilon$. Thus, the Poincar\'e maps for both the systems corroborate the observations presented in subsection (\ref{sec:orbits}). 

\section{Lyapunov Characteristic Exponents}

A very popular indicator of chaos in dynamical systems is the calculation of the Lyapunov Characteristic Exponents (LCE), which has been extensively applied to the study of chaos in celestial dynamics especially in the context of the three-body problem (\citet{Gueron01, Dubeibe16, Dubeibe17}; \citet*{Wu03, Wu06b}). It is a measure of the exponential divergence of two neighbouring trajectories in phase space. The rate of separation of the two trajectories is dependent on the initial separation vector. For a pair of trajectories, the number of exponents for the system is equal to dimension of its phase space. However, the largest exponent dominates in the limit $t\to\infty$. The largest Lyapunov exponent, called the Maximal Lyapunov Exponent (MLE), is defined by,
\begin{equation}
\Lambda_{max}=\lim_{t\to\infty}\frac{1}{t}\log\frac{||\Upsilon(t)||}{||\Upsilon(0)||}
\label{eqn:LyapunovVar}
\end{equation}
where $\Upsilon(t)$ is the solution to the variational equations for the potential under consideration \citep*{Tancredi01}. Such a computation mechanism for the MLE is called the variational method and is the most accurate. However, for systems such as the one under consideration where computation of the variational equations are cumbersome, an alternative was introduced in \citet*{Benettin76}. The equation (\ref{eqn:LyapunovVar}) is thus replaced by the following:
\begin{equation}
\Lambda_{max}=\lim_{t\to\infty}\frac{1}{t}\log\frac{||\mathbf{\delta x}(t)||}{||\mathbf{\delta x}(0)||}
\label{eqn:Lyapunov}
\end{equation}
where, the deviation vector between the two trajectories is $\mathbf{\delta x(t)}$, with $\mathbf{\delta x(0)}\rightarrow0$. The mean rate of deviation of the two trajectories is given by:
\begin{align}
\Lambda_{max}=&\frac{1}{n\tau}\sum_{k=1}^n\log\frac{||\mathbf{\delta x}(k\tau)||}{||\mathbf{\delta x}(0)||}
\end{align}

\begin{figure}
	\includegraphics[scale=0.2]{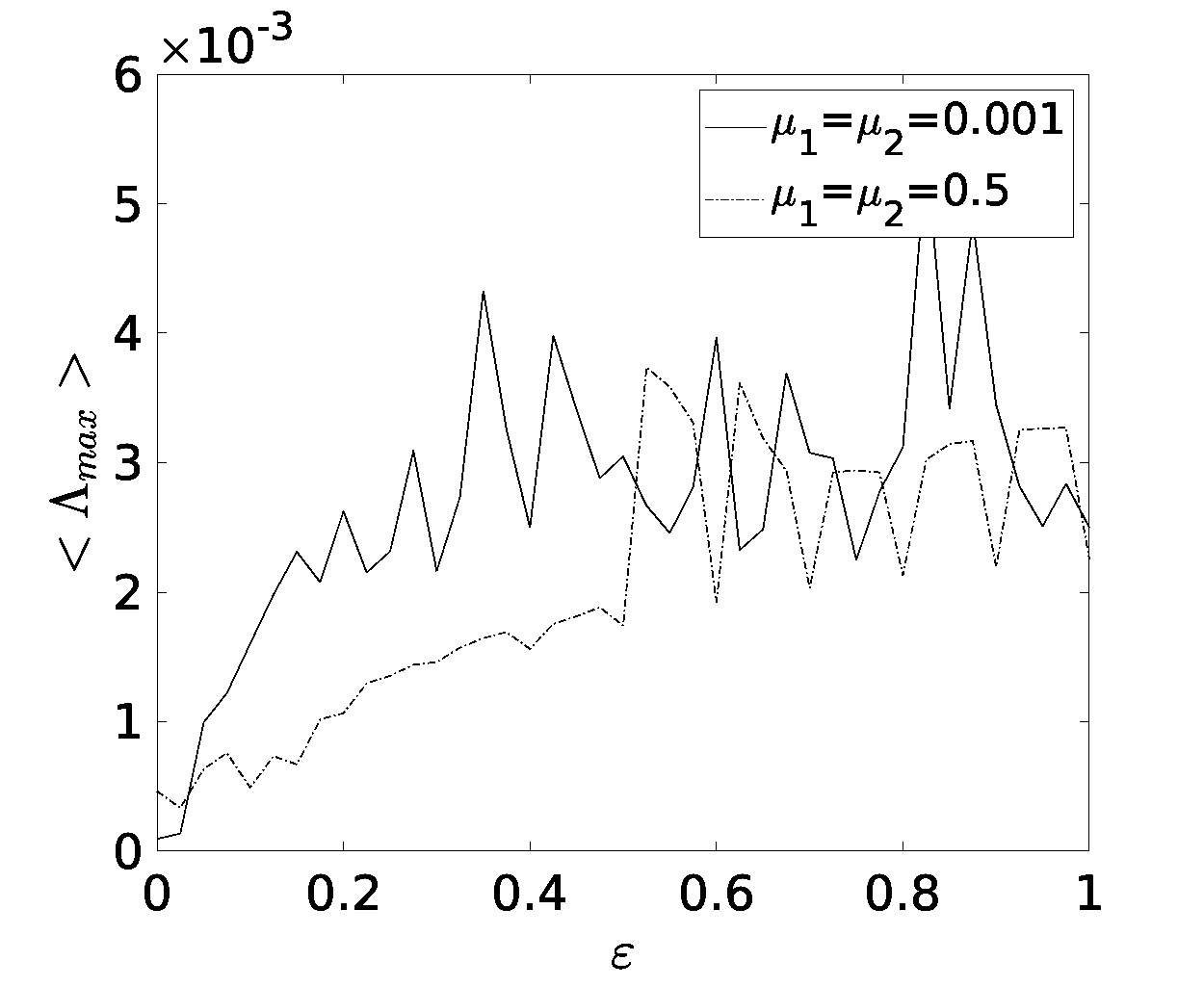}
	\caption{Plot of the Lyapunov Characteristic Exponent (LCE) vs $\epsilon$.} 
	\label{fig:LyapunovPlot}
\end{figure}

This method is called the two-particle method and is the one we utilize to calculate the MLE for each system. The result is accurate as long as the two trajectories are in the immediate neighbourhood of each other in phase space and the machine used for computation has enough precision. As concluded in the work by \citet{Tancredi01}, we have also taken the initial separation between the two trajectories to be $\mathbf{\delta x(0)}=10^{-8}$ and have integrated the system in double precision for $n=10^{5}$ iterations, each of time step $\tau=0.1$. Numerical integration diverges rapidly unless the deviation vectors are re-normalized periodically. The two trajectories are evolved separately and the deviation vector is re-normalized using the Gram-Schmidt re-normalization after each time step. To get a quantitative representation of the chaos in the system, the MLE is averaged over the entire phase space. But, as a preliminary investigation of system, we use the initial conditions: $x_{0}=[3.0,3.5,3.75,4.0,4.25,4.5,4.75,5.0,5.25,5.5,6.0]$, $y_{0}=0.0$ and $\dot{x}_{0}=0.0$, with $C_{j}=4.0$ for the biased-mass and Copenhagen systems. If trajectories are stable, the value of the MLE remains very small, usually less than $5\times10^{-4}$ (low value of MLE). But for chaotic trajectories, the deviations are exponential and the value of the MLE increases rapidly with time. After $10^5$ iterations, its value is usually greater than $5\times10^{-4}$ (high value of MLE). The MLE for initial conditions $x_0=[3.0,3.5,3.75,4.0,4.25,4.5,4.75,5.0,5.25,5.5,6.0]$ are calculated and averaged for each value of $\epsilon$ and is called the Lyapunov Characteristic Exponent (LCE) for the particular value of $\epsilon$ \citep*{Nag17}. The LCE provides a qualitative measure of the amount of chaos in the system, even for the few initial conditions chosen for the study (see \citet{Nag17} as an example). Figure (\ref{fig:LyapunovPlot}) is the plot of the LCE against $\epsilon$ for both the biased-mass and Copenhagen systems.

Both for the biased-mass system and the Copenhagen system, the total chaos in the system for small $\epsilon$ is low. For the biased-mass system, the LCE for all the initial conditions are $<5\times10^{-4}$ for values of $\epsilon=0.0$ indicating stable orbits. The same is true for the Copenhagen system, except for $x_0=6.0$ which gives an LCE of $4.22\times10^{-03}$. For the biased-mass system, the LCE for all initial conditions are $<5\times10^{-4}$ for $x_0=[3.5,3.75,4.0,4.5]$, implying stable orbits. Some initial conditions, like $x_0=[4.25,4.75,5.0,5.25,6.0]$ for the biased mass system and $x_0=5.5$ for the Copenhagen system, the system shows high values of LCE for intermediate values of $\epsilon$, but low values of LCE for higher values of $\epsilon$. The most interesting among these is the initial condition $x_0=6.0$, which shows low values of LCE only for $\epsilon=0.0$ and $1.0$. This reaffirms the conclusion drawn from the Poincar\'e maps that the chaos in the system is maximum for intermediate values of $\epsilon$. For the Copenhagen system, values of LCE for $x_0=[3.5,3.75,4.0,4.5,4.75]$ are low for all values of $\epsilon$. The initial condition $x_0=6.0$ show high values of LCE for values of $\epsilon$. The initial condition $x_0=5.5$ shows high values of LCE for all values of $\epsilon$ except for $\epsilon=0.0$.

Figure (\ref{fig:LyapunovPlot}) shows that the chaos in the system is low for both the biased-mass system and the Copenhagen system, as indicated by low values of the averaged LCE. Its value rises rapidly for the biased mass system but much slower for the Copenhagen system. Both the systems show maximum values of the averaged LCE for intermediate values of epsilon, which for the biased mass system is at $\epsilon=0.825$ and $\epsilon=0.525$ for the Copenhagen. This re-iterates the observations made from the orbital evolution and the Poincar\'e maps of the systems.

\begin{figure*}
	\begin{tabular}{cc}
		\includegraphics[scale=0.4]{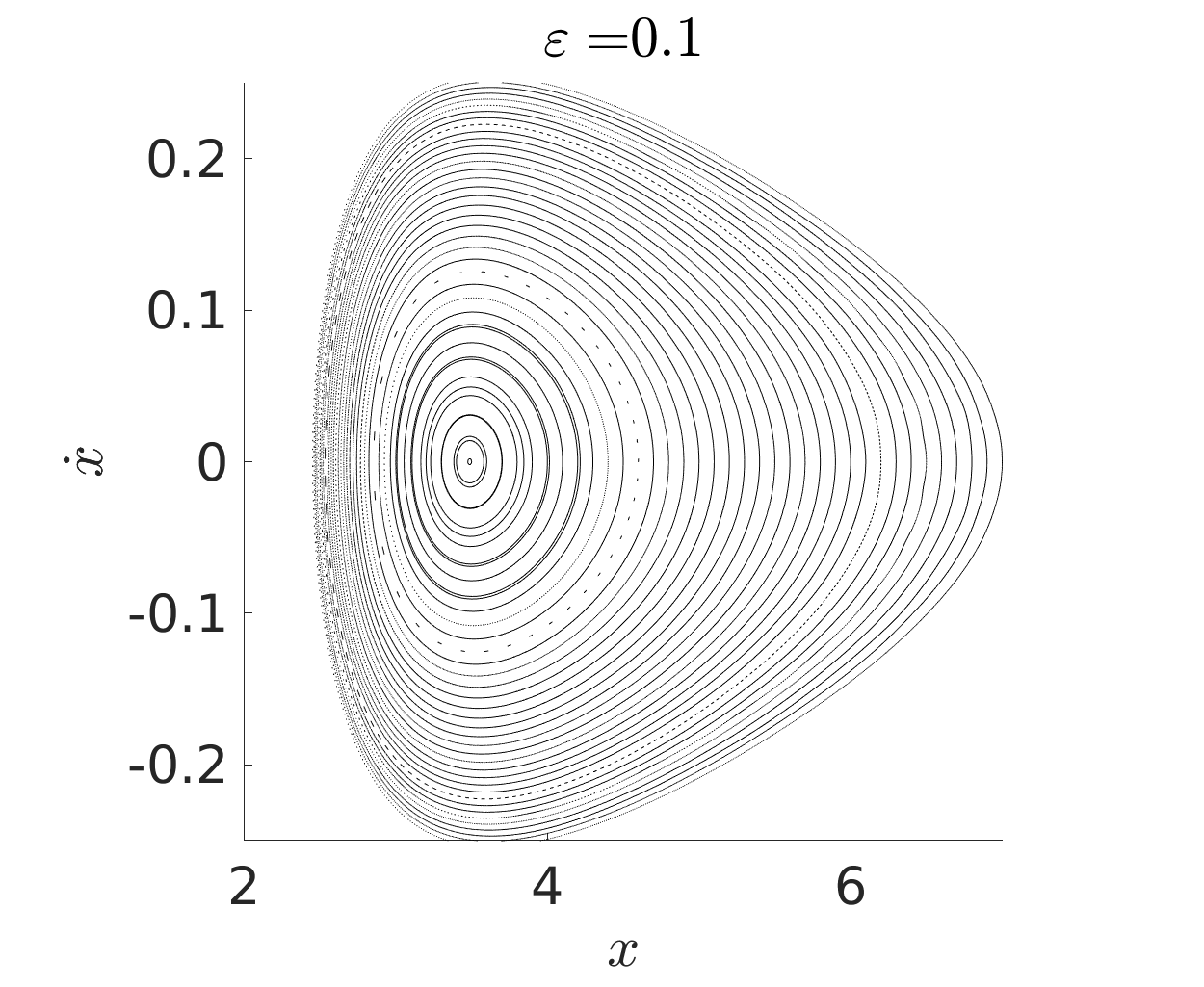}&\includegraphics[scale=0.4]{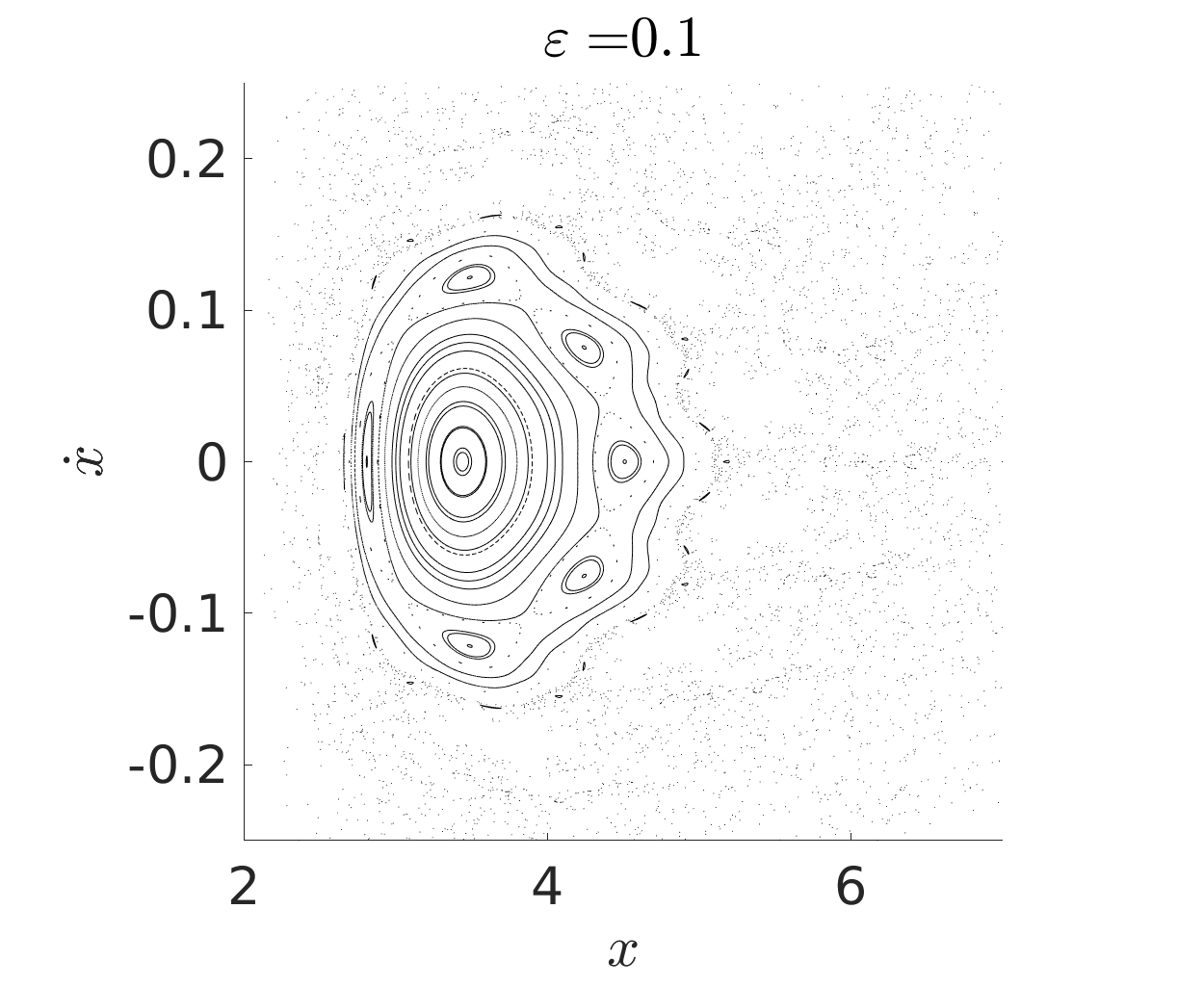} \\
		\includegraphics[scale=0.4]{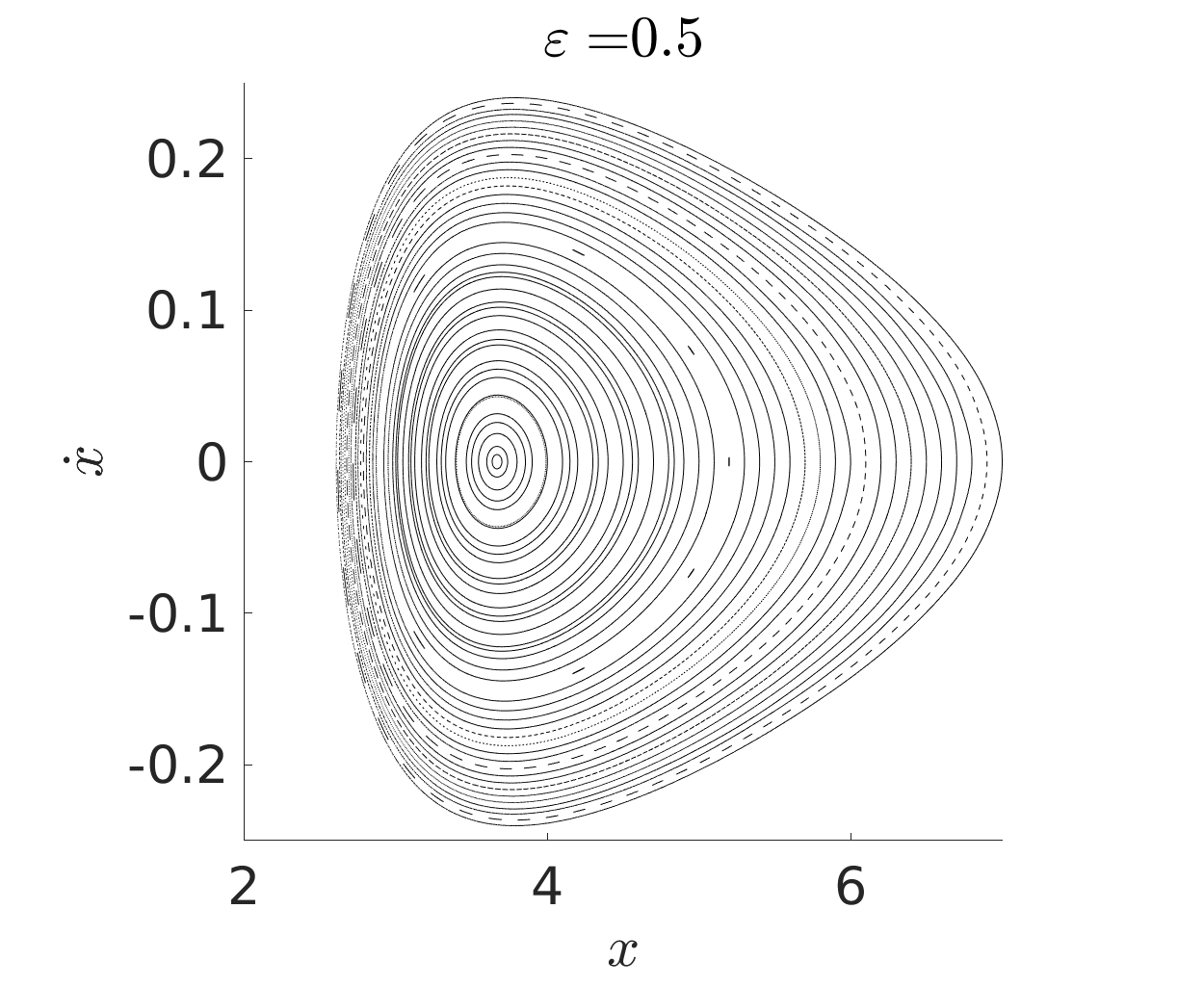}&\includegraphics[scale=0.4]{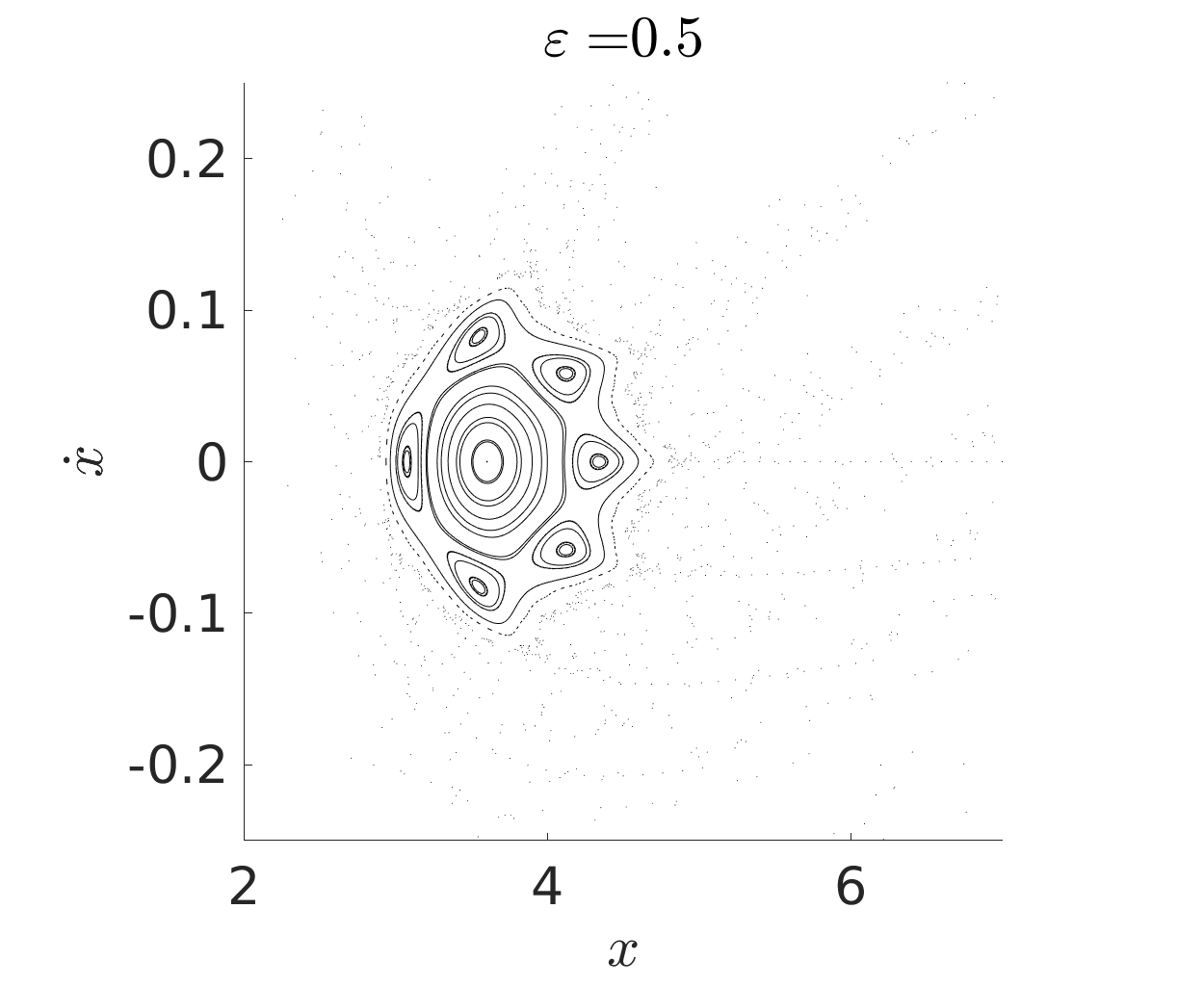} \\
		\includegraphics[scale=0.4]{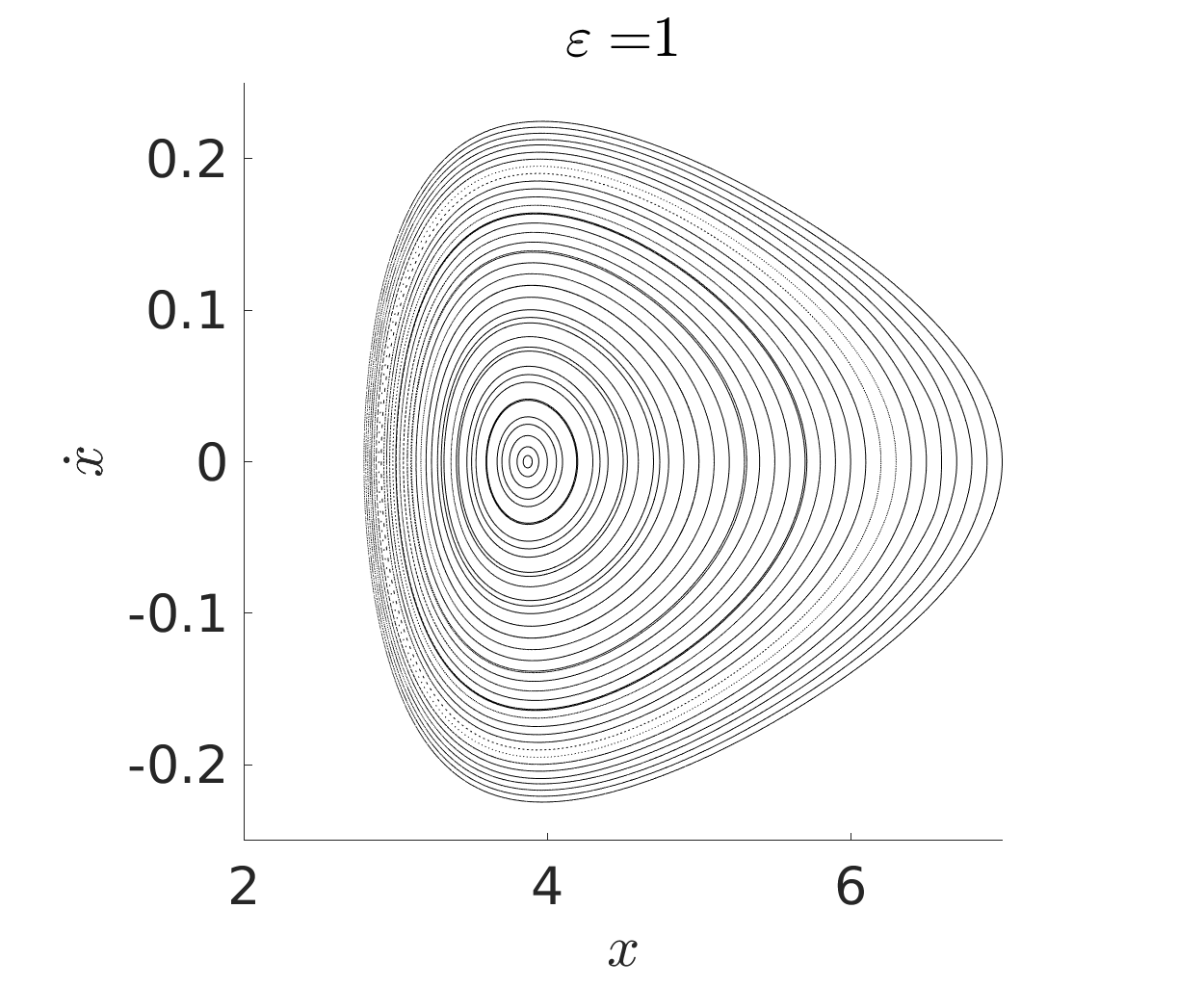}&\includegraphics[scale=0.4]{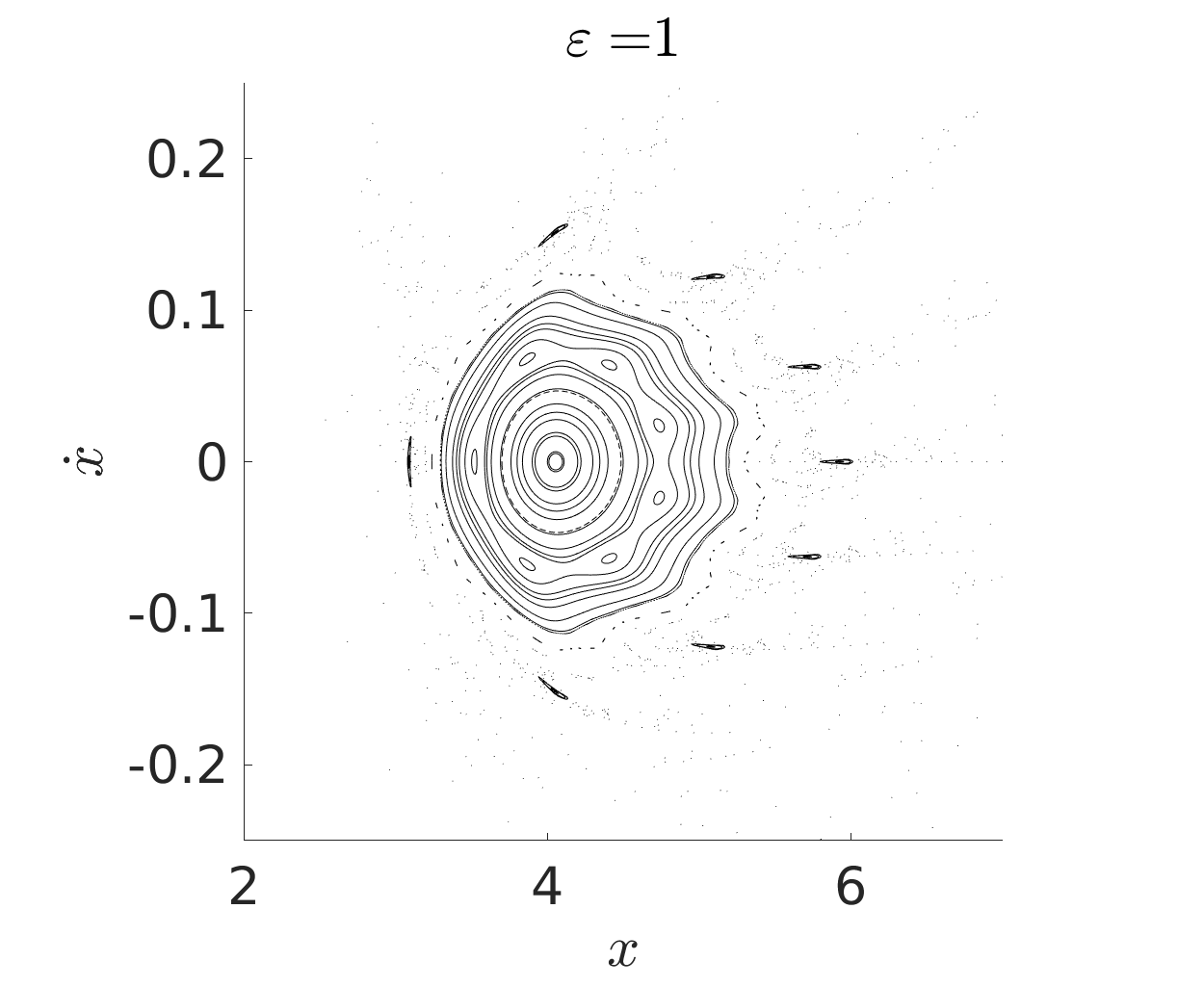} \\
	\end{tabular}
	\caption{Poincar\'e map of section for $y=0$ and $\dot{y}<0.0$ for the biased-mass systems (mass ratio of the primaries equals to 0.001) for different values of $\epsilon$. The figures on the left are maps for the system with Schwarzschild-like primaries while those on the right are for the system with Kerr-like primaries.}
	\label{fig:kerr_pesudokerr0001}
\end{figure*}

\begin{figure*}
	\begin{tabular}{cc}
		\includegraphics[scale=0.4]{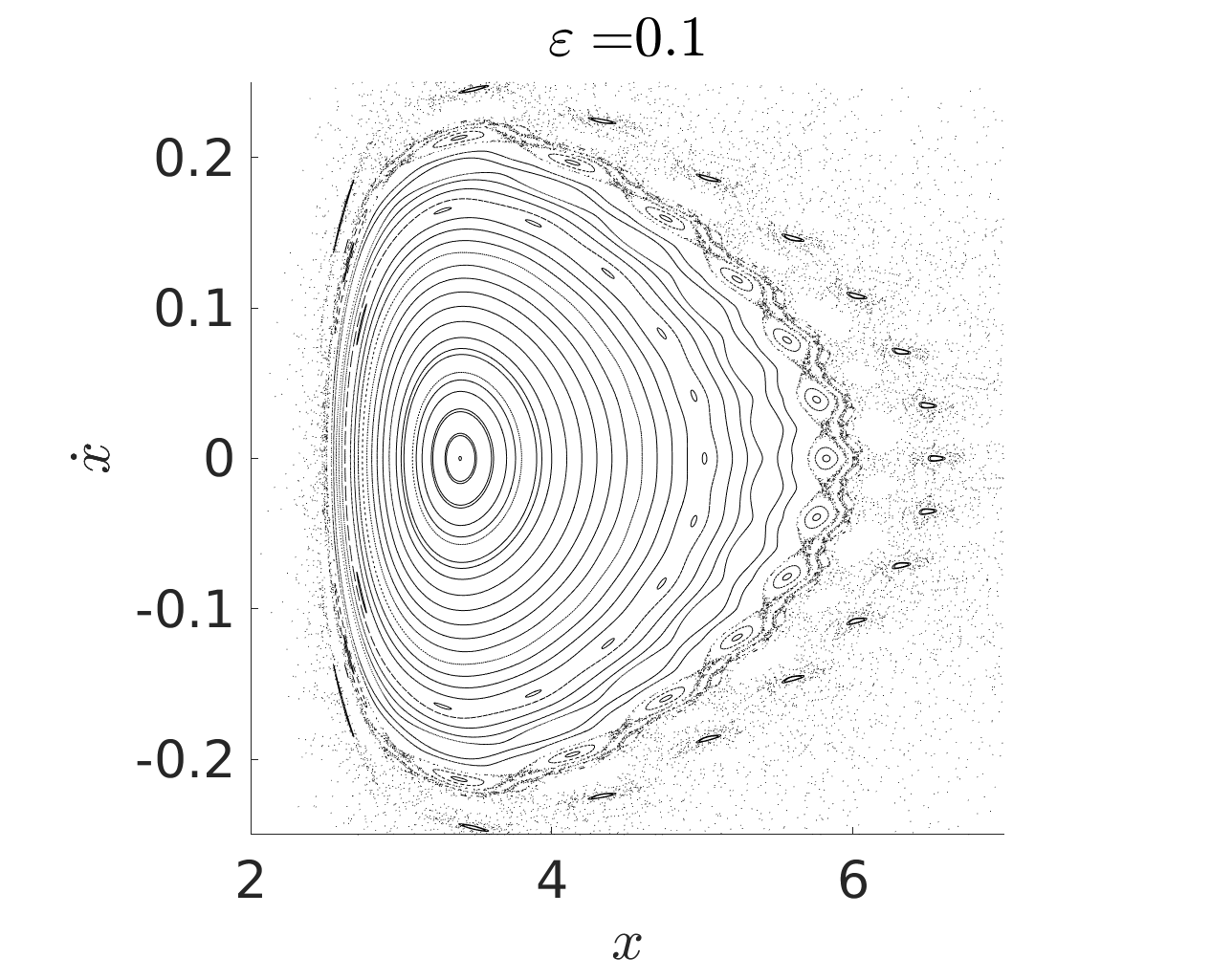}&\includegraphics[scale=0.4]{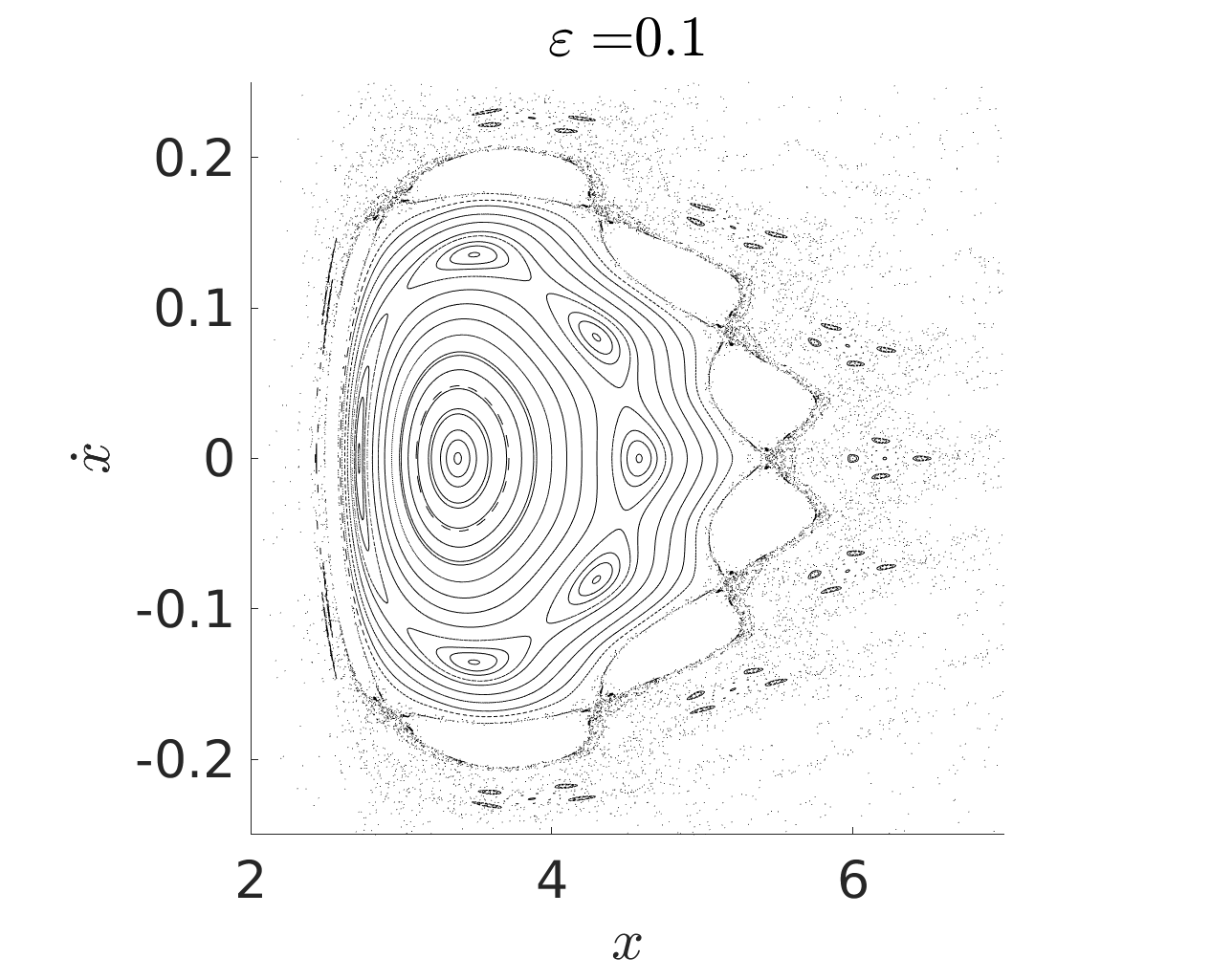} \\
		\includegraphics[scale=0.4]{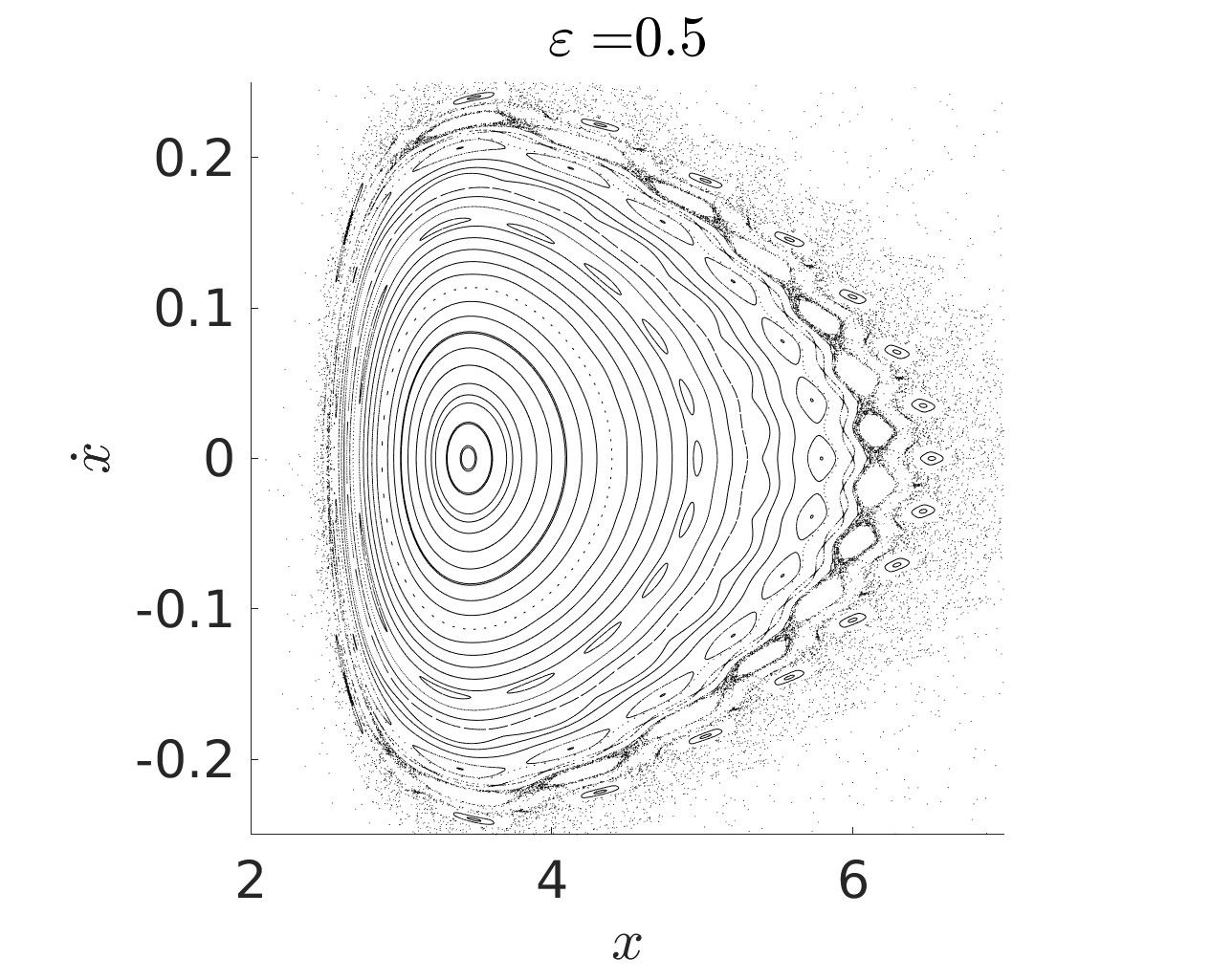}&\includegraphics[scale=0.4]{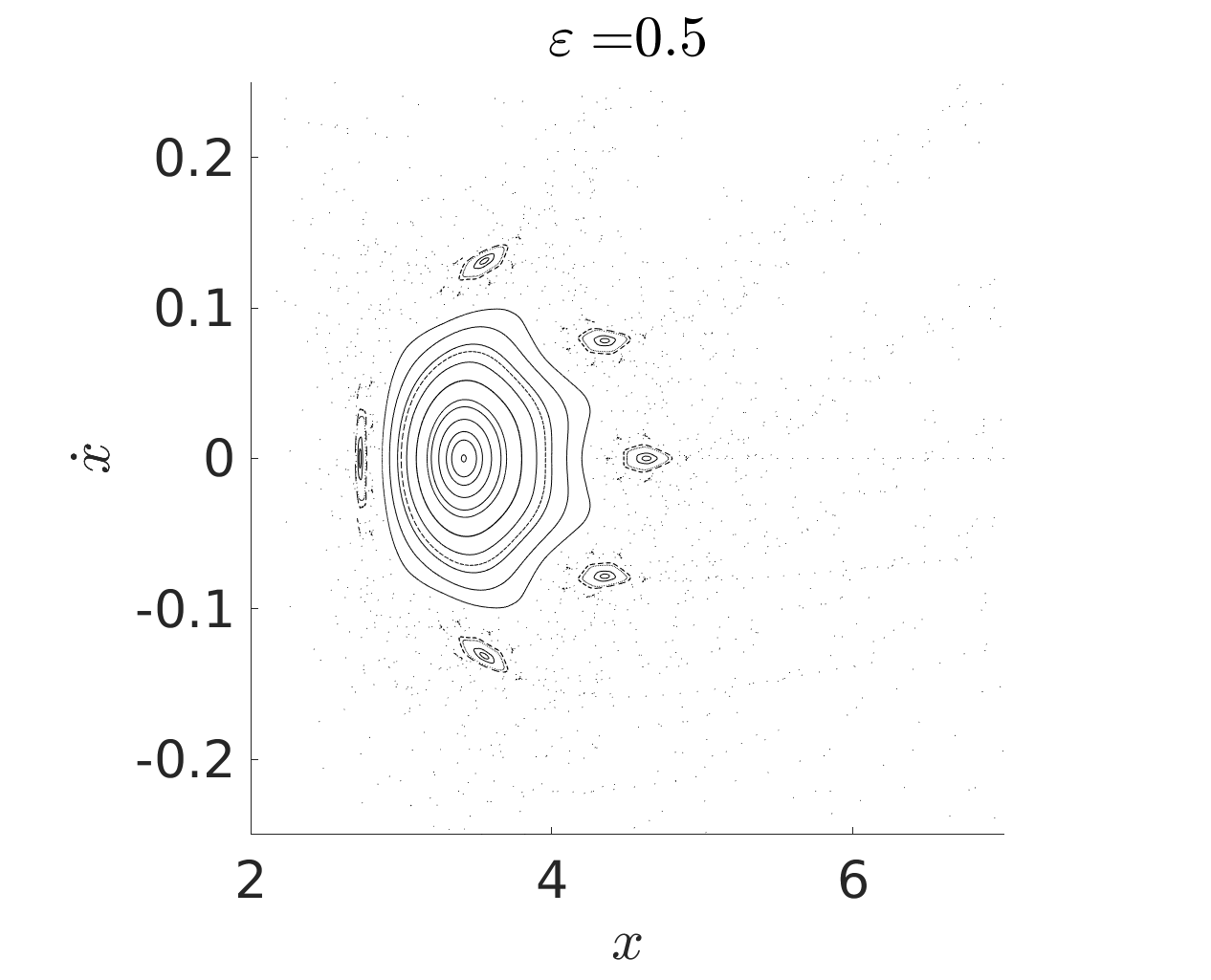} \\
		\includegraphics[scale=0.4]{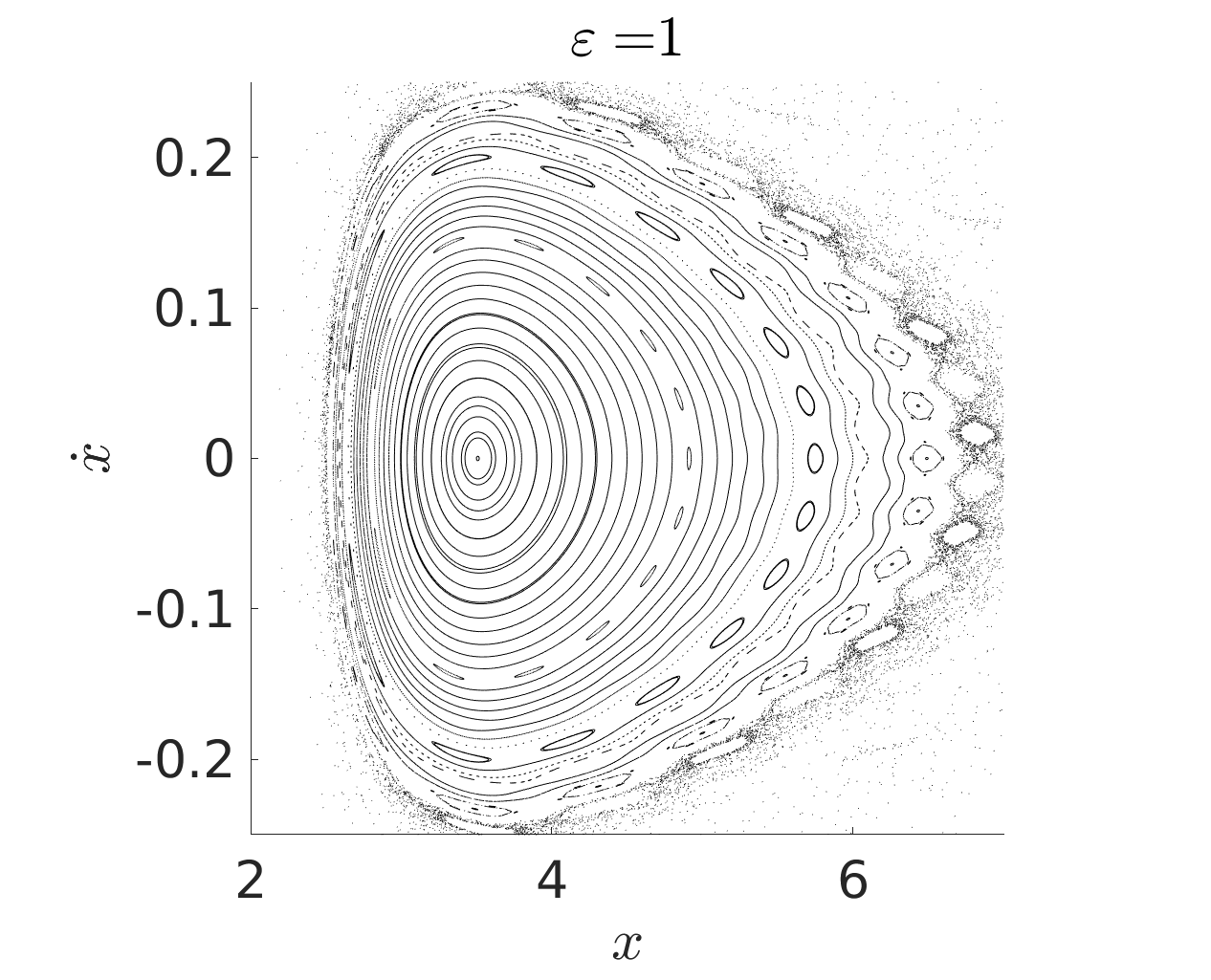}&\includegraphics[scale=0.4]{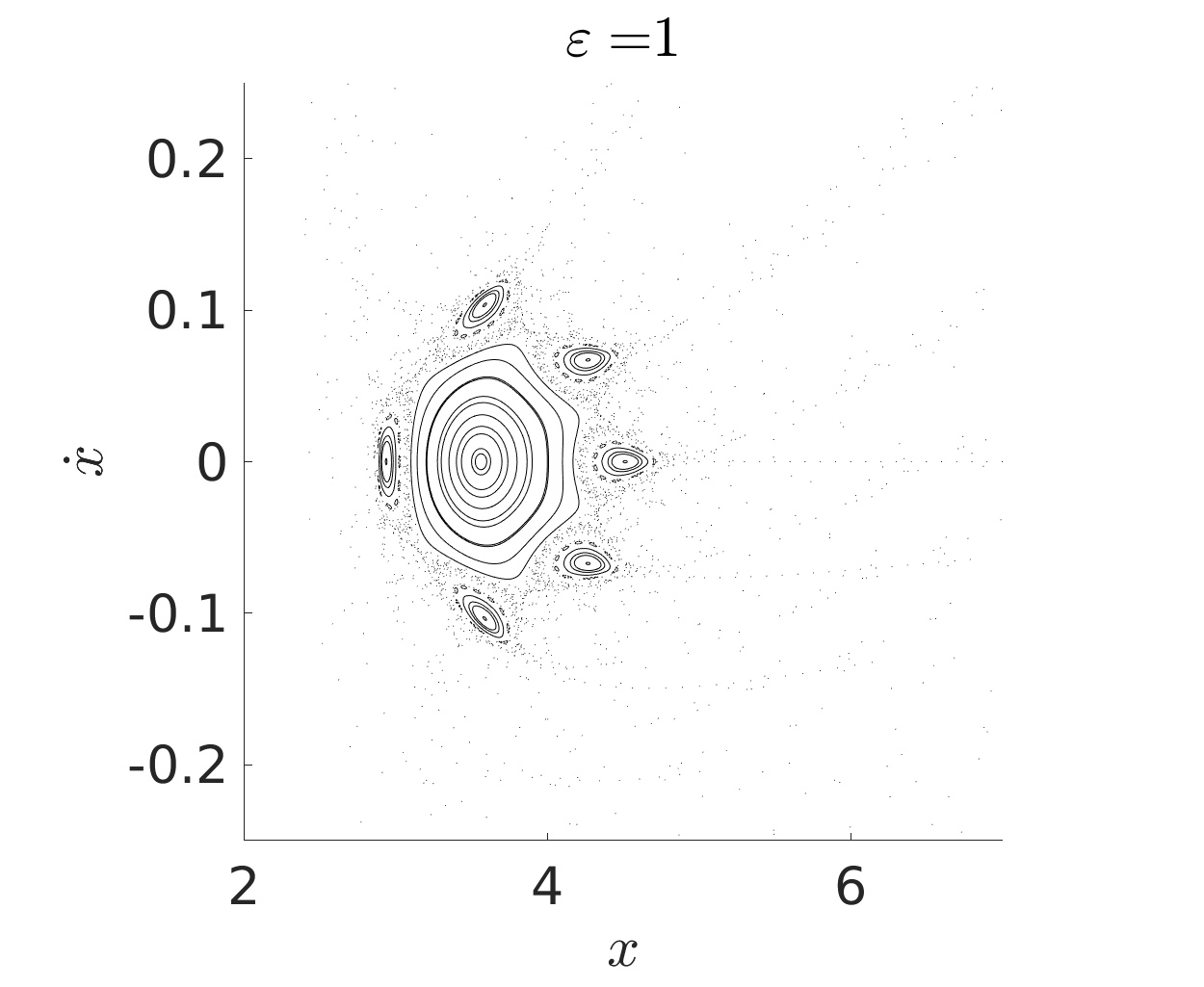} \\
	\end{tabular}
	\caption{Poincar\'e map of section for $y=0.0$ and $\dot{y}<0.0$ for the Copenhagen systems (mass ratio of the primaries equals to 0.5) for different values of $\epsilon$. The figures on the left are maps for the system with Schwarzschild-like primaries while those on the right are for the system with Kerr-like primaries.}
	\label{fig:kerr_pesudokerr05}
\end{figure*}

\section{Schwarzschild and Kerr primaries: a comparison}
In order to examine the effect of the spin of the primaries on the system, we present a comparison to a system with two Schwarzschild like primaries. Using the potential described in \citet{Dubeibe16} and \citet{Zotos17b}, we construct a set of Poincar\'e maps of section for $\epsilon=[0.1,0.5,1.0]$. We evolve each orbit for 3000 iterations in time-steps of $\tau=10$ and plot the section of the orbit for $y=0.0$ and $\dot{y}<0.0$. Figures (\ref{fig:kerr_pesudokerr0001}) and (\ref{fig:kerr_pesudokerr05}) show Poincar\'e maps of the biased-mass and Copenhagen systems respectively. For both the systems, the plots on the left are for the system with Schwarzschild like primaries, while those on the right are for the system with Kerr like primaries. 

While for $\epsilon=0.0$ both systems reduce to the Newtonian CRTBP, it is apparent that even for small perturbations to the Newtonian system, as represented by $\epsilon=0.1$, the introduction of the spin destabilizes a number of initial conditions. For the biased-mass case, the Schwarzschild system shows all chosen initial conditions to be stable and quasi-periodic, with the Poincar\'e map showing KAM tori for all values of $\epsilon$. The Poincar\'e map for the Kerr system differs radically from its Schwarzschild counterpart even for $\epsilon=0.1$, showing a large sea of chaotic points surrounding an island of stable initial conditions. The island of stability grows smaller as $\epsilon$ is increased, as has already been discussed in section (\ref{sec:PS}). For the Copenhagen case, the Poincar\'e maps for both the Schwarzschild and Kerr systems feature an island of stability surrounded by a sea of chaos for all three values of $\epsilon$. The Poincar\'e maps for both the systems look alike, implying a similar number of stable initial conditions. As $\epsilon$ is increased, the number of stable initial conditions for the Kerr system decreases rapidly, as evident from the smaller islands of stability on the Poincar\'e maps of the system for $\epsilon=0.5$ and $\epsilon=1.0$. However, for the same values of $\epsilon$, the number of stable initial conditions for the Schwarzschild system remains approximately the same.

Thus, for both mass ratios, we observe that the introduction of spin in the CRTBP with Schwarzschild-like primaries destabilizes a number of initial conditions, with the amount of chaos in the system growing with increase in $\epsilon$.

\section{Conclusions}

In the present paper, we present a beyond-Newtonian potential for the planar circular restricted three-body problem with Kerr like primaries. This is achieved by using the Fodor-Hoenselaers-Perj\'es procedure to expand the Kerr metric and by retaining corrections up to the first non-Newtonian term. The system is conservative, with the Hamiltonian being time independent. The parameter $\epsilon\in[0.0,1.0]$ is introduced in order to facilitate the observation of the system as it transitions from the Newtonian to the beyond-Newtonian regime. The dynamics of a test particle in this potential for $\mu_{1}=\mu_{2}=0.001$ (or the biased-mass system) and for $\mu_{1}=\mu_{2}=0.5$ (or the Copenhagen system), are inspected for a Jacobi constant $C_{j}=4.0$. For an initial investigation of the system, orbits for a few selected initial conditions are plotted. A short analysis of the fixed points of the systems and their stability is undertaken. A purely Newtonian CRTBP system is known to have five Lagrange points, as seen for $\epsilon\,=\,0$ in our case. However, number of Lagrange points is not constant as the system transitions from the Newtonian to the beyond-Newtonian regime. It is observed that the number of fixed points strongly depends on the parameter $\epsilon$ as does their stability. Next, the stability of the orbits is also examined through the use of the Poincar\'e map of section for different values of $\epsilon$. The Poincar\'e maps for all non-zero values of $\epsilon$ show islands of stability constructed of concentric Kolmogorov-Arnold-Moser (KAM) tori, embedded in a sea of chaos.

Thus we note that the introduction of the parameter $\epsilon$ helps us to conclude that even small perturbations to the Newtonian CRTBP destabilizes the system for both the cases.  If we track the evolution of the system keeping the Jacobian constant fixed, a stable orbit in the Newtonian system is observed to become either chaotic or sometimes even remain regular in the beyond-Newtonian limit. In the limits $\epsilon\,=\,0$ and $\epsilon\,=\,1$, the phase space is seen to be filled mostly with periodic orbits, rarely interspersed with chaotic ones. However, as $\epsilon$ departs even slightly from zero, trajectories that were stable in the Newtonian system become unstable. It is seen that in most of the cases (for a given set of initial conditions) whose phase space is bounded in the classical regime, correspond to unbounded trajectories in the non-Newtonian regime. This implies that both systems become largely unstable for intermediate values of $\epsilon$. The instability of the orbits can possibly be linked to the observed lack of stable fixed points in both the systems. This is also confirmed by the Lyapunov Characteristic Exponent, calculated for each value of $\epsilon$, which is in accordance to the conclusions made by several authors earlier for different systems \citep{Huang14a, Dubeibe16, Nag17}.  In conclusion, we may say that even the smallest corrections to the Newtonian circular restricted three-body problem could drastically change the stability and the dynamics of the system.

In addition, we would like to note that an in-depth study of the phase space using more rapid indicators of chaos, like Fast Lyapunov Indicators (FLI) (\citet*{Froeschle97, Froeschle00}; \citet{Wu06b}), Small Alignment Index (SALI) \citep*{Skokos01} and General Alignment Index (GALI) \citep*{Skokos07} will facilitate a much more detailed analysis of the evolution of the Lagrange points of the proposed potential. Coupled with this, a detailed linear stability analysis of the Lagrange points as a function of the parameter $\epsilon$ and an analysis of the basins of convergence is expected to reveal more information about the attractors of the system. Further, we would also like to investigate the degree of equivalence of the potential constructed in our paper with the pseudo-Newtonian potential formulation of a binary with spinning primaries, for example that of a system modelled by the superposition of two Artemova potentials \citep{Artemova96}. This would allow us to reproduce features like the Innermost Stable Circular Orbit (ISCO), maximally stable orbits, and the horizon radius, in our chosen scalings and units. This would in turn facilitate the calculation of physically relevant distances, for example, the coordinates of fixed points for different values of $\epsilon$ and primary masses in real physical units, thereby allowing us to predict real astrophysical scenarios using our present model (for a recent example refer to \citet*{Yi20}). Thus, we would like to explore these issues in greater depth as part of our future work.

We also note that the current formalism is strictly valid for particles whose motion is restricted to the plane containing the primaries. However, a more general model for accreting particles should also include a study of the dynamics of such off-axis motion. Thus, we would like to direct our future studies to incorporate such effects for off-axis halo particles in a generalized beyond-Newtonian framework. 
\section*{Acknowledgements}

The authors would like to thank Dr. Sankhasubhra Nag for his helpful suggestions and discussions. The authors would also like to thank Ms. Pratyusha Banerjee for her help with computational work. In addition, they would also like to acknowledge Dr. Tanaya Bhattacharyya for taking time out to go through the manuscript meticulously. Last but not the least, the authors would like to acknowledge the anonymous referee for his/her valuable comments and suggestions. We would like to dedicate this paper to all the warriors fighting the COVID-19 pandemic across the globe. 

\section*{Data Availability}
No new data were generated or analysed in support of this research.








\appendix
\section{Table for Orbit Classification for biased-mass system}
\label{tab:Details0001} 
\begin{tabular}{p{0.2cm}p{2.4cm}p{2.05cm}p{2.07cm}}
\hline \textbf{$\epsilon$} & \textbf{$x_0$} & \textbf{Orbit Type} &	\textbf{Poincar\'e Map}\\ \hline
0.0 &   all                 &   Regular     & Quasi-periodic \\		0.1 &   [5.25,5.5,6]        &   Sticky      & Chaotic\\
0.2 &   [4.25,5,5.25,5.5,6] &   Sticky/Escaping &Chaotic\\
0.3 &   [5,5.25,5.5,6]      &   Sticky/Escaping &Chaotic\\
0.4 &   [4.75,5,5.25,5.5,6] &   Sticky/Escaping &Chaotic\\
0.5 &   [5,5.25,5.5,6]      &   Sticky/Escaping &Chaotic\\
0.6 &   [3,5,5.25,5.5,6]    &   Sticky/Escaping &Chaotic\\
0.7 &   [3.5,5.25,5.5,6]    &   Sticky/Escaping &Chaotic\\
0.8 &   [3,5.25,5.5,6]      &   Sticky/Escaping &Chaotic\\
0.9 &   [3,5.25,5.5,6]      &   Sticky/Escaping &Chaotic\\
1.0 &   [3,5.5]	            &   Sticky/Escaping &Chaotic\\
\hline
\end{tabular}

\section{Table for Orbit Classification for Copenhagen system}
\label{tab:Details05} 
\begin{tabular}{p{0.2cm}p{2.4cm}p{2.05cm}p{2.07cm}}
\hline \textbf{$\epsilon$} & \textbf{$x_0$} & \textbf{Orbit Type} &	\textbf{Poincar\'e Map}\\ \hline
0.0 &   [6.0]               &   Sticky      & Chaotic\\
0.1 &   [5.5,6]             &   Sticky/Escaping & Chaotic\\
0.2 &   [5,5.25,5.5,6]      &   Sticky/Escaping &Chaotic\\
0.3 &   [5,5.25,5.5,6]      &   Sticky/Escaping &Chaotic\\
0.4 &   [5,5.25,5.5,6]      &   Sticky/Escaping &Chaotic\\
0.5 &   [4.25,5,5.25,5.5,6]      &   Sticky/Escaping &Chaotic\\
0.6 &   [4.25,5,5.25,5.5,6]    &   Sticky/Escaping &Chaotic\\
0.7 &   [4.25,5,5.25,5.5,6]    &   Sticky/Escaping &Chaotic\\
0.8 &   [4.25,5,5.25,5.5,6]      &   Sticky/Escaping &Chaotic\\
0.9 &   [3,4.25,5,5.25,5.5,6]      &   Sticky/Escaping &Chaotic\\
1.0 &   [4.25,5,5.25,5.5,6]  &   Sticky/Escaping &Chaotic\\
\hline
\end{tabular}

\section{Table for co-ordinates of the Fixed Points for the Biased-Mass system}
\label{tab:FixedPnts0001}
\begin{tabular}{p{1.5cm}p{5.5cm}}
\hline
$\epsilon$ & Coordinates of the Fixed Points $(x_0,y_0)$ \\ \hline
0.0 & \begin{tabular}[c]{@{}l@{}}$(-1.004,0.0)$\\ $(0.931,0.0)$\\ $(1.07,0.0)$\\ $(0.499,0.866)$\\ $(0.499,-0.866)$\end{tabular} \\ \hline
0.3 & \begin{tabular}[c]{@{}l@{}}$(-1.065,0.0)$\\ $(-0.412,0.0)$\\ $(1.03,0.0)$\\ $(-0.001,0.007)$\\ $(-0.001,-0.007)$\end{tabular} \\ \hline
0.5 & \begin{tabular}[c]{@{}l@{}}$(-0.001,0.001)$\\ $(-0.001,-0.001)$\\ $(-0.888,0.0)$\\ $(-0.829,0.0)$\\ $(-0.741,0.748)$\\ $(-0.741,-0.748)$\\ $(1.021,0.0)$\\ $(1.001,0.0)$\\ $(0.998,0.0)$\end{tabular} \\ \hline
0.7 & \begin{tabular}[c]{@{}l@{}}$(-0.529,0.910)$\\ $(-0.529,-0.910)$\\ $(-0.001,0.005)$\\ $(-0.001,-0.005)$\\ $(1.015,0.0)$\\ $(1.001,0.0)$\end{tabular} \\ \hline
1.0 & \begin{tabular}[c]{@{}l@{}}$(-0.529,-0.910)$\\ $(-0.529,-0.910)$\\ $(-0.001,0.005)$\\ $(-0.001,-0.005)$\\ $(1.015,0.0)$\\ $(1.001,0.0)$\end{tabular} \\ \hline
\end{tabular}

\section{Table for co-ordinates of the Fixed Points for the Copenhagen system}
\label{tab:FixedPnts05}
\begin{tabular}{p{1.5cm}p{5.5cm}}
\hline
$\epsilon$ & Coordinates of the Fixed Points $(x_0,y_0)$\\ \hline
0.0 & \begin{tabular}[c]{@{}l@{}}$(0.0,0.0)$\\ $(-1.198,0.0)$\\ $(1.198,0.0)$\\ $(0.0,-0.866)$\\ $(0.0,0.866)$\end{tabular} \\ \hline 
0.3 & \begin{tabular}[c]{@{}l@{}}$(-1.256,0.0)$\\ $(0.703,0.0)$\\ $(-0.140,0.304)$\\ $(-0.140,-0.304)$\\ $(0.045,0.851)$\\ $(0.045,-0.851)$\\ $(0.299,0.0)$\\ $(0.860,0.548)$\\ $(0.860,-0.548)$\\ $(-0.501,0.019)$\\ $(-0.501,-0.019)$\\ $(0.499,0.110)$\\ $(0.499,-0.110)$\end{tabular} \\ \hline
0.5 & \begin{tabular}[c]{@{}l@{}}$(-1.245,0.0)$\\ $(-0.849,0.0)$\\ $(0.054,0.818)$\\ $(0.054,-0.818)$\\ $(-0.092,0.493)$\\ $(-0.092,-0.493)$\\ $(0.174,0.0)$\\ $(0.643,0.734)$\\ $(0.643,-0.734)$\\ $(-0.500,-0.006)$\\ $(0.500,-0.006)$\\ $(0.500,0.009)$\\ $(0.500,-0.009)$\end{tabular} \\ \hline
0.7 & \begin{tabular}[c]{@{}l@{}}$(-1.139,0.0)$\\ $(-1.073,0.0)$\\ $(-1.149,0.081)$\\ $(-1.149,-0.081)$\\ $(-0.50,0.002)$\\ $(-0.50,-0.002)$\\ $(0.105,0.0)$\\ $(0.509,0.790)$\\ $(0.509,-0.790)$\\ $(0.501,0.044)$\\ $(0.501,-0.044)$\end{tabular} \\ \hline
1.0 & \begin{tabular}[c]{@{}l@{}}$(-1.01,0.497)$\\ $(-1.01,-0.497)$\\ $(-0.5,0.013)$\\ $(-0.5,-0.013)$\\ $(0.063,0.0)$\\ $(0.5,0.020)$\\ $(0.5,-0.020)$\\ $(0.422,0.782)$\\ $(0.422,-0.782)$\end{tabular} \\ \hline
\end{tabular}


\bsp	
\label{lastpage}
\end{document}